\documentclass[a4paper,11pt]{article}
\usepackage{jheppub} % for details on the use of the package, please see the JINST-author-manual
\usepackage{lineno}

\usepackage{tikz} %drawings
\usepackage{tikzscale} %extends the \includegraphics command to support tikzpictures
\usetikzlibrary{patterns} %allows to use patterns in figures
\usepackage[figurename=\sffamily{\bfseries{Figure}}]{caption} %customization of captions
\usepackage{microtype} %subliminal refinements towards typographical perfection

\usepackage{calrsfs}
    	\DeclareMathAlphabet{\mathcal}{OMS}{zplm}{m}{n} 
\usepackage{amsmath}
\usepackage{amssymb}
\usepackage{systeme}
\usepackage{mathtools}
\usepackage{physics}
\usepackage{upgreek}
\usepackage{tensor} %tensor-style super- and subscripts with offsets between successive indices
\usepackage{url} %insert URL links
\usepackage{cancel} %allows for \cancel
\usepackage{cases} %more 'cases' options, allows to separately label equations in the same 'cases' environment
\usepackage{anyfontsize}
\usepackage{wrapfig}
\usepackage{verbatim}
\usepackage{amsfonts}
\usepackage{bm} %bold math symbols
\usepackage{bold-extra} %new bold fonts (es: \texttt in boldface)
\usepackage{multicol} %text on multiple columns
\usepackage{comment}
\usepackage{pgfplots} \pgfplotsset{compat=1.18} %more plot options, allows to use plots from Matlab
\usepackage{titletoc,tocloft} %more customization in table of content
\usepackage{lmodern} %loads the 'Latin Modern Sans Serif' font (used in titles)
\usepackage{titlesec} %allows for customization of titles
	\titleformat*{\section}{\Large\bfseries\sffamily\color{black}}
	\titleformat*{\subsection}{\large\bfseries\sffamily\color{black}}	
\usepackage{standalone} %allows to individually compile chapters, figures, ecc

\DeclareMathAlphabet{\mathbbold}{U}{bbold}{m}{n}

\tikzset{every picture/.style={line width=0.75pt}}

\newcommand{\diff}[2][]{\text{d}^{#1}#2\,}

\title{\boldmath {Quantum JT Gravity in a box as a P\"oschl--Teller Scattering Problem}}

\author[a]{Luca Griguolo,}
\author[b]{Jacopo Papalini,}
\author[c]{Lorenzo Russo,}
\author[c]{Domenico Seminara}
\author[a]{and Alex Tarana}

\affiliation[a]{Dipartimento SMFI, Universit\`a di Parma and INFN Gruppo Collegato di Parma,\\
Viale G. P. Usberti 7/A, 43100 Parma, Italy}

\affiliation[b]{Department of Physics and Astronomy,\\
Ghent University, Krijgslaan 281-S9, 9000 Gent, Belgium}

\affiliation[c]{Dipartimento di Fisica, Universit\`a di Firenze and INFN Sezione di Firenze,\\
Via G. Sansone 1, 50019 Sesto Fiorentino, Italy}

\emailAdd{luca.griguolo@unipr.it}
\emailAdd{jacopo.papalini@ugent.be}
\emailAdd{lorenzo.russo@unifi.it}
\emailAdd{domenico.seminara@unifi.it}
\emailAdd{alex.tarana@unipr.it}

\abstract{We present a canonical quantization of Jackiw--Teitelboim gravity with finite Dirichlet boundary conditions, using the geodesic length between the two boundaries and its conjugate momentum as reduced phase space variables. The dynamics is recast as the scattering problem of a nonrelativistic particle in a repulsive P\"oschl--Teller potential, naturally embedded within a hyperbolic reduction of the $\mathfrak{sl}(2,\mathbb{R})$ Casimir. We obtain exact wavefunctions of the universe and the disk partition function, interpreted as a transition matrix element between states of vanishing bare length. In the asymptotic limit, the theory reduces to Liouville quantum mechanics and reproduces the standard Schwarzian spectral density. At finite cutoff, however, the spectral measure exhibits genuinely nonperturbative corrections, absent in existing $T\bar T$ approaches.
These contributions admit an interpretation in terms
of pinched disk geometries and are responsible for a controlled violation of the Wheeler--DeWitt equation.
We also obtain closed form expressions for thermal two-point functions in terms of Wilson functions and propose diagrammatic rules for time- and out-of-time-ordered four-point functions.
We further address the issue of the branch cut singularity of the quasi-local energy and propose a UV completion of the model in which the Brown--York charge is extended beyond the black hole horizon. This extension naturally generalizes the scattering problem to configurations that foliate the black hole interior.
}

\begin{document}
\maketitle
\flushbottom

\section{Introduction}
\label{sec:introduction}
Two-dimensional dilaton gravity has played a central role in recent progress on quantum gravity.  Its simplest representative, Jackiw--Teitelboim (JT) gravity \cite{Jackiw:1984je,Teitelboim:1983ux}, involves a linear potential and has attracted a considerable amount of attention in recent years thanks to the unprecedented degree of analytical control. The bulk path integral was shown to be dual to the Schwarzian quantum mechanics \cite{Maldacena:2016upp,Jensen:2016pah,Engelsoy:2016xyb,Stanford:2017thb} which describes the reparameterization modes of the asymptotic boundary. This limit has led to an exceptionally detailed understanding of the spectral density, thermal correlation functions, quantum chaos \cite{Maldacena:2015waa}, and the relation among JT gravity, the low-energy regime of the SYK model \cite{Sachdev:1992fk,Maldacena:2016hyu,Kitaev:2017awl}, and random matrix theory \cite{Saad:2019lba,Stanford:2019vob}.  Much less is known, however, about the same theory when the gravitational boundary is not pushed to infinity.  In that case the boundary conditions are imposed at finite value of the dilaton and the system should be thought of as dilaton gravity in a box \cite{Iliesiu:2020zld,Goel:2020yxl}.

Finite cutoff gravity is interesting for several related reasons.
First, it provides a controlled way to ask which properties of the asymptotic theory are robust and which ones depend on having sent the boundary to infinity. Indeed, from the bulk perspective, finite Dirichlet walls give access to gravitational dynamics at shorter distance scales than those captured by the asymptotic theory. Second, it probes a regime in which the geodesic length between the two boundaries is an actual finite observable and no renormalization is required.  Third, it is the natural two-dimensional analogue of the finite-radius constructions that underlie many discussions of holography at finite cutoff and of irrelevant deformations of boundary dynamics.
A common strategy is to start from the Schwarzian theory and deform it, in close analogy with the role of $T\bar T$ deformations in higher-dimensional examples \cite{Zamolodchikov:2004ce,Smirnov:2016lqw,Cavaglia:2016oda,McGough:2016lol,Gross:2019ach}\footnote{Other interesting $T\bar{T}$-inspired works in this context are \cite{Ebert:2022ehb, AliAhmad:2025kki, Aguilar-Gutierrez:2024nst, Callebaut:2025thw, Blacker:2024rje, Morone:2024ffm,Aguilar-Gutierrez:2024oea,Aguilar-Gutierrez:2026ogo}, while a different bulk approach is pursued in \cite{Ferrari:2024kpz,Chaudhuri:2024yau}.}.
This point of view is useful, but it has the limitation of treating the finite cutoff theory as a deformation of the asymptotic Hilbert space, instead of quantizing the finite cutoff gravitational phase space directly \cite{Harlow:2018tqv,Iliesiu:2019xuh,Blommaert:2018oro}.

The purpose of this paper is to follow the second route.
We perform a canonical quantization of JT gravity with finite Dirichlet boundary conditions. The reduced phase space is described by two diffeomorphism-invariant variables: the length $L$ of a spacelike geodesic connecting the two boundaries and its conjugate momentum $P$.  The finite cutoff Hamiltonian can be written as
\begin{equation}\label{eq:intro_phih}
    H(L,P) = \Phi_b - \sqrt{\Phi_b^2 - \Phi_h^2(L,P)}, \qquad \text{where} \qquad
    \Phi_h^2(L,P)=P^2+\frac{\Phi_b^2}{\cosh^2(L/2)}.
\end{equation}
Here $\Phi_b$ is the boundary value of the dilaton and $\Phi_h$ is the value of the dilaton at the horizon.
Thus, although the Brown--York energy involves a square root, the basic spectral problem is that of a particle scattering off a repulsive P\"oschl--Teller potential.
This type of potential has already appeared in relation to the dynamics on AdS$_2$ backgrounds in the literature \cite{Correa_2009}.
The ordinary Liouville quantum mechanics of asymptotic JT gravity is recovered by taking the large cutoff limit while keeping the renormalized length fixed \cite{Mertens:2017mtv,Mertens:2018fds}.

Quantizing the P\"oschl--Teller problem \eqref{eq:intro_phih} leads to an ordering ambiguity. The most direct ordering gives a well-defined scattering problem, but its spectral density contains features whose origin is not especially transparent.
We therefore focus on a second ordering which is distinguished by representation theory. Upon setting $\hbar = 1$ and introducing the variables
\begin{equation}
  s \equiv \Phi_h, \qquad \nu \equiv \Phi_b,
\end{equation}
this ordering gives the Schr\"odinger equation
\begin{equation}
  -\frac{\dd ^2}{\dd L^2} \psi_s(L)
  + \left( \frac{\nu^2}{\cosh^2(L/2)}
  -\frac{1}{4 \sinh^2(L)} \right) \, \psi_s(L)
  =
  s^2 \, \psi_s(L).
  \label{eq:intro_quantum_PT}
\end{equation}
The second term in the potential vanishes in the classical limit and is therefore a genuine quantum correction. We choose the Friedrichs self-adjoint extension \cite{ReedSimonII,Case:1950an}, for which the physical scattering states behave as $\psi_s(L)\sim \sqrt{L/2}$ near the origin and no bound states are present.
The resulting wavefunctions are hypergeometric functions and are normalized with the measure
\begin{equation}
   \dd \mu(s)=\frac{2s\sinh(2\pi s)}{\cosh(2\pi s)+\cosh(2\pi\nu)}\, \dd s.
  \label{eq:intro_measure}
\end{equation}

We propose a prescription to compute the disk partition function as the open-channel amplitude
\begin{equation}\label{intro:partition}
    Z(\beta) = \langle L=0|e^{2 \pi \nu - \beta \hat{H}}|L=0\rangle,
\end{equation}
where $\beta$ corresponds to the inverse Tolman temperature \cite{McGough:2016lol}.
As a result, the partition function can be written as an integral over the spectral parameter $s$
\begin{equation}\label{eq:gravenergy}
    Z(\beta) = e^{2 \pi \nu} \int_0^\nu \text{d}\mu(s) \, e^{-\beta E(s)}, \qquad\text{where}\qquad E(s)=
    \nu-\sqrt{\nu^2-s^2},
    \qquad 0<s<\nu,
\end{equation}
and the energy spectrum matches the perturbative solution of the dimensionally reduced $T\bar T$ flow equation \cite{Gross:2019ach}. 

In implementing the prescription \eqref{intro:partition}, we can choose to restrict the spectral parameter to the range $0<s<\nu$, which corresponds to retaining only configurations $\Phi_h<\Phi_b$ where the finite cutoff boundary remains outside the black hole horizon.  
In the large cutoff limit, with $\beta_{\rm JT} = \beta/\Phi_b$ fixed, the quasi-local energy reduces to the large cutoff JT gravity expression, $E_{\mathrm{JT}}=s^2/2$, while the spectral density associated with \eqref{eq:intro_measure} reduces to the familiar JT/Schwarzian density proportional to $s\sinh(2\pi s)$.  At finite cutoff, however, the denominator in \eqref{eq:intro_measure} modifies the spectral density by contributions that are nonperturbative in the cutoff parameter. This constitutes one of the main differences between the direct canonical quantization developed here and approaches based on deforming the asymptotic Schwarzian theory.

This difference is also the main lesson of our comparison with previous finite cutoff proposals \cite{Iliesiu:2020zld,Griguolo:2021wgy}.
A $T\bar T$-inspired deformation of Schwarzian quantum mechanics changes the relation between the actual energy and the undeformed one, but it leaves the asymptotic eigenstates and spectral density essentially unchanged.
By contrast, a boundary at finite distance changes the Hilbert space since, for instance, the state $|L_{\rm ren}=-\infty\rangle$ used in the asymptotic computation is not part of the Hilbert space at finite cutoff.  This distinction is invisible in perturbation theory around the large cutoff limit, but it affects the nonperturbative spectral measure. Our result nevertheless exhibits the expected semiclassical behavior.

We also study matter correlation functions in this canonical language. They are computed as matrix elements of the form
\begin{equation}
    \bra{L=0} e^{-(\beta-\tau)\hat{H}} G_\Delta(\hat{L}) e^{-\tau\hat{H}} \ket{L=0},
\end{equation}
where the insertion consists of the operator version of the AdS$_2$ scalar propagator
\begin{equation}
    G_\Delta(\hat{L}) = \left( \frac{\Phi_b}{\cosh\!{(\hat{L}/2)}} \right)^{2\Delta} {}_2F_1\!\left( \Delta, \Delta, 2\Delta; \frac{1}{\cosh^2\!{(\hat{L}/2)}} \right).
\end{equation}
This operator reduces, after the usual large cutoff length renormalization, to the standard JT geodesic operator $e^{-\Delta \hat{L}_{\rm ren}}$.
Its matrix elements between $L=0$ states can be evaluated exactly and are expressed in terms of Wilson functions \cite{Groenevelt:2003,Groenevelt:2005}.  This leads to integral expressions for thermal two-point functions and to diagrammatic rules for higher-point functions. 

We further examine the semiclassical limit of the thermal two-point function and show that it localizes on the boundary-to-boundary geodesic length in the Euclidean hyperbolic disk at finite cutoff. To the best of our knowledge, this is the first proposal in the literature that not only admits a closed-form expression but also reproduces the expected semiclassical behavior. Finally, we propose finite cutoff expressions for time-ordered and out-of-time-ordered four-point functions, where the latter involves the same fusion kernel that appears in the asymptotic JT analysis \cite{Mertens:2017mtv,Blommaert:2018oro}.

The representation-theoretic origin of the construction is clarified by performing a hyperbolic reduction of the $\mathfrak{sl}(2,\mathbb R)$ Casimir.
We parameterize group elements locally as
\begin{equation}
    g = e^{aX} e^{L H} e^{bX}, \qquad X = \frac{E+F}{2},
\end{equation}
and fix the left and right eigenvalues associated with the hyperbolic generator $X$.
The radial part of the Casimir, after conjugation to the flat measure $\text{d}L$, is precisely the quantum P\"oschl--Teller Hamiltonian \eqref{eq:intro_quantum_PT}.
This provides a natural explanation for the ordering used above.
% It also explains how ordinary JT gravity emerges at large cutoff: the hyperbolic reduction degenerates into the familiar parabolic reduction of the Casimir \cite{Blommaert:2018oro,Iliesiu:2019xuh,Mertens:2018fds}, and the P\"oschl--Teller scattering problem becomes the Liouville scattering problem \cite{Mertens:2017mtv,Mertens:2018fds}.

Finally, we address the issue associated with the presence of a square root in the Hamiltonian \eqref{eq:intro_phih}, which would become complex when the finite cutoff boundary reaches the black hole horizon $\Phi_h > \Phi_b$.
This pathological behavior is the same as that found in $T\bar{T}$-deformed spectra \cite{Cavaglia:2016oda,Smirnov:2016lqw}.
As anticipated, a way to avoid this regime is to restrict the allowed energies by hand and only retain states that satisfy $\Phi_h < \Phi_b$.
Nevertheless, this approach might seem ad hoc, and 
% While this restriction still leads to a well-defined partition function, it comes at a cost:  the sharply localized length state $|L=0\rangle$, which appears in the standard disk-state prescription, ceases to admit a transparent geometric interpretation.
we therefore propose an alternative UV completion of the theory
% One that appears natural from the perspective of the Brown--York quasi-local charge. Rather than excluding configurations for which the finite cutoff boundary crosses the black hole horizon,
that consists of defining a Hamiltonian with two branches
\begin{equation}\label{eq:introbranches}
H(\Phi_h) = 
\begin{cases} 
    \Phi_b-\sqrt{\Phi_b^2-\Phi_h^2}, &\qquad \Phi_h<\Phi_b,\\
    \Phi_b+\sqrt{\Phi_h^2-\Phi_b^2}, &\qquad \Phi_h>\Phi_b. 
\end{cases} 
\end{equation}
The second branch corresponds to configurations with $\Phi_h>\Phi_b$, for which the cutoff curve lies behind the black hole horizon. To describe this regime, we continue to foliate the Penrose diagram by horizontal Cauchy slices and the resulting expression of $\Phi_h^2(L,P)$ in terms of their length is once again given by a P\"oschl--Teller Hamiltonian.

Crucially, the cutoff boundary becomes a spacelike curve in this regime and therefore the Brown--York charge must be defined for a surface with timelike normal.
% and we interpret the second branch in \eqref{eq:introbranches} as the continuation of the same Lorentzian Brown--York energy into the interior region.
Another consequence is that the evolution generated by the quasi-local charge becomes Euclidean for these configurations.
This prescription provides a natural extension of the finite cutoff theory beyond the horizon and establishes a connection with the framework of the $T\bar{T}+\Lambda$ deformation \cite{AliAhmad:2025kki, Gorbenko:2018oov,Torroba:2022jrk, Chang:2025ays, Shyam:2021ciy}.

The results of this paper should be viewed as the first step in a broader program. The canonical quantization method used here
%---identify finite cutoff gauge-invariant phase space variables, reduce the gravitational dynamics to a one-dimensional scattering problem, quantize the corresponding Hamiltonian, and compute disk and matter amplitudes from length-state matrix elements---
is not tied in principle to the linear JT dilaton potential.
In a companion paper \cite{futurepaper} we will extend this approach to sinh and sine dilaton gravities \cite{Blommaert:2023wad,Blommaert:2024whf,Bossi:2024tvh}.
These models provide natural deformations of the JT potential and should give a useful testing ground for the universality, or non-universality, of some features of the theory such as the spectral measure at finite cutoff.
They may also clarify which aspects of the present construction are special to the $\text{SL}(2,\mathbb R)$ structure of JT gravity and which persist for more general dilaton potentials \cite{Witten:2020ert,Blommaert:2023wad}.

Let us finally summarize the organization of the paper.
In Section \ref{sec:classical_JT_gravity} we review the classical finite cutoff JT solutions, we derive the reduced Hamiltonian in terms of the geodesic length and its conjugate momentum and reformulate the evolution of the universe in terms of a P\"oschl--Teller scattering problem.
Section \ref{sec:canonical_quantization} is devoted to the quantization of the associated P\"oschl--Teller quantum mechanics, discussing the ordering ambiguity and the choice of self-adjoint extension.
In Section \ref{sec:partition_function} we compute the partition function of the theory and analyze its features in detail.
Next, in Section \ref{sec:correlation_functions}, we turn to matter correlation functions and formulate diagrammatic rules for higher-point functions at finite cutoff. In Section \ref{sec:grouptheory} we explain the relation between the preferred operator ordering of the Hamiltonian and the hyperbolic reduction of the $\text{SL}(2,\mathbb R)$ Casimir.
In Section \ref{subsec:boundary_behind_horizon} we discuss the consequences of a sharp cutoff in the energy spectrum and propose a UV completion of the theory that includes configurations where the finite cutoff boundary can cross the black hole horizon.  

Finally, in Section \ref{sec:conclusions} we compare the resulting finite cutoff theory with $T\bar T$-inspired deformations and with other approaches to JT gravity with non-asymptotic boundaries. We also add some more speculative remarks and comment on related future directions.

% The main conclusion is that finite cutoff JT gravity is not obtained simply by deforming the Schwarzian energy spectrum: the finite wall modifies the Hilbert space and, nonperturbatively in the cutoff, the spectral density itself.

\section{Classical JT gravity}\label{sec:classical_JT_gravity}
As the main goal of the first part of the paper is to perform a Hamiltonian quantization of JT gravity, we begin by quickly reviewing the classical theory and by highlighting the major results that will be needed in what follows.

\subsection{Geometry of spacetime}\label{sec:geometry_of_spacetime}

The Lorentzian action of JT gravity is given by 
\begin{equation}\label{eq:JT_action_lorentzian}
	S_\text{JT}[g,\Phi] = \frac{1}{16 \pi G_\text{N}} \left[ \int_\mathcal{M} \diff[2]{x} \sqrt{-g} \, \Phi \left( R + \frac{2}{\ell^2} \right) + 2 \int_{\partial\mathcal{M}} \dd y \sqrt{\abs{h}} \, \Phi \left( K - \frac{1}{\ell} \right) \right],
\end{equation}
where $\ell$ is the length scale of the theory, $\mathcal{M}$ is a two-dimensional manifold, and $\partial\mathcal{M}$ is its one-dimensional boundary. In the following, we will choose units such that $\ell=1$ and $8 \pi G_\text{N}=1$, and we will reintroduce them explicitly only when needed. In the action \eqref{eq:JT_action_lorentzian}, we recognize two fields: the metric tensor $g_{\mu \nu}$ and a real scalar field $\Phi$, the dilaton.
The quantity $K$ that appears in the boundary term is the trace of the extrinsic curvature of $\partial\mathcal{M}$ and makes the variational principle well defined when Dirichlet boundary conditions are imposed on the fields. It is defined as
\begin{equation}
    K \equiv h^{\mu\nu} \nabla_\mu n_\nu,
\end{equation}
where $n_\nu$ is the unit vector normal to $\partial\mathcal{M}$, which, in our convention, is taken to be outward-pointing.

The equations of motion of the theory are
\begin{equation}\label{eq:eom_dilaton}
	R = - 2, \qquad \tensor{\nabla}{_\mu}\tensor{\nabla}{_\nu} \Phi = \tensor{g}{_\mu_\nu} \Phi,
\end{equation}
and we supplement them with Dirichlet boundary conditions for both fields
\begin{equation}\label{eq:boundary_conditions}
        \Phi\big\lvert_{\partial\mathcal{M}} = \Phi_b, \qquad g_{\mu \nu}\, \dd x^{\mu } \dd x^{\nu}\big\lvert_{\partial\mathcal{M}} = \rm const.
\end{equation}
In the ordinary JT case one takes both $\Phi_b$ and $\sqrt{-g\lvert_{\partial\mathcal{M}}}$ to be $\sim 1/\varepsilon$ and studies the theory in the $\varepsilon \to 0$ limit. Here, instead, we keep $\Phi_b$ finite and study JT gravity for generic values of the boundary dilaton and metric. 

As a remark, we notice that the contribution $\sim 1/\ell$ in the boundary term of the action \eqref{eq:JT_action_lorentzian} is a holographic counterterm needed to render the on-shell JT gravity action finite when $\varepsilon \to 0$. Therefore, such a contribution to the action is only strictly needed in that limit. Nevertheless, we are still allowed to add such a term to the action in order for the resulting theory to be continuously related to ordinary JT gravity when $\varepsilon \to 0$.

The classical dilaton and metric configurations are obtained by solving the equations \eqref{eq:eom_dilaton} and, in the Schwarzschild gauge, read\footnote{The boundary condition on the metric \eqref{eq:boundary_conditions} in the Schwarzschild gauge we adopt here reads
\begin{equation}\label{eq:boundary_condition_metric}
    g \big\lvert_{\partial\mathcal{M}} = - (\Phi_b^2 - \Phi_h^2) \, \dd t^2,
\end{equation}
where $\Phi_h$ is some real parameter that we recognize to be the location of the black hole horizon. Here we are implicitly assuming that the boundary data $(\Phi_h,\Phi_b)$ are such that $\Phi_h < \Phi_b$. Nevertheless, in what follows we will be able to relax this assumption.}
\begin{equation}\label{eq:classical_configuration_schwarzschild}
    \dd s^2 = -(r^2-\Phi_h^2) \, \dd t^2 + \frac{1}{r^2-\Phi_h^2} \, \dd r^2,
    \qquad
    \Phi = r,
    \qquad \text{where} \qquad
    \begin{cases}
        \Phi_h<r<\Phi_b,\\
        -\infty<t<+\infty,
    \end{cases}
\end{equation}
where $r=\Phi_b$ is the position of the boundary curve and $r=\Phi_h$ is the location of the black hole horizon.
The two parameters $\Phi_h$ and $\Phi_b$ are fixed by the choice of Dirichlet boundary conditions \eqref{eq:boundary_conditions}.
As usual, the spacetime described by \eqref{eq:classical_configuration_schwarzschild} is locally hyperbolic, but differently from the ordinary case, now it has a boundary at finite distance in correspondence of the curve $(t,r=\Phi_b)$.

Of course, Schwarzschild coordinates are not suited to describe the classical spacetime beyond the horizon. That is why it will be particularly useful to introduce the global coordinates $(\nu,\sigma)$ defined as
\begin{equation}\label{eq:map_BH_exterior}
    \begin{split}
        \frac{1}{\tan^2\!\sigma} &= \frac{r^2-\Phi_h^2}{\Phi_h^2} \cosh^2(\Phi_h t),\\
        \tan^2\!\nu &= \frac{r^2-\Phi_h^2}{r^2} \sinh^2(\Phi_h t),
    \end{split}
\end{equation}
in terms of which the classical metric and dilaton take the form
\begin{equation}\label{eq:classical_configuration_global}
    \dd s^2 = \frac{-\dd\nu^2+\dd\sigma^2}{\sin^2\!\sigma},
    \qquad
    \Phi = \Phi_h \frac{\cos \nu}{\sin \sigma},
    \qquad \text{where} \qquad
    \begin{cases}
        0<\sigma<\pi,\\
        \displaystyle{-\frac{\pi}{2}<\nu<\frac{\pi}{2}},
    \end{cases}
\end{equation}
where the range of the global time $\nu$ descends from requiring the positivity of the dilaton field $\Phi \ge 0$.
We mention that starting from the global solutions \eqref{eq:classical_configuration_global} and performing the coordinate transformation
\begin{equation}\label{eq:map_BH_interior}
    \begin{split}
        \frac{1}{\tan^2\!\sigma} &= \frac{\Phi_h^2-r^2}{\Phi_h^2} \sinh^2(\Phi_h t),\\
        \tan^2\!\nu &= \frac{\Phi_h^2-r^2}{r^2} \cosh^2(\Phi_h t),
    \end{split}
\end{equation}
one is able to cover the black hole interior, namely the region where $r<\Phi_h$, with a metric in the Schwarzschild coordinates
\begin{equation}\label{eq:classical_metric_BH_interior}
\dd s^2 = (\Phi_h^2 - r^2) \, \dd t^2 - \frac{1}{\Phi_h^2 - r^2} \, \dd r^2,
    \qquad
    \Phi = r,
    \qquad \text{where} \qquad
    \begin{cases}
        0 < r < \text{min}(\Phi_b,\Phi_h),\\
        -\infty < t < +\infty.
    \end{cases}
\end{equation}
The metric \eqref{eq:classical_metric_BH_interior} is suitable to describe the classical solutions also when the boundary data are such that $\Phi_h > \Phi_b$, as the boundary consists of the curve $(t,r=\Phi_b < \Phi_h)$ in that case.
Here the role of the coordinates is interchanged: the $t$-direction becomes spacelike, whereas the $r$-direction becomes timelike.
\begin{figure}
    \centering
    \includegraphics[width=\linewidth]{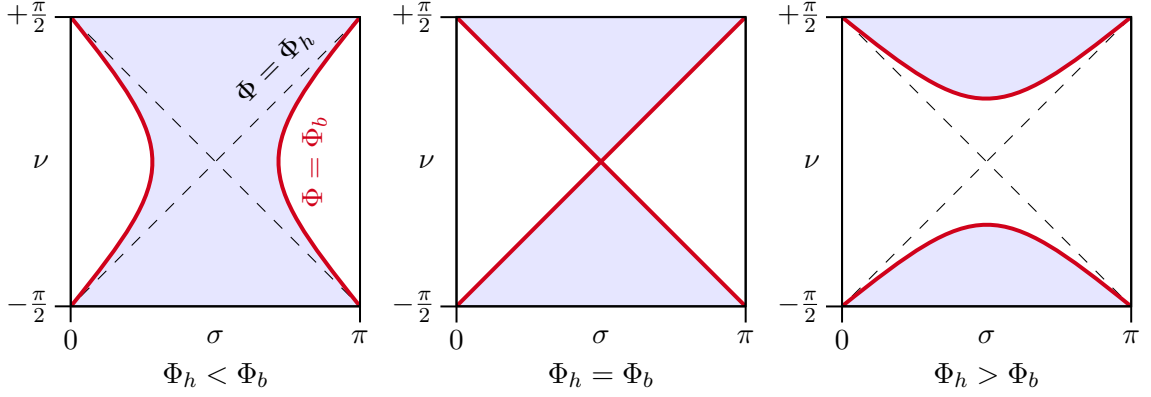}
    \caption{The Penrose diagram for the maximally extended JT gravity spacetime \eqref{eq:classical_configuration_global}. The shaded region is the spacetime bulk defined by the condition $0 \leq \Phi \leq \Phi_b$. The red lines represent the boundary curves at $\Phi = \Phi_b$ and the dashed lines correspond to the black hole horizons at $\Phi = \Phi_h$.}
    \label{fig:Penrose_diagram}
\end{figure}
The global metric \eqref{eq:classical_configuration_global} is conformally flat, so it is straightforward to draw the Penrose diagram for the maximally extended classical JT spacetime, as depicted in Figure \ref{fig:Penrose_diagram}. From the global expression of the dilaton \eqref{eq:classical_configuration_global} we learn that the boundary curves obeying \eqref{eq:boundary_conditions} are qualitatively different depending on the hierarchy between the boundary data $\Phi_h$ and $\Phi_b$:
\begin{itemize}
    \item $\Phi_h < \Phi_b$ (Figure \ref{fig:Penrose_diagram}, left panel): the boundary consists of two disconnected timelike pieces, and the spacetime region that interpolates between them includes an external region, a black hole horizon and a black hole interior.
    \item $\Phi_h = \Phi_b$ (Figure \ref{fig:Penrose_diagram}, central panel): the boundary curves are null curves.
    \item $\Phi_h > \Phi_b$ (Figure \ref{fig:Penrose_diagram}, right panel): the boundary consists of two disconnected spacelike pieces and the spacetime itself consists of two disconnected regions that are fully beyond the black hole horizon.
\end{itemize}
In principle, all three cases discussed above are allowed. Nevertheless, throughout most of this paper we will focus on spacetimes such that $\Phi_h < \Phi_b$. A discussion on the case $\Phi_h > \Phi_b$ is presented in Section \ref{subsec:boundary_behind_horizon}.

\subsection{JT gravity as a scattering problem}\label{sec:JT_scattering}
Having discussed the classical solutions of the theory, we now turn to the dynamics of JT gravity. The main goal of this section is to formulate JT gravity in a box as a one-dimensional Hamiltonian system, along the lines of the large cutoff case studied in \cite{Harlow:2018tqv}. This reformulation is expected to be possible because the phase space of ordinary JT gravity is two-dimensional \cite{Louis-Martinez:1993bge}, and this conclusion is not affected by the finite cutoff boundary conditions considered here.
Mimicking what the authors of \cite{Harlow:2018tqv} did in the large cutoff setting, we will regard $\Phi_b$ as a datum of the theory, whereas $\Phi_h$, which is determined by the Dirichlet boundary condition on the metric \eqref{eq:boundary_condition_metric}, will be seen as a quantity that labels the on-shell geometries depicted in Figure \ref{fig:Penrose_diagram}. The difference with respect to \cite{Harlow:2018tqv} consists of $\Phi_b$ being finite for us.

Specifically, we are now going to establish a correspondence between the on-shell geometries discussed previously and the classical trajectories in the phase space of the Hamiltonian system. This correspondence establishes an equivalence between the two theories, at least at the classical level. The advantages of working with a one-dimensional Hamiltonian mechanics are twofold: we know how to handle its quantization and it exhibits no gauge redundancies.

\begin{figure}
    \centering
    \includegraphics[width=0.325\linewidth]{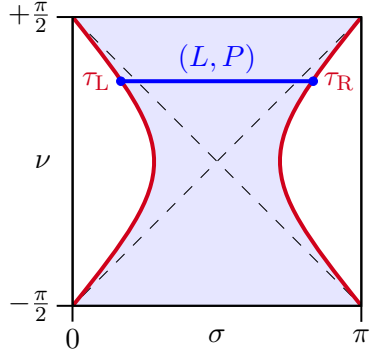}
    \caption{The Penrose diagram for some value of $\Phi_h$ such that $\Phi_h < \Phi_b$. The blue curve is a Cauchy slice on which we specify the state of the system. The endpoints of the slice touch the two boundaries at points labeled by the proper times $\tau_\text{L}$ and $\tau_\text{R}$.}
    \label{fig:Penrose_diagram_exterior}
\end{figure}

To construct the dual model, we first foliate the Penrose diagram in Figure \ref{fig:Penrose_diagram} by one-dimensional Cauchy slices and identify each of them with the state of the Hamiltonian system at a given time. For convenience, we choose these slices to be horizontal boundary-to-boundary spacelike geodesics as depicted in Figure \ref{fig:Penrose_diagram_exterior}.
%As it will be useful in the following, we are also interested in horizontal geodesics of the spacetime that connect the two boundaries in the case $\Phi_h < \Phi_b$ see Figure \ref{fig:Penrose_diagram}.
Guided by diffeomorphism invariance, we consider the proper times along the left and right boundary curves,
\begin{equation}\label{eq:proper_time}
    \tau_{\text{L}/\text{R}}
    \equiv
    \sqrt{\Phi_b^2-\Phi_h^2} \, t_{\text{L}/\text{R}},
\end{equation}
and we choose, as the time parameter on phase space that labels the geodesics, the quantity $\tau \equiv \tau_\text{L} + \tau_\text{R}$.
The left and right Hamiltonians associated with the evolution in $\tau_\text{L}$ and $\tau_\text{R}$ are naturally identified with the conserved charges associated with translations  generated by the Killing vector fields
\begin{equation}\label{eq:Killing_vector_field}
    \partial_{\tau_{\text{L}/\text{R}}}
    =
    \frac{\partial_{t_{\text{L}/\text{R}}}}{\sqrt{\Phi_b^2 - \Phi_h^2}}.
\end{equation}
The Hamiltonians $H_\text{L}$ and $H_\text{R}$ are obtained by contracting  the Brown--York stress-energy tensor \cite{Brown:1992br} evaluated at the respective boundaries with \eqref{eq:Killing_vector_field}.\footnote{We recall that for a timelike boundary
\[
    T^{\text{L}/\text{R}}_{\alpha\beta}
    \equiv
    - \frac{2 }{\sqrt{\abs{h}}}
    \frac{\delta S_{\text{JT}}}{\delta h^{\alpha\beta}} \bigg\lvert_{\partial\mathcal{M}_{\text{L}/\text{R}}}
        =
        \left( n^\mu \nabla_\mu \Phi - \Phi \right)\bigg\lvert_{\partial\mathcal{M}_{\text{L}/\text{R}}}
        h_{\alpha\beta}
        =
        - H_{\text{L}/\text{R}} h_{\alpha\beta},
\]
where $n^\mu$ was taken to be the outward-pointing unit normal.}
% \textcolor{red}{Here we choose to define to Brown-York tensor with an opposite sign depending on the signature of the boundary. How legal is this choice? Were we not to flip the sign, the partition function would not localize on an Euclidean on-shell disk geometry in the semiclassical limit, as discussed in Appendix \ref{app:???}.}}.
% \begin{equation}\label{eq:Brown_York_tensor}
%         T^{\L/\R}_{\alpha\beta}
%         \equiv
%         n^2 \frac{2}{\sqrt{\abs{h}}}
%         \frac{\delta S_{\text{JT}}}{\delta h^{\alpha\beta}}
%         =
%         -n^2 \left( n^\mu \nabla_\mu \Phi - \Phi \right)\bigg\lvert_{\partial\mathcal{M}}
%         h_{\alpha\beta}
%         =
%         \begin{cases}
%             \left( +\sqrt{\Phi_b^2 - \Phi_h^2} - \Phi_b \right) h_{\alpha\beta},
%             &\qquad \Phi_h < \Phi_b,\\
%             \left( +\sqrt{\Phi_h^2 - \Phi_b^2} + \Phi_b \right) h_{\alpha\beta},
%             &\qquad \Phi_h > \Phi_b,
%         \end{cases}
% \end{equation}
% where $n^\mu$ was taken to be outward-pointing when $\Phi_h < \Phi_b$ and inward-pointing when $\Phi_h > \Phi_b$.
% By contracting \eqref{eq:Brown_York_tensor} with \eqref{eq:Killing_vector_field} twice, 
One obtains
\begin{equation}\label{eq:conserved_charge}
    H_\text{L} = H_\text{R}
    \equiv
    \Phi_b - \sqrt{\Phi_b^2 - \Phi_h^2}.
    % \\
    % &=
    % \Phi_b
    % -
    % \operatorname{sign}(\Phi_b^2-\Phi_h^2)
    % \sqrt{\abs{\Phi_b^2-\Phi_h^2}}.
\end{equation}
Therefore, the evolution in $\tau$ is described by the Hamiltonian\footnote{We note that our conventions for \(\tau\) and \(H\) differ slightly from
those of Ref.~\cite{Harlow:2018tqv}. These conventions are chosen so that the
partition function obtained by using the Hamiltonian \eqref{eq:hamiltonian} can be
interpreted as the gravitational path integral on a disk with boundary length
\(\beta\).}
\begin{equation}\label{eq:hamiltonian}
    H(\Phi_h)
    \equiv
    \frac{H_\text{L} + H_\text{R}}{2} = \Phi_b - \sqrt{\Phi_b^2 - \Phi_h^2}.
    % =
    % \begin{cases}
    %     \Phi_b - \sqrt{\Phi_b^2 - \Phi_h^2},
    %     &\qquad \Phi_h < \Phi_b,\\
    %     \Phi_b + \sqrt{\Phi_h^2 - \Phi_b^2},
    %     &\qquad \Phi_h > \Phi_b.
    % \end{cases}
\end{equation}
In \eqref{eq:hamiltonian} the Hamiltonian depends only on the conserved quantity
\(\Phi_h\) and on the boundary datum \(\Phi_b\). In terms of the canonical pair
\((\tau,H)\), with \(\{\tau,H\}=1\), the phase-space dynamics is therefore trivial:
\(H\) is constant along each orbit and \(\tau\) merely labels the position along it.
Before turning to the quantization of the theory, it is useful to replace
\((\tau,H)\) with a more convenient set of phase-space coordinates, in which the
Hamiltonian becomes a non-trivial function on phase space. Again
guided by diffeomorphism invariance, we consider as the first variable the
length \(L\) of the Cauchy surfaces shown in Figure \ref{fig:Penrose_diagram_exterior}. These geodesics have a very natural parameterization in terms of global coordinates since they are curves of constant $\nu$, and their length in the metric \eqref{eq:classical_configuration_global} can be computed as
\begin{equation}\label{eq:geodesic_length}
    L(\tau,\Phi_h) = 2 \int_{\sigma_b}^{\pi/2} \frac{\dd\sigma}{\sin \sigma} = 2 \sinh^{-1}\!\left[\frac{1}{\tan \sigma_b}\right]
    = 2 \sinh^{-1}\!\left[ \frac{\sqrt{\Phi_b^2 - \Phi_h^2} }{\Phi_h} \cosh\!\left( \frac{\Phi_h}{\sqrt{\Phi_b^2-\Phi_h^2}} \frac{\tau}{2} \right) \right],
\end{equation}
where $\tan \sigma_b$ is given by evaluating the map \eqref{eq:map_BH_exterior} at $r=\Phi_b$ and, for horizontal geodesics, $t_\text{L} = t_\text{R} = \frac{\tau}{2 \sqrt{\Phi_b^2 - \Phi_h^2}}$.

\begin{figure}
    \centering
    \includegraphics{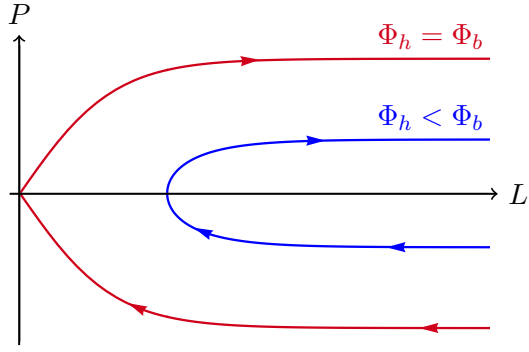}
    \caption{The classical phase space of the Hamiltonian system. The displayed curves are solutions of Equation \eqref{eq:P_computation} with $\Phi_h < \Phi_b$ (blue) and $\Phi_h = \Phi_b$ (red). These curves can be thought of as the trajectories dual to the classical geometries on the left and at the center of Figure \ref{fig:Penrose_diagram}.}
    \label{fig:phase_space}
\end{figure}

The second phase space variable, $P$, is the conjugate momentum to $L$ and it is readily found by recalling that Hamilton's equation must hold for classical trajectories\footnote{Equivalently, we choose the symplectic form of the reduced model to be
\begin{equation*}\label{eq:symplectic_form}
    \omega = \dd\tau \wedge \dd H = \dd L \wedge \dd P.
\end{equation*}
}
\begin{equation}\label{eq:Hamilton_equation}
    \dv{L}{\tau}
    =
    \pdv{H}{P}
    =
    \dv{H}{\Phi_h}
    \pdv{\Phi_h}{P}.
\end{equation}
Since all quantities in \eqref{eq:Hamilton_equation} are known except for $P$, this relation can be inverted to obtain
\begin{equation}\label{eq:P_computation}
        \Phi_h^2
        =
        P^2
        +
        \frac{\Phi_b^2}{\cosh^2\!{(L/2)}}.
\end{equation}
Equation \eqref{eq:P_computation} allows us to draw the phase space trajectories of the Hamiltonian system for different values of $\Phi_h$, see Figure \ref{fig:phase_space}. Each of these trajectories corresponds to a classical geometry in Figure \ref{fig:Penrose_diagram_exterior}.
Plugging the expression of $L(\tau,\Phi_h)$ into \eqref{eq:P_computation} yields the following expression of $P$ as a function of $\tau$
\begin{equation}\label{eq:momentum}
    P(\tau,\Phi_h)=
    \frac{\Phi_h \sqrt{\Phi_b^2 - \Phi_h^2} \sinh\!\left( \frac{\Phi_h}{\sqrt{\Phi_b^2 - \Phi_h^2}} \frac{\tau}{2} \right)}{\sqrt{\Phi_b^2 + (\Phi_b^2 - \Phi_h^2) \sinh^2\!\left( \frac{\Phi_h}{\sqrt{\Phi_b^2 - \Phi_h^2}} \frac{\tau}{2} \right)}}.
\end{equation}
Finally, using the expression for $\Phi_h^2$ in terms of $H$ \eqref{eq:hamiltonian} and inverting Equation \eqref{eq:P_computation}, we find
\begin{equation}\label{eq:hamiltonian_LP}
    H(L,P)
    =
        \Phi_b
        -
        \sqrt{
        \Phi_b^2
        -
        \bigg(
        P^2
        +
        \frac{\Phi_b^2}{\cosh^2\!{(L/2)}}
        \bigg)}.
\end{equation}
Under the square root in the Hamiltonian \eqref{eq:hamiltonian_LP} we recognize a one-dimensional P\"oschl--Teller potential. The analysis and quantization of the latter will be one of the main themes of the present paper.

Before turning to the quantization of \eqref{eq:hamiltonian_LP}, let us comment on how ordinary large cutoff JT gravity is recovered within this framework. In what follows, we will refer to the limit where
\begin{equation}\label{eq:large_cutoff_limit}
    \Phi_b \to \infty
\end{equation}
and the renormalized length\footnote{This regularization is motivated by the behavior of \eqref{eq:geodesic_length} as $\Phi_b \to \infty$. Subtracting $2\log(2\Phi_b)$ from $L$ isolates the finite renormalized length $L_{\text{ren}}$. This is the same regularization procedure adopted in \cite{Harlow:2018tqv}.}
\begin{equation}\label{eq:geodesic_length_renormalized}
    L_{\text{ren}}
    \equiv
    L - 2\log(2\Phi_b)
\end{equation}
is kept finite as the large cutoff limit of the theory.
In this limit, the Hamiltonian \eqref{eq:hamiltonian_LP} reduces to
\begin{equation}\label{eq:hamiltonian_Liouville}
    H(L_{\text{ren}},P)
    =
    \frac{1}{2\Phi_b}
    \left(
    P^2 + e^{-L_{\text{ren}}}
    \right).
\end{equation}
The right-hand side is, up to some constant prefactors, the Liouville Hamiltonian that appears in the canonical quantization of JT gravity with asymptotic boundaries \cite{Harlow:2018tqv}, and also in earlier work on the SYK model \cite{Bagrets:2016cdf,Bagrets:2017pwq}.
In what follows, we will regard \eqref{eq:hamiltonian_Liouville} as the large cutoff limit of the more general finite cutoff Hamiltonian \eqref{eq:hamiltonian_LP}. This limit will serve as an important consistency check on our results.

\section{Canonical quantization at finite cutoff}
\label{sec:canonical_quantization}
Having discussed the classical Hamiltonian structure of JT gravity, we now turn to the quantization of the theory. As usual, the phase space variables $(L,P)$ are promoted to operators satisfying the canonical commutation relation
\begin{equation}\label{eq:canonical_commutation}
    [\hat{L},\hat{P}] = i \hbar.
\end{equation}
The time-independent Schr\"odinger equation is then
\begin{equation}\label{eq:eigenvalue_problem}
    \hat{H}(\hat{L},\hat{P}) \ket{\Phi_h}
    = E(\Phi_h) \ket{\Phi_h},
\end{equation}
where we labeled the energy eigenstates by $\Phi_h$, in accordance with the classical relation \eqref{eq:hamiltonian} between the energy and the value of the dilaton at the horizon:
\begin{equation}\label{eq:energy}
    E(\Phi_h) =
        \Phi_b - \sqrt{\Phi_b^2 - \Phi_h^2}.
\end{equation}

The quantum Hamiltonian $\hat{H}(\hat{L},\hat{P})$ appearing in \eqref{eq:eigenvalue_problem} is an operator version of the classical Hamiltonian \eqref{eq:hamiltonian_LP}. Formally, it may be viewed as a function of the operator $\hat{\Phi}_h^2(\hat{L},\hat{P})$,
\begin{equation}\label{eq:hamiltonian_quantum}
    \hat{H}(\hat{L}, \hat{P})
    =
    \Phi_b - \sqrt{\Phi_b^2 - \hat{\Phi}_h^2(\hat{L}, \hat{P})},
\end{equation}
where $\hat{\Phi}_h^2(\hat{L},\hat{P})$ is the quantum counterpart of the classical P\"oschl--Teller operator
\begin{equation}\label{eq:Phi_h^2}
    \Phi_h^2(L,P)
    =
    P^2 + \frac{\Phi_b^2}{\cosh^2(L/2)}.
\end{equation}
Throughout this work, rather than solving the eigenvalue problem \eqref{eq:eigenvalue_problem} directly, we focus on the spectrum of the simpler operator $\hat{\Phi}_h^2(\hat{L},\hat{P})$. This approach avoids the technical complications associated with the square root in the original Hamiltonian \eqref{eq:hamiltonian_LP}.
%\textcolor{red}{The price one pays is the assumption that $\hat{H}$ and $\hat{\Phi}_h^2$ share the same eigenvectors.
%In other words, we assume that every eigenvector of $\hat{H}$ is also an eigenvector of $\hat{\Phi}_h^2$, although the corresponding spectra are related non-trivially.}

Before turning to the P\"oschl--Teller eigenvalue problem, let us emphasize a simple point. A single classical Hamiltonian can give rise to several inequivalent quantum Hamiltonians, all of which may be legitimate operator realizations of the same classical function $H(L,P)$.

These ambiguities arise because the classical phase space variables commute, whereas their quantum counterparts do not. As a result, one must choose an ordering prescription when constructing the operator $\hat{H}(\hat{L},\hat{P})$. Different orderings can lead to both qualitative and quantitative differences in the predictions of the quantum theory. In ordinary physical applications, the appropriate ordering may ultimately be selected by comparison with experiment. In the present context, however, no such criterion is available and we must therefore rely on a different principle: we anticipate that, in Section~\ref{sec:grouptheory}, we will establish a connection between the current canonical approach to JT gravity and the representation theory of $\mathfrak{sl}(2,\mathbb{R})$.
The latter instructs us to adopt a rather unconventional operator ordering for the classical P\"oschl--Teller operator $\Phi_h^2(L,P)$ \eqref{eq:Phi_h^2} given by
\begin{equation}\label{eq:hamiltonian_Phi_h^2}
    \hat{\Phi}_h^2(\hat{L},\hat{P})
    \equiv
\frac{1}{2}\left(
\frac{1}{\sqrt{\sinh\!{\hat{L}}}}\,\hat{P}\sqrt{\sinh\!{\hat{L}}}\,\hat{P}
+
\hat{P}\sqrt{\sinh\!{\hat{L}}}\,\hat{P}\frac{1}{\sqrt{\sinh\!{\hat{L}}}}
\right)
+
\frac{\Phi_b^2}{\cosh^2\!{(\hat{L}/2)}} .
\end{equation}
The trivial operator ordering of the Hamiltonian \eqref{eq:Phi_h^2} is discussed in Appendix \ref{appendix:flat}, where we also point out that it leads to unclear results.
%\textcolor{red}{Moreover, in a companion paper~\cite{futurepaper}, we will show that this more natural ordering prescription also emerges directly from the large %central charge limit of the finite cutoff formulation of Liouville gravity~\cite{Mertens:2020hbs}, viewed through its dilaton gravity realization.}

\begin{figure}
    \centering
    \includegraphics{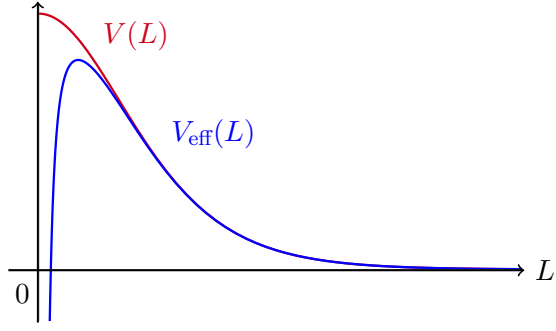}
    \caption{We display the classical P\"oschl--Teller potential \eqref{eq:hamiltonian_LP} in red, while the blue curve is the effective quantum potential \eqref{eq:potential_quantum}. One can think of having an infinite barrier at $L = 0$ that enforces the condition \eqref{eq:condition_L_positive}.}
    \label{fig:potential}
\end{figure}

Let us begin by writing down the operator \eqref{eq:hamiltonian_Phi_h^2} in differential form:
\begin{equation}
-\hbar^2\dv[2]{}{L}+\frac{\Phi_b^2}{\cosh^2(L/2)}-\frac{\hbar^2}{4\sinh^2 L} .
\end{equation}
This operator describes the dynamics of a nonrelativistic particle subject to the effective quantum potential
\begin{equation}\label{eq:potential_quantum}
    V_{\text{eff}}(L)
    =
    \frac{\Phi_b^2}{\cosh^2(L/2)}
    -
    \frac{\hbar^2}{4 \sinh^2(L)}
\end{equation}
shown in Figure \ref{fig:potential}.
We notice that the non-trivial operator ordering produces an additional term in the effective potential.
We refer to the additional term in \eqref{eq:potential_quantum} as ``quantum'' because it is proportional to $\hbar^2$ and therefore has no classical analogue. As a result, although the classical potential in \eqref{eq:Phi_h^2} is regular and purely repulsive, the effective quantum potential becomes singular and attractive near the origin:
\begin{equation}
    V_{\text{eff}}(L)
    \overset{L\to 0}{\sim}
    -\frac{\hbar^2}{4L^2}.
\end{equation}
This observation provides a useful way to interpret the role of the ordering ambiguity. The terms generated by different operator orderings are quantum corrections, and therefore become relevant precisely in the regime where $L$, the length of the universe, is of order $\hbar$. From this perspective, different admissible orderings may be regarded as different all-energy, or ultraviolet, completions of the same classical JT dynamics. These differences could not be appreciated in the ordinary large cutoff treatment of JT gravity, where the renormalization of the geodesic length \eqref{eq:geodesic_length_renormalized} hides the small-$L$ region in which these ordering effects become important.

Let us now turn to the spectral problem of \eqref{eq:hamiltonian_Phi_h^2}.
For notational convenience, we set $\hbar = 1$ from now on and denote the parameters as
\begin{equation}
\label{eq:s_nu_def}
    s \equiv \Phi_h,
    \qquad
    \nu \equiv \Phi_b.
\end{equation}
With these choices, the Schr\"odinger equation takes the form
\begin{equation}\label{eq:Schroedinger_equation}
    -\dv[2]{\psi_s(L)}{L}
    +
    \left(
    \frac{\nu^2}{\cosh^2(L/2)}
    -
    \frac{1}{4 \sinh^2(L)}
    \right)
    \psi_s(L)
    =
    s^2 \, \psi_s(L),
\end{equation}
where $\psi_s(L) \equiv \bra{L}\ket{s}$. The solutions of Equation \eqref{eq:Schroedinger_equation} are obtained in Appendix \ref{app:solutions_of_the_Schroedinger_equation}. This eigenvalue problem must be
supplemented by the physical requirement that $L$, being a geodesic length, is
non-negative. At the quantum level, we impose it by requiring the physical
wavefunctions to obey
\begin{equation}\label{eq:condition_L_positive}
    \psi(L \leq 0) = 0.
\end{equation}
However, due to the singular behavior of the effective potential near $L=0$, both solutions of \eqref{eq:Schroedinger_equation} vanish at the origin and the condition \eqref{eq:condition_L_positive} is not enough to uniquely select a combination of them.
Therefore, we must choose a self-adjoint extension among the one-parameter family of those admitted by the operator \eqref{eq:hamiltonian_Phi_h^2} \cite{inoue2025scattering}, i.e. we must specify the allowed leading behavior of the wavefunctions near $L=0$. This choice is parameterized by a real number $\alpha$ through
\begin{equation}
    \psi_s(L)
    \overset{L\to 0}{\sim}
    \sqrt{L/2}
    +
    \alpha \sqrt{L/2} \log\!\left( L/2 \right) .
\end{equation}
In this work, we choose the Friedrichs self-adjoint extension, corresponding to
\(
    \alpha=0.
\)
We stress, however, that the other self-adjoint extensions are also admissible at this stage. Different choices lead to theories with both qualitatively and quantitatively different physical predictions, and in particular our choice is the only one that excludes bound states from the spectrum of the theory for all values of the boundary dilaton $\Phi_b$ (see Appendix \ref{app:bound_states}).
This choice singles out the eigenfunctions
\begin{equation}\label{eq:eigenfunction}
    \psi_{s}(L)
    =
    \frac{\sqrt{\tanh(L/2)}}{\cosh(L/2)^{2is}} \, _2F_1\!\left( \frac{1}{2} + is+ i\nu, \frac{1}{2} + is - i\nu, 1 ; \tanh^2(L/2)  \right).
\end{equation}
As discussed in Appendix \ref{app:solutions_of_the_Schroedinger_equation}, these wavefunctions are real and obey the orthogonality condition
\begin{equation}\label{eq:orthogonality}
    \bra{s'}\ket{s}
    =
    \int_0^\infty
    \dd L\,
    \psi_{s'}^*(L)\psi_{s}(L)
    =
    \frac{\delta(s-s')}{\rho(s)},
    \qquad
    \rho(s) = 
    \frac{2 s\sinh(2\pi s)}
    {\cosh(2\pi s)+\cosh(2\pi\nu)}.
\end{equation}
As a consequence of \eqref{eq:orthogonality}, we have the following resolution of the identity
\begin{equation}\label{eq:identity}
    1 = \int_0^\infty \dd s \, \rho(s) \ket{s}\bra{s}.
\end{equation}

\subsection{Quantum observables from transition matrix elements}
\label{sec:quantum_observables}
The analysis developed so far is naturally formulated in a JT gravity setup with two spacetime boundaries. In this section, we argue that the formalism developed above can nevertheless be used to compute observables that belong to a single-sided description, such as the disk partition function $Z(\beta)$ and matter correlators. The core idea is simple and has appeared repeatedly in recent years, having been applied to JT gravity at large cutoff \cite{Lin:2022zxd, Saad:2019pqd} and its supersymmetric extensions \cite{Belaey:2024dde}.

The natural starting point is to identify the partition function with the
transition amplitude between two slices of vanishing length separated by a
Euclidean time $\beta$,
\begin{equation}\label{eq:partition_function_prescription}
    Z(\beta) = \bra{L=0} e^{-\beta \hat{H}} \ket{L=0}
    \quad = \quad \vcenter{\hbox{\begin{tikzpicture}[x=0.75pt,y=0.75pt,yscale=-1,xscale=1]
%uncomment if require: \path (0,184); %set diagram left start at 0, and has height of 184

%Shape: Chord [id:dp22347345871302193] 
\draw  [draw opacity=0][fill={rgb, 255:red, 230; green, 230; blue, 255 }  ,fill opacity=1 ] (150,100) .. controls (150,100) and (150,100) .. (150,100) .. controls (150,122.09) and (132.09,140) .. (110,140) .. controls (87.91,140) and (70,122.09) .. (70,100) -- cycle ;
%Shape: Arc [id:dp6054503487036216] 
\draw  [draw opacity=0][line width=1.5]  (149.99,99.3) .. controls (150.38,121.39) and (132.79,139.61) .. (110.7,139.99) .. controls (88.61,140.38) and (70.39,122.79) .. (70.01,100.7) -- (110,100) -- cycle ; \draw  [color={rgb, 255:red, 208; green, 2; blue, 27 }  ,draw opacity=1 ][line width=1.5]  (149.99,99.3) .. controls (150.38,121.39) and (132.79,139.61) .. (110.7,139.99) .. controls (88.61,140.38) and (70.39,122.79) .. (70.01,100.7) ;  
%Shape: Chord [id:dp5898309633305231] 
\draw  [draw opacity=0][fill={rgb, 255:red, 230; green, 230; blue, 255 }  ,fill opacity=1 ] (70,90) .. controls (70,90) and (70,90) .. (70,90) .. controls (70,67.91) and (87.91,50) .. (110,50) .. controls (132.09,50) and (150,67.91) .. (150,90) -- cycle ;
%Shape: Arc [id:dp9167676043308838] 
\draw  [draw opacity=0][line width=1.5]  (70,90) .. controls (70,90) and (70,90) .. (70,90) .. controls (70,67.91) and (87.91,50) .. (110,50) .. controls (132.09,50) and (150,67.91) .. (150,90) -- (110,90) -- cycle ; \draw  [color={rgb, 255:red, 208; green, 2; blue, 27 }  ,draw opacity=1 ][line width=1.5]  (70,90) .. controls (70,90) and (70,90) .. (70,90) .. controls (70,67.91) and (87.91,50) .. (110,50) .. controls (132.09,50) and (150,67.91) .. (150,90) ;  
%Shape: Circle [id:dp09762004638403732] 
\draw  [color={rgb, 255:red, 0; green, 0; blue, 255 }  ,draw opacity=1 ][fill={rgb, 255:red, 0; green, 0; blue, 255 }  ,fill opacity=1 ] (107.37,140) .. controls (107.37,138.55) and (108.55,137.37) .. (110,137.37) .. controls (111.45,137.37) and (112.63,138.55) .. (112.63,140) .. controls (112.63,141.45) and (111.45,142.63) .. (110,142.63) .. controls (108.55,142.63) and (107.37,141.45) .. (107.37,140) -- cycle ;
%Shape: Circle [id:dp1598658512218839] 
\draw  [color={rgb, 255:red, 0; green, 0; blue, 255 }  ,draw opacity=1 ][fill={rgb, 255:red, 0; green, 0; blue, 255 }  ,fill opacity=1 ] (107.37,50) .. controls (107.37,48.55) and (108.55,47.37) .. (110,47.37) .. controls (111.45,47.37) and (112.63,48.55) .. (112.63,50) .. controls (112.63,51.45) and (111.45,52.63) .. (110,52.63) .. controls (108.55,52.63) and (107.37,51.45) .. (107.37,50) -- cycle ;
%Shape: Arc [id:dp4396138631896195] 
\draw[<->]  [draw opacity=0] (145,100) .. controls (145,119.36) and (129.33,135.06) .. (110,135.06) .. controls (90.67,135.06) and (75,119.36) .. (75,100) -- (110,100) -- cycle ; \draw[<->] (144.94,102.02) .. controls (143.9,120.44) and (128.65,135.06) .. (110,135.06) .. controls (91.25,135.06) and (75.94,120.29) .. (75.04,101.73) ; %\draw [shift={(75,100)}, rotate = 79.83] [color={rgb, 255:red, 0; green, 0; blue, 0 }  ][line width=0.75]    (6.56,-2.94) .. controls (4.17,-1.38) and (1.99,-0.4) .. (0,0) .. controls (1.99,0.4) and (4.17,1.38) .. (6.56,2.94)   ; %\draw [shift={(145,100)}, rotate = 100.17] [color={rgb, 255:red, 0; green, 0; blue, 0 }  ][line width=0.75]    (6.56,-2.94) .. controls (4.17,-1.38) and (1.99,-0.4) .. (0,0) .. controls (1.99,0.4) and (4.17,1.38) .. (6.56,2.94)   ;
%Shape: Arc [id:dp5998882567638605] 
\draw[<->]  [draw opacity=0] (145,90) .. controls (145,70.64) and (129.33,54.94) .. (110,54.94) .. controls (90.67,54.94) and (75,70.64) .. (75,90) -- (110,90) -- cycle ; \draw[<->]    (144.94,87.98) .. controls (143.9,69.56) and (128.65,54.94) .. (110,54.94) .. controls (91.25,54.94) and (75.94,69.71) .. (75.04,88.27) ; %\draw [shift={(75,90)}, rotate = 280.17] [color={rgb, 255:red, 0; green, 0; blue, 0 }  ][line width=0.75]    (6.56,-2.94) .. controls (4.17,-1.38) and (1.99,-0.4) .. (0,0) .. controls (1.99,0.4) and (4.17,1.38) .. (6.56,2.94)   ; %\draw [shift={(145,90)}, rotate = 259.83] [color={rgb, 255:red, 0; green, 0; blue, 0 }  ][line width=0.75]    (6.56,-2.94) .. controls (4.17,-1.38) and (1.99,-0.4) .. (0,0) .. controls (1.99,0.4) and (4.17,1.38) .. (6.56,2.94)   ;

% Text Node
\draw (110,143.4) node [anchor=north] [inner sep=0.75pt]  [font=\scriptsize]  {$\textcolor[rgb]{0,0,1}{\bra{L=0}}$};
% Text Node
\draw (110,46.6) node [anchor=south] [inner sep=0.75pt]  [font=\scriptsize]  {$\textcolor[rgb]{0,0,1}{\ket{L=0}}$};
% Text Node
\draw (110,128.6) node [anchor=south] [inner sep=0.75pt]  [font=\scriptsize]  {$\beta /2$};
% Text Node
\draw (110,61.4) node [anchor=north] [inner sep=0.75pt]  [font=\scriptsize]  {$\beta /2$};

\end{tikzpicture}}}.
\end{equation}
What differs in the present finite cutoff setup with respect to \cite{Lin:2022zxd, Saad:2019pqd, Belaey:2024dde} is the length eigenstate appearing at the endpoints of the amplitude: here $\ket{L= 0}$ genuinely corresponds to a zero-length geodesic in spacetime, whereas in the large cutoff JT gravity treatment the necessary renormalization \eqref{eq:geodesic_length_renormalized} effectively forces one to consider the state $\ket{L_\text{ren} = -\infty}$, which is formally associated with a geodesic of ``negative infinite length.''

Before proceeding, we must discuss some further details about the prescription \eqref{eq:partition_function_prescription}.
First, we notice that the quantum Hamiltonian in \eqref{eq:partition_function_prescription} is of the form \eqref{eq:hamiltonian_quantum}, which involves a square root. This means that whenever the eigenvalues of the operator $\hat{\Phi}_h(\hat{L},\hat{P})$ are such that $\Phi_h > \Phi_b$, the corresponding eigenvalues of the full Hamiltonian become complex. This fact might compromise the unitarity of the theory, hence we propose to restrict its Hilbert space to the states that obey the condition $\Phi_h < \Phi_b$. On this space, the resolution of the identity takes the form of the projector
\begin{equation}\label{eq:projector}
    \hat{\Pi} = \int_0^{\nu} \dd s \, \rho(s) \ket{s}\bra{s},
\end{equation}
and the time evolution operator $e^{-\beta \hat{H}}$ appearing in \eqref{eq:partition_function_prescription} should actually be understood as its projection by the latter
\begin{equation}\label{eq:prescription_evolution_operator}
    e^{-\beta \hat{H}} \quad \to \quad \hat{\Pi} e^{-\beta \hat{H}} \hat{\Pi}.
\end{equation}
The role of this projected evolution operator is to prevent the time evolution from reaching configurations in which the entire spacetime lies behind the black hole horizon, see Figure \ref{fig:Penrose_diagram}.

Next, we discuss the role of the state $\ket{L=0}$. Taken literally, the
prescription \eqref{eq:partition_function_prescription} would give a vanishing
partition function, since inserting a complete set of eigenstates makes it
proportional to $|\psi_s(0)|^2$, which vanishes for every $s$ because of the
condition \eqref{eq:condition_L_positive}.

Nevertheless, the amplitude \eqref{eq:partition_function_prescription} can still
be given a non-trivial physical meaning. Its vanishing is due to an overall
factor that does not depend on the energy, and the physical information contained
in the partition function is insensitive to such factors, which simply shift the
entropy by a constant. Likewise, normalized correlation functions are ratios of
the form\footnote{From now on, the quantum amplitudes are understood in the
projected sense \eqref{eq:prescription_evolution_operator}.}
\begin{equation}\label{eq:observable}
    \frac{
    \bra{L=0} e^{-(\beta-\tau)\hat{H}}
    \hat{\mathcal{O}}
    e^{-\tau\hat{H}} \ket{L=0}
    }{
    \bra{L=0} e^{-\beta \hat{H}} \ket{L=0}
    },
\end{equation}
in which numerator and denominator vanish at the same rate as $L\to 0$. Since the
small-$L$ behavior of the eigenfunctions \eqref{eq:eigenfunction} does not depend
on $s$,
\begin{equation}
    \psi_s(L) \overset{L \to 0}{\sim} \sqrt{L/2},
\end{equation}
we adopt the following prescription: whenever $\ket{L=0}$ appears, we replace it
with
\begin{equation}\label{eq:prescription_L=0}
    \ket{L = 0} \quad \to \quad \lim_{L \to 0} \frac{\ket{L}}{\sqrt{L/2}}.
\end{equation}
In other words, this prescription amounts to extracting the common leading vanishing behavior by applying l'Hôpital's rule to the ratio of amplitudes \eqref{eq:observable}. This issue is not specific to the finite cutoff theory and a closely related problem also appears in ordinary JT gravity, as discussed in Appendix I of \cite{Lin:2022zxd}.

Lastly, we comment on the fact that the canonical approach of Section \ref{sec:classical_JT_gravity} does not capture a topological contribution to the Euclidean action of JT gravity
\begin{equation}\label{eq:JT_action_euclidean}
    I_\text{JT} = -\frac{1}{2} \int_\mathcal{M} \diff[2]{x} \sqrt{g} \Phi (R+2) - \int_{\partial\mathcal{M}} \diff{y} \sqrt{h} \Phi (K-1) - \Phi_b \sum_j \left( \pi - \theta_j \right).
\end{equation}
The last term takes into account contributions from possible corners along $\partial\mathcal{M}$ (the red arrow specifies the orientation of the boundary):
\begin{center}
    \includegraphics[width=0.15\linewidth]{images/corner.tikz}
\end{center}
Indeed, using the Gauss--Bonnet theorem, we can rewrite the boundary term of \eqref{eq:JT_action_euclidean} as
\begin{equation}
    -\int_{\partial\mathcal{M}} \diff{y} \sqrt{h} \Phi (K-1) = \Phi_b \beta + \frac{\Phi_b}{2} \int_{\mathcal{M}} \dd^2 x \, \sqrt{g} R - 2 \pi \Phi_b \chi(\mathcal{M}).
\end{equation}
The topological term $-2 \pi \Phi_b$\footnote{Here we used that the Euler characteristic of a smooth disk is $\chi(\mathcal{M}) = 1$.} gets lost in the process of reconstructing the boundary Hamiltonian \eqref{eq:conserved_charge} from integrating  
\eqref{eq:Hamilton_equation}, and indeed the quantum mechanical action reproduces only the first two contributions.
Hence, we include this term in the computation of $Z(\beta)$ by prescribing
\begin{equation}\label{pippo}
    Z(\beta) \equiv \left( \lim_{L \to 0} \frac{\bra{L}}{\sqrt{L/2}} \right) \hat{\Pi} \, e^{2 \pi \Phi_b - \beta \hat{H}} \, \hat{\Pi} \left( \lim_{L' \to 0} \frac{\ket{L'}}{\sqrt{L'/2}} \right).
\end{equation}
This last addition will be required in the next sections in order to obtain finite results in the large cutoff limit.

\section{Partition function}\label{sec:partition_function}
\begin{figure}
    \centering
    \includegraphics{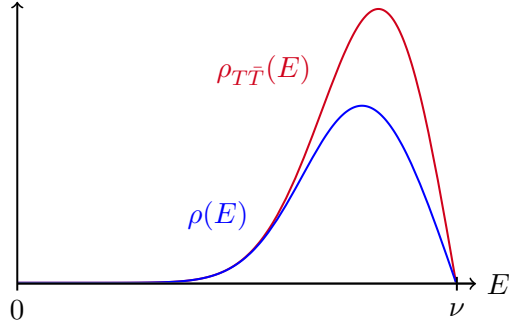}
    \caption{The blue curve is a plot of the spectral density \eqref{eq:spectral_density}. We compare it to the red curve $\rho_{T\bar{T}}(E)$ depicting the leading order contribution in the expansion \eqref{eq:spectralexpansion} (upon performing the appropriate change of variable). The latter represents the result obtained by $T\bar{T}$ approaches (see Section \ref{sec:TT}) and does not include higher-order nonperturbative corrections in the cutoff.}
    \label{fig:spectral_density}
\end{figure}

We are finally in a position to use the fixed-energy wavefunctions of the universe \eqref{eq:eigenfunction} together with the prescription \eqref{pippo} to determine the disk partition function of JT gravity at finite cutoff. We obtain
\begin{equation}\label{eq:partition_function}
    \begin{split}
        Z(\beta)
        &= e^{2 \pi \nu}
        \int_0^\nu \diff{s}\,
        \frac{2 s\sinh(2\pi s)}
        {\cosh(2\pi s)+\cosh(2\pi\nu)}
        e^{-\beta (\nu - \sqrt{\nu^2 - s^2})}.
    \end{split}
\end{equation}
Next, we change variables from $s$ to the energy so that the partition function can be brought to standard form
\begin{equation}
    Z(\beta)
    =
    \int_0^\nu \dd E \,
    \rho(E) \, e^{-\beta E}.
\end{equation}
The corresponding spectral density is
\begin{equation}\label{eq:spectral_density}
   \boxed{ \rho(E) =
        e^{2 \pi \nu}
        \displaystyle
        \frac{
        2 (\nu-E) \,
        \sinh\!\left(
        2\pi\sqrt{E\,(2 \nu - E)}
        \right)}
        {
        \cosh\!\left(
        2\pi\sqrt{E\,(2 \nu - E)}
        \right)
        +
        \cosh(2\pi\nu)
        }}.
\end{equation}
As depicted in Figure \ref{fig:spectral_density}, we see that the spectral density \eqref{eq:spectral_density} vanishes at $E=0$ and at $E=\nu$, and by construction has a compact support.

As a consistency check, let us examine the large cutoff limit of the partition function \eqref{eq:partition_function}, meaning that we take $\Phi_b \to \infty$ while keeping the renormalized inverse temperature
\begin{equation}\label{eq:beta_renormalized}
    \beta_{\text{JT}}
    \equiv
    \beta/\Phi_b
\end{equation}
finite.
The introduction of $\beta_\text{JT}$ mimics the renormalization procedure that is necessary in the large cutoff setup: here $\beta_\text{JT}$ is understood as the amount of (Euclidean) Schwarzschild time along the boundary of the disk, whereas its proper length $\beta$ is identified with the inverse Tolman temperature of the black hole felt by an observer at the boundary.
Using the results derived in Appendix \ref{app:large_cutoff_asymptotics}, we obtain the following large cutoff partition function
\begin{equation}\label{eq:partition_function_large_cutoff}
    Z(\beta_{\text{JT}})
    =
    \int_0^\infty
    \dd s \,
    4 s \sinh(2\pi s) \,
    e^{-\beta_{\text{JT}} s^2/2}.
\end{equation}
This result is consistent with the ordinary JT partition function obtained in the literature with several independent approaches \cite{Stanford:2017thb,Mertens:2017mtv,Iliesiu:2019xuh,Yang:2018gdb,Griguolo:2023aem}.

Several comments are now in order. First, the theory obtained from this quantization procedure shares some features with the representation theory of the universal cover of $\text{SL}(2,\mathbb{R})$. For instance, the spectral density $\rho(s)$ appearing in Equation \eqref{eq:orthogonality} coincides with the Plancherel measure of the group \cite{Iliesiu:2019xuh, Kitaev:2017hnr, Kitaev:2018wpr}
\begin{equation}
    \frac{2s\sinh(2\pi s)}
    {\cosh(2\pi s)+\cos(2\pi\mu)}
\end{equation}
after the formal identification $ \nu = i\mu$,  where $\mu$ labels the central element of the group. We will elaborate on the relation with $\mathfrak{sl}(2,\mathbb{R})$ in Section \ref{sec:grouptheory}.

Moreover, using the identity
\begin{equation}
    \abs{\Gamma\!\left(\frac{1}{2}+iz\right)}^2
    =
    \frac{\pi}{\cosh(\pi z)},
\end{equation}
we can express \eqref{eq:partition_function} in the equivalent form\footnote{Here we use the shorthand notation
\[
    \Gamma(x\pm iy\pm iz)
    \equiv
    \Gamma(x+iy+iz)
    \Gamma(x-iy+iz)
    \Gamma(x+iy-iz)
    \Gamma(x-iy-iz).
\]
}
\begin{equation}
\label{eq:partition_function_Gamma}
    Z(\beta)
    =
    e^{2 \pi \nu} \int_0^\nu \diff{s}\,
    \frac{s\sinh(2\pi s)}{\pi^2} \,
    \Gamma\!\left(\frac{1}{2}\pm is\pm i\nu\right)
    e^{-\beta (\nu - \sqrt{\nu^2 - s^2})}.
\end{equation}
An interesting observation is that the resulting expression closely resembles the integral representation of the two-point function in large cutoff JT gravity \cite{Mertens:2017mtv,Iliesiu:2019xuh}
%\begin{equation}\label{eq:partastwopoint}
%    \langle \mathcal{O}_\Delta(0,\tau) \rangle_{\beta_\text{JT}}
%    =
%    \frac{\int_0^\infty \dd k_1 \, \frac{k_1 \sinh(2 \pi k_1)}{\pi^2} \int_0^\infty \dd k_2 \, \frac{k_2 \sinh(2 \pi k_2)}{\pi^2} \frac{\Gamma(\Delta \pm ik_1 \pm k_2)}{\Gamma(2 \Delta)} \, e^{-(\beta_\text{JT} - \tau) \, k_2^2/2} \, e^{-\tau \, k_1^2/2}}{\int_0^\infty \dd k \, \frac{k \sinh(2 \pi k)}{\pi^2} e^{-\beta_\text{JT} k^2/2}}
%\end{equation}
\begin{equation}\label{eq:partastwopoint}
\begin{split}
    \langle \mathcal{O}_\Delta(0,\tau) \rangle_{\beta_\text{JT}}
    =&\frac{1}{Z(\beta_{\mathrm{JT}})}
    \int_0^\infty  \dd k_1 \, \frac{k_1 \sinh(2 \pi k_1)}{\pi^2} \int_0^\infty \dd k_2 \, \frac{k_2 \sinh(2 \pi k_2)}{\pi^2} \frac{\Gamma(\Delta \pm ik_1 \pm i k_2)}{\Gamma(2 \Delta)} \times \\
    & \times e^{-(\beta_\text{JT} - \tau) \, k_2^2/2} \, e^{-\tau \, k_1^2/2}
\end{split}
\end{equation}
for an operator of conformal dimension $\Delta=1/2$ and external energies, or momenta, $k_1=s$ and $k_2=\nu$ fixed. This kind of structure will also emerge for the finite cutoff two-point function: we will comment on a possible interpretation in Section \ref{sec:outlook}. 

Finally, we discuss the semiclassical behavior of the partition function \eqref{eq:partition_function}. To do so, we must first reintroduce $\hbar$ by sending $s \to s/\hbar$ and $\nu \to \nu/\hbar$ and then take $\hbar \to 0$. The computation is performed in detail in Appendix \ref{app:semiclassical_partition_function}, where we find that
\begin{equation}\label{eq:partition_function_semiclassical}
    Z(\beta) \overset{\hbar \to 0}{\sim}
    2 \nu^{3/2} \frac{\beta}{2 \pi} \left[ 1 + \left( \frac{\beta}{2 \pi} \right)^2 \right]^{-5/4} \, e^{\nu \left( \sqrt{4 \pi^2 + \beta^2} - \beta \right)}
\end{equation}
at leading order.
Remarkably, this computation matches the one-loop result obtained in~\cite{Griguolo:2025kpi}. We want to stress that choosing a different operator ordering for the P\"oschl--Teller Hamiltonian \eqref{eq:hamiltonian_Phi_h^2} would have led to a different one-loop determinant, as discussed in Appendix \ref{appendix:flat} in the case of the trivial ordering.

As a sanity check, we should compare the semiclassical partition function computed with our prescription \eqref{eq:partition_function_semiclassical} with the
Euclidean gravitational path integral of the theory \eqref{eq:JT_action_euclidean} on a manifold with disk topology
\begin{equation}\label{eq:euclidean_path_integral}
    \int_\text{disk} [\mathcal{D}g] [\mathcal{D}\Phi] \; e^{-I_\text{JT}},
\end{equation}
with Dirichlet boundary conditions fixing the boundary dilaton to a constant, $\Phi \lvert_{\partial\mathcal{M}} = \Phi_b$,
and the length of the boundary to $\oint_{\partial\mathcal{M}} \dd s = \beta$. In the semiclassical limit, the main contribution to the path integral \eqref{eq:euclidean_path_integral} comes from the Euclidean on-shell geometry
\begin{equation}\label{eq:classical_solution_euclidean}
\begin{split}
    \dd s^2 &= \left( r^2 - \frac{\Phi_b^2}{1 + \left(\frac{\beta}{2 \pi}\right)^2} \right) \dd t_\text{E}^2 + \left( r^2 - \frac{\Phi_b^2}{1 + \left(\frac{\beta}{2 \pi}\right)^2} \right)^{-1} \dd r^2, \quad
    \begin{cases}
    \displaystyle{\frac{\Phi_b}{\sqrt{1 + \left(\frac{\beta}{2 \pi}\right)^2}}} \leq r \leq \Phi_b,\\
    t_\text{E} \sim t_\text{E} + \displaystyle{\frac{2\pi}{\Phi_b}} \sqrt{1 + \left(\frac{\beta}{2 \pi}\right)^2},
    \end{cases}
    \\
    \Phi &= r,
\end{split}
\end{equation}
where we Wick-rotated the Lorentzian Schwarzschild solution \eqref{eq:classical_configuration_schwarzschild} and expressed the black hole horizon $\Phi_h$ in terms of the inverse Tolman temperature $\beta$.
Evaluating $I_\text{JT}$ \eqref{eq:JT_action_euclidean} on the solution \eqref{eq:classical_solution_euclidean} yields
\begin{equation}\label{eq:semiclassicalZ}
    \int_\text{disk} [\mathcal{D}g] [\mathcal{D}\Phi] \; e^{-I_\text{JT}} \overset{\hbar \to 0}{\sim} e^{\nu \left( \sqrt{4 \pi^2 + \beta^2} - \beta \right)}
\end{equation}
at exponential leading order. This result is indeed compatible with the semiclassical behavior of the partition function obtained with our prescription \eqref{eq:partition_function_semiclassical}, validating the latter as a plausible proposal for the partition function of JT gravity at finite cutoff.

\subsection{P\"oschl--Teller instantons}\label{sec:Poschl_Teller_instantons}
In this section, we discuss the partition function \eqref{eq:partition_function} in detail and provide a geometric interpretation for the denominator term.
Let us start by rewriting the spectral density in a suggestive way \cite{Yang:2018gdb}
\begin{equation}\label{eq:spectralexpansion}
    e^{2 \pi \nu} \rho(s) = 
    e^{2 \pi \nu} \frac{2 s\sinh(2\pi s)}
    {\cosh(2\pi s)+\cosh(2\pi\nu)}
    = 4 s \sum_{n=1}^{\infty} (-1)^{n-1} e^{-2\pi (n-1) \nu} \sinh(2\pi n s).
\end{equation}
The $n = 1$ term in the sum clearly corresponds to the large cutoff spectral density \eqref{eq:partition_function_large_cutoff}, while all the corrections to the latter are nonperturbative both in the cutoff parameter $1/\Phi_b$ and in $\hbar$.
Furthermore, these corrections look like contributions coming from classical saddles of the theory, motivating us to look for their geometric interpretation.

First, let us consider the auxiliary Hamiltonian system introduced in Section \ref{sec:canonical_quantization}. In Lorentzian signature, we established a one-to-one correspondence between each on-shell JT geometry \eqref{eq:classical_configuration_global} and a scattering trajectory in the P\"oschl--Teller potential \eqref{eq:Phi_h^2}. Studying the corresponding problem in Euclidean signature requires additional care: it is no longer true that on-shell JT geometries on a disk topology are in one-to-one correspondence with classical trajectories in a (reversed) P\"oschl--Teller potential. To build some intuition, one can write the amplitude \eqref{eq:partition_function_prescription} as\footnote{Here the functional integration measure $[\mathcal{D} L]$ has to be chosen according to the operator ordering chosen for the canonical quantization of $\hat{H}$. We can overlook this detail as long as we only discuss semiclassical features of the theory.}
\begin{equation}\label{eq:amplitude_path_integral}
    \bra{L=0} e^{2 \pi \nu -\beta \hat{H}} \ket{L=0} = \int_{L(0)=0}^{L(\beta)=0} [ \mathcal{D} L ] \, e^{- \frac{\Phi_b}{\hbar} \left[ \int_0^\beta \dd \tau_\text{E} \left( 1 - \tanh(L/2) \, \sqrt{1 -
    \dot{L}^2} \right) - 2 \pi \right]},
\end{equation}
where we used the fact that the Lagrangian corresponding to \eqref{eq:hamiltonian_LP} is
\begin{equation}\label{eq:Poschl_Teller_lagrangian}
    \mathcal{L} = \Phi_b \left( \tanh(L/2) \, \sqrt{1 + \dot{L}^2} - 1 \right)
\end{equation}
and performed the Wick rotation $\tau \to - i \tau_\text{E}$.
From Equation \eqref{eq:amplitude_path_integral} it is clear that the amplitude picks up contributions from all the classical saddles corresponding to trajectories that solve the Euler--Lagrange equation
\begin{equation}
    \ddot{L} \, \sinh{L} + 1 - \dot{L}^2 = 0
\end{equation}
and are compatible with the conditions $L(0)=0=L(\beta)$ as well as the boundary constraint $L \geq 0$. One can check that such solutions are those of the form
\begin{equation}
    L_n(\tau_\text{E}) = 2 \sinh^{-1}\!\left[ \frac{\beta}{2 \pi n} \abs{\sin\!\left( \frac{\pi n}{\beta} \tau_\text{E} \right)} \right], \qquad n=1,2,\dots,
\end{equation}
and their corresponding action reads
\begin{equation}\label{eq:on_shell_action_Poschl_Teller}
    I[L_n] = - \Phi_b \left( \sqrt{4 \pi^2 n^2 + \beta^2} - \beta - 2 \pi (n-1) \right).
\end{equation}
Consequently, picking up the contribution from each of these trajectories yields the formal semiclassical expression for the amplitude
\begin{equation}\label{eq:amplitude_semiclassical}
    \bra{L=0} e^{2 \pi \nu -\beta \hat{H}} \ket{L=0} \sim \sum_{n \geq 1} (-1)^{n-1} e^{\nu \left( \sqrt{4 \pi^2 n^2 + \beta^2} - \beta - 2 \pi (n-1) \right)}.
\end{equation}
This path integral approach allows us to build physical intuition for each of these contributions (see Figure \ref{fig:necklace}). They correspond to instanton trajectories in a reversed/Euclidean P\"oschl--Teller potential that bounce $n-1$ times against the Dirichlet barrier at $L=0$ at intermediate times. Heuristically, the sign $(-1)^{n-1}$ in \eqref{eq:amplitude_semiclassical} can be traced to the $n-1$ bounces of the
trajectory against the wall: each reflection
comes with a factor $-1$, as in the method of images. Importantly, their gravitational interpretation is that of hyperbolic disks whose thermal boundaries are pinched $n-1$ times, resulting in a ``necklace'' of disks as in Figure \ref{fig:necklace}. This fact can be verified by computing the gravitational on-shell action \eqref{eq:JT_action_euclidean} for each necklace\footnote{Recall that each pinch contributes a corner term $2 \pi \Phi_b$ to the on-shell action.}
\begin{equation}\label{eq:on_shell_action_gravitational}
    I[n\text{-necklace}] = -\Phi_b \left( \sqrt{4 \pi^2 n^2 + \beta^2} - \beta - 2 \pi (n-1) \right).
\end{equation}
Remarkably, the gravitational on-shell actions \eqref{eq:on_shell_action_gravitational} match their P\"oschl--Teller counterparts \eqref{eq:on_shell_action_Poschl_Teller}.
To summarize, the only classical trajectory that is consistent with a smooth boundary for the disk is given by $n=1$.

We acknowledge that a closely related discussion is presented in Section 4 of \cite{Yang:2018gdb}, where the same higher-order corrections to the density of states are given the geometrical interpretation of disks whose boundaries wind $n$ times around the thermal circle. The interpretation we provide here is different, but close in spirit. A discussion about finite cutoff JT gravity with a self-avoiding boundary is presented in \cite{Stanford:2020qhm}.
\begin{figure}
    \centering
    \includegraphics[width=\linewidth]{images/instantons.tikz}
    \caption{The necklace saddles contributing to \eqref{eq:amplitude_semiclassical}. On the left, $n$ hyperbolic disks of boundary length $\beta/n$ pinched at $n-1$ points. On the right, the corresponding Euclidean trajectory in
    the reversed P\"oschl--Teller potential, which bounces $n - 1$ times against the wall at $L = 0$.}
    \label{fig:necklace}
\end{figure}

Finally, we can compare these results with the exact partition function of the theory \eqref{eq:partition_function}. Employing the representation \eqref{eq:spectralexpansion} of the spectral density and discarding the exponentially decaying term of the sinh, we can perform a saddle point approximation of each contribution in the series
\begin{equation}\label{eq:saddle_series}
    \begin{split}
        Z(\beta)
        &\overset{\hbar \to 0}{\sim} 2 \sum_{n \geq 1} (-1)^{n-1} e^{-2 \pi (n-1) \nu} \int_0^\nu \dd s \, s \, e^{2 \pi n s - \beta (\nu - \sqrt{\nu^2 - s^2})}\\
        &\overset{\hbar \to 0}{\sim} 2 \nu^{3/2} \sum_{n \geq 1} (-1)^{n-1} \frac{1}{\sqrt{n}} \frac{\beta}{2 \pi n} \left[ 1 + \left(  \frac{\beta}{2 \pi n} \right)^2 \right]^{-5/4} e^{\nu \left( \sqrt{4 \pi^2 n^2 + \beta^2} - \beta - 2 \pi (n-1) \right)}.
    \end{split}
\end{equation}
Here we also computed the one-loop determinant associated with each term in the sum but, at exponential order, we notice a match between each contribution in \eqref{eq:saddle_series} and the on-shell actions \eqref{eq:on_shell_action_Poschl_Teller} and \eqref{eq:on_shell_action_gravitational}, hence confirming our geometrical interpretation.

\section{Correlation functions}\label{sec:correlation_functions}
Having obtained the partition function, we now turn to other observables of the
theory: thermal correlation functions of a matter field coupled to JT gravity.
Specifically, we consider a free massive scalar coupled
minimally to the metric and not to the dilaton,
\begin{equation}\label{eq:JT_matter_action}
    I[g,\Phi,\varphi] = I_\text{JT}[g,\Phi] + I_\text{mat}[g,\varphi], \qquad
    I_\text{mat}[g,\varphi] = \frac{1}{2} \int_{\mathcal{M}} \text{d}^2 x \sqrt{g}\,
    \left( g^{\mu\nu} \partial_\mu \varphi \partial_\nu \varphi + m^2 \varphi^2 \right),
\end{equation}
where $I_\text{JT}$ is given by \eqref{eq:JT_action_euclidean}. Since $\varphi$
does not couple to $\Phi$, the dilaton equation of motion still imposes $R=-2$,
and the matter propagates on a locally AdS$_2$ background.
Hence, the Green's function of $\varphi$ is obtained as the solution of
\begin{equation}\label{eq:Green_equation}
    \left( - \frac{1}{\sqrt{g}} \partial_\mu \sqrt{g} g^{\mu\nu} \partial_\nu + m^2 \right) G_\Delta(x,x') = \frac{\delta^2(x-x')}{\sqrt{g}},
\end{equation}
where $g$ is the AdS$_2$ metric.
If we parameterize the mass as $m^2 = \Delta (\Delta - 1)$, then the solution of \eqref{eq:Green_equation} is given by \cite{dhoker2002supersymmetricgaugetheoriesadscft}
\begin{equation}\label{eq:AdS_propagator}
    G_\Delta(L) =
    %\frac{\Gamma(\Delta)^2}{4 \pi \Gamma(2 \Delta)} 
    \left( \frac{\Phi_b}{\cosh(L/2)} \right)^{2\Delta} {}_2F_1\!\left( \Delta, \Delta, 2\Delta; \frac{1}{\cosh^2(L/2)} \right),
\end{equation}
which only depends on the geodesic distance $L = d(x,x')$ between the two points and the normalization is chosen for later convenience.
Importantly, when computing the Green's function for two points on the asymptotic boundary of spacetime, one must take the large cutoff limit \eqref{eq:large_cutoff_limit} together with a renormalization of the length \eqref{eq:geodesic_length_renormalized}, yielding
\begin{equation}\label{eq:AdS_propagator_large_cutoff}
    G_\Delta(L_\text{ren}) \overset{\Phi_b \to \infty}{\sim} e^{-\Delta L_\text{ren}}.
\end{equation}

Now, let us turn our attention to how the formalism developed in Section \ref{sec:quantum_observables} can be used to compute thermal matter correlators in the theory at finite cutoff.

At large cutoff, a scalar field coupled to JT gravity acts as the source associated with a specific type of bi-local operator in the dual Schwarzian theory \cite{Maldacena:2016upp}
\begin{equation}\label{eq:bilocal_operator_schwarzian}
    \mathcal{O}_\Delta(\tau_1,\tau_2) = \left( \frac{\sqrt{f'(\tau_1) f'(\tau_2)}}{\frac{\beta_\text{JT}}{\pi} \sin\!\left( \frac{\pi}{\beta_\text{JT}} (f(\tau_1) - f(\tau_2)) \right)} \right)^{2 \Delta},
\end{equation}
where $\Delta$ has the interpretation of conformal dimension and $f(\tau)$ is the boundary reparameterization mode.
The thermal expectation value of this operator was computed exactly in \cite{Mertens:2017mtv,Mertens:2018fds}, and an explicit computation shows that it coincides with the following expectation value in the auxiliary Hamiltonian system
\begin{equation}\label{eq:bilocal_operator_schwarzian_thermal}
    \langle \mathcal{O}_\Delta(\tau_1,\tau_2) \rangle_{\beta_\text{JT}} = \frac{\bra{L_\text{ren} = -\infty} e^{-(\beta - \tau_2 + \tau_1) \hat{H}_\text{JT}} \, e^{-\Delta \hat{L}_\text{ren}} \, e^{-(\tau_2 - \tau_1) \hat{H}_\text{JT}} \ket{{L_\text{ren} = -\infty}}}{\bra{L_\text{ren} = -\infty} e^{-\beta_\text{JT} \hat{H}_\text{JT}} \ket{{L_\text{ren} = -\infty}}}.
\end{equation}
Here we have inserted the operator version of the Green's function \eqref{eq:AdS_propagator_large_cutoff} between points on the asymptotic boundary.

Since an explicit microscopic dual boundary theory generalizing the Schwarzian at finite cutoff is not available\footnote{An exact boundary action for the finite cutoff Schwarzian mode is so far not available in closed form, although it is known to be a solution of an exact Riccati-type differential equation, as shown in \cite{Griguolo:2025kpi}. }, we do not have an operator such as \eqref{eq:bilocal_operator_schwarzian} at our disposal in the current setting.
Yet, the goal of this section is to generalize the right-hand side of Equation \eqref{eq:bilocal_operator_schwarzian_thermal} at finite cutoff, and propose it as a prescription for computing the thermal expectation value on the left-hand side in the one-dimensional theory dual to JT gravity at finite cutoff.
An analogous proposal for the time-ordered and out-of-time-order four-point functions will be presented as well.

To conclude, we must discuss the role the parameter $\Delta$ plays in the finite cutoff setting. At large cutoff, the dual boundary theory is conformally invariant and thus $\Delta$ is interpreted as the conformal dimension of the bi-local operator. At finite cutoff, conformal invariance of the dual boundary theory might be lost in principle, thus we will regard $\Delta$ simply as a label in what follows.
\begin{figure}
    \centering
    \includegraphics[width=0.6\linewidth]{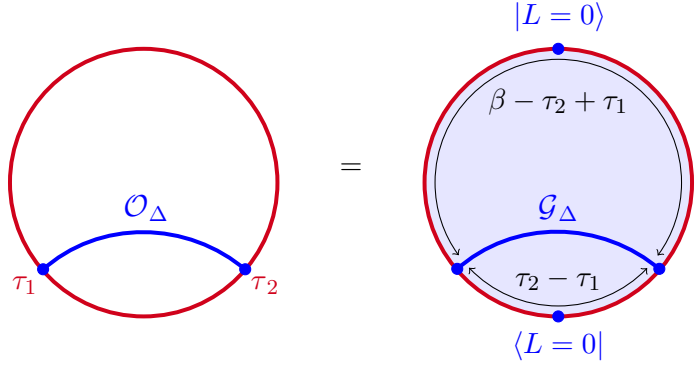}
    \caption{Graphical representation of Equation \eqref{eq:2pt_function}. The left-hand side denotes a thermal two-point function of operators separated by Euclidean time $\tau_2 - \tau_1$. In the gravitational description, this is represented by evolving the state $\ket{L=0}$ for Euclidean time $\tau_2-\tau_1$, inserting the non-local operator $G_\Delta(\hat{L})$ on the corresponding Cauchy slice, and then evolving for a further time $\beta - \tau_2 + \tau_1$ before projecting back onto $\ket{L=0}$.}
    \label{fig:2pt_function}
\end{figure}

\subsection{Two-point function}
\label{sec:2_point_function}
Motivated by the known result at large cutoff \eqref{eq:bilocal_operator_schwarzian_thermal}, we propose to compute the JT gravity two-point function at finite cutoff as\footnote{As always, \eqref{eq:2pt_function} is understood together with the prescriptions \eqref{eq:prescription_evolution_operator} and \eqref{eq:prescription_L=0}.
% In particular, for any insertion of the time evolution operator $e^{-\tau \hat{H}}$ we prescribe to include a factor of $e^{2 \pi \nu}$ in order to recover the expected results at large cutoff.
}
\begin{equation}\label{eq:2pt_function}
\begin{split}
    \langle \mathcal{O}_\Delta(\tau_1,\tau_2) \rangle_\beta
    &=
    \frac{
    \bra{L=0}
    e^{-(\beta-\tau_2+\tau_1)\hat{H}}\,
    G_{\Delta}(\hat{L})\,
    e^{-(\tau_2-\tau_1)\hat{H}}
    \ket{L=0}
    }
    {
    \bra{L=0} e^{-\beta\hat{H}} \ket{L=0}},
\end{split}
\end{equation}
for some operator $G_\Delta(\hat{L})$, see Figure \ref{fig:2pt_function}.
Since we lack an explicit boundary theory, we will not attempt to derive $G_\Delta(\hat{L})$ from first principles. Our only consistency requirement is that, in the large cutoff limit \eqref{eq:large_cutoff_limit}, such operator should reduce to $e^{-\Delta \hat{L}_\text{ren}}$.
% For instance, two reasonable candidates are given by
% \begin{equation}\label{eq:2pt_candidates}
%     \left( \frac{\Phi_b}{\cosh\!{(\hat{L}/2)}} \right)^{2\Delta}
%     \qquad \text{and} \qquad
%     \left( \frac{\Phi_b}{\cosh\!{(\hat{L}/2)}} \right)^{2\Delta} {}_2F_1\!\left( \Delta, \Delta, 2\Delta; \frac{1}{\cosh^2\!{(\hat{L}/2)}} \right).
% \end{equation}
% The most straightforward possibility is to consider
% \begin{equation}\label{eq:bulk_operator_exp}
%     G_\Delta(\hat{L})
%     =
%     \Phi_b^{2\Delta} e^{-\Delta\hat{L}},
% \end{equation}
% which indeed has the desired large cutoff limit.
A first option could be to insert the operator
\begin{equation}\label{eq:bulk_operator_cosh}
    \tilde{G}_\Delta(\hat{L}) = V(\hat{L})^\Delta = \left( \frac{\Phi_b}{\cosh\!{(\hat{L}/2)}} \right)^{2\Delta},
\end{equation}
where $V(\hat{L})$ is the potential term appearing in \eqref{eq:hamiltonian_LP}.
The rationale behind choosing \eqref{eq:bulk_operator_cosh} comes from the fact that the large cutoff two-point function is obtained by inserting an operator given by the $\Delta$-th power of the potential appearing in the corresponding Hamiltonian formulation \eqref{eq:hamiltonian_Liouville}
\begin{equation}
    G_\Delta(\hat{L}_\text{ren}) = V(\hat{L}_\text{ren})^{\Delta}, \qquad V(\hat{L}_\text{ren}) = e^{-\hat{L}_\text{ren}}.
\end{equation}
%Remarkably, \eqref{eq:bulk_operator_cosh} reduces to $e^{-\Delta \hat{L}_\text{ren}}$ at large cutoff.
Nevertheless, a more natural candidate would be given by
\begin{equation}\label{eq:bulk_operator_2F1}
    G_\Delta(\hat{L}) = \left( \frac{\Phi_b}{\cosh\!{(\hat{L}/2)}} \right)^{2\Delta} {}_2F_1\!\left( \Delta, \Delta, 2\Delta; \frac{1}{\cosh^2\!{(\hat{L}/2)}} \right),
\end{equation}
which is nothing but the operator version of the Green's function \eqref{eq:AdS_propagator}.
Moreover, an additional motivation for \eqref{eq:bulk_operator_2F1} comes from the representation theory of $\mathfrak{sl}(2,\mathbb{R})$ and will be discussed in Section \ref{sec:grouptheory}, where its structure is shown to emerge in relation to the discrete irreducible representations of the latter.

We notice that the Green's function operator can be expanded in a power series as
\begin{equation}\label{eq:Green_function_expansion}
    G_\Delta(\hat{L}) = \sum_{k = 0}^\infty \frac{(\Delta)_k^2 \, \Phi_b^{-2k}}{(2\Delta)_k \, k!} \left( \frac{\Phi_b}{\cosh\!{(\hat{L}/2)}} \right)^{2(\Delta+k)} = \sum_{k = 0}^\infty \frac{(\Delta)_k^2 \, \Phi_b^{-2k}}{(2\Delta)_k \, k!} \, \tilde{G}_{\Delta+k}(\hat{L}),
\end{equation}
meaning that inserting \eqref{eq:bulk_operator_2F1} amounts to inserting an infinite series of operators of the form \eqref{eq:bulk_operator_cosh} with increasing weights.
This motivates us to start by considering insertions of $\tilde{G}_\Delta(\hat{L})$, for which the computation of the two-point function \eqref{eq:2pt_function} boils down to performing the integral
\begin{equation}\label{eq:2pt_function_preliminary}
    \begin{split}
     \langle \tilde{\mathcal{O}}_\Delta(\tau_1,\tau_2) \rangle_\beta
     &= \frac{{e^{2 \pi \nu}}}{Z(\beta)} \bra{L=0} e^{-(\beta-\tau_2+\tau_1) \hat{H}} \, \left( \frac{\Phi_b}{\cosh\!{(\hat{L}/2)}} \right)^{2\Delta} \, e^{-(\tau_2-\tau_1) \hat{H}} \ket{L=0}\\
     &= \frac{{e^{2 \pi \nu}}}{Z(\beta)} \int_0^\nu \dd s_1 \, \dd s_2 \, \rho(s_1) \, \rho(s_2) \, e^{-(\beta-\tau_2+\tau_1) E(s_2)} \, e^{-(\tau_2-\tau_1) E(s_1)}  \, 
     \gamma_\Delta(s_1,s_2),
\end{split}
\end{equation}
where we defined 
\begin{equation}\label{eq:Delta_integral}
\begin{split}
    \gamma_\Delta(s_1,s_2) &\equiv \bra{s_2} \tilde{G}_\Delta\ket{s_1}
    = \Phi_b^{2\Delta} \int_0^\infty \dd L \, \cosh(L/2)^{-2\Delta} \, \psi_{s_1}(L) \, \psi_{s_2}^*(L)\\
    &= \Phi_b^{2\Delta} \, \Gamma(\Delta \pm is_1 \pm is_2) \, \mathbb{W}(s_2,\nu;\textstyle{\frac{1}{2}}+is_1, \textstyle{\frac{1}{2}}-is_1, \Delta - i\nu, \Delta + i\nu).
\end{split}
\end{equation}
The integral \eqref{eq:Delta_integral} is symmetric under $s_1 \leftrightarrow s_2$ and it is computed in Appendix \ref{appendix:exact2point}, where we also discuss the definition and some properties of the Wilson function $\mathbb{W}(\lambda,x;a,b,c,d)$ appearing in its closed-form expression.
It is interesting to notice that the form of \eqref{eq:2pt_function_preliminary} suggestively resembles the form of an out-of-time-ordered four-point function of operators of dimension $\Delta_A = \Delta$ and $\Delta_B = 1/2$ in JT gravity at large cutoff. This is in line with the previous observation that the partition function of the theory seemed to behave as a two-point function in large cutoff JT gravity for the insertion of $\Delta=1/2$ operators.

If we conveniently introduce the measure
\begin{equation}
    \dd\mu(s) \equiv \rho(s) \, \dd s = \frac{2 s \sinh(2 \pi s)}{\cosh(2 \pi s) + \cosh(2 \pi \nu)} \dd s,
\end{equation}
then our proposal for the two-point function of JT gravity at finite cutoff with the insertion of \eqref{eq:bulk_operator_2F1} takes the compact form
\begin{equation}\label{eq:2pt_function_exact}
\begin{split}
     \vcenter{\hbox{\begin{tikzpicture}[x=0.75pt,y=0.75pt,yscale=-1,xscale=1]
%uncomment if require: \path (0,108); %set diagram left start at 0, and has height of 108

%Shape: Ellipse [id:dp17060354802358102] 
\draw   (30,60) .. controls (30,43.43) and (43.43,30) .. (60,30) .. controls (76.57,30) and (90,43.43) .. (90,60) .. controls (90,76.57) and (76.57,90) .. (60,90) .. controls (43.43,90) and (30,76.57) .. (30,60) -- cycle ;
%Shape: Circle [id:dp033560362345279815] 
\draw  [color={rgb, 255:red, 75; green, 19; blue, 254 }  ,draw opacity=1 ][fill={rgb, 255:red, 75; green, 19; blue, 254 }  ,fill opacity=1 ] (81.23,79.72) .. controls (82.07,79.72) and (82.74,80.4) .. (82.74,81.23) .. controls (82.74,82.07) and (82.07,82.74) .. (81.23,82.74) .. controls (80.4,82.74) and (79.72,82.07) .. (79.72,81.23) .. controls (79.72,80.4) and (80.4,79.72) .. (81.23,79.72) -- cycle ;
%Shape: Ellipse [id:dp9986871344751356] 
\draw  [color={rgb, 255:red, 75; green, 19; blue, 254 }  ,draw opacity=1 ][fill={rgb, 255:red, 75; green, 19; blue, 254 }  ,fill opacity=1 ] (38.79,79.73) .. controls (39.62,79.73) and (40.3,80.41) .. (40.3,81.24) .. controls (40.3,82.08) and (39.62,82.76) .. (38.79,82.76) .. controls (37.95,82.76) and (37.28,82.08) .. (37.28,81.24) .. controls (37.28,80.41) and (37.95,79.73) .. (38.79,79.73) -- cycle ;
%Shape: Arc [id:dp3310402754319589] 
\draw  [draw opacity=0] (38.79,81.24) .. controls (44.22,75.82) and (51.72,72.46) .. (60,72.46) .. controls (68.28,72.46) and (75.78,75.82) .. (81.21,81.24) -- (60,102.46) -- cycle ; \draw  [color={rgb, 255:red, 75; green, 19; blue, 254 }  ,draw opacity=1 ] (38.79,81.24) .. controls (44.22,75.82) and (51.72,72.46) .. (60,72.46) .. controls (68.28,72.46) and (75.78,75.82) .. (81.21,81.24) ;  

% Text Node
\draw (36.79,84.64) node [anchor=north east] [inner sep=0.75pt]  [font=\scriptsize]  {$\tau _{1}$};
% Text Node
\draw (83.21,84.64) node [anchor=north west][inner sep=0.75pt]  [font=\scriptsize]  {$\tau _{2}$};
% Text Node
\draw (83.21,35.39) node [anchor=south west] [inner sep=0.75pt]  [font=\scriptsize,color={rgb, 255:red, 255; green, 255; blue, 255 }  ,opacity=1 ]  {$\tau _{3}$};
% Text Node
\draw (36.79,35.39) node [anchor=south east] [inner sep=0.75pt]  [font=\scriptsize,color={rgb, 255:red, 255; green, 255; blue, 255 }  ,opacity=1 ]  {$\tau _{4}$};
% Text Node
\draw (60,71.6) node [anchor=south] [inner sep=0.75pt]  [font=\scriptsize,color={rgb, 255:red, 75; green, 19; blue, 254 }  ,opacity=1 ]  {$\Delta $};

\end{tikzpicture}}}
     &= \langle \mathcal{O}_\Delta(\tau_1,\tau_2) \rangle_\beta
     =
     \sum_{k = 0}^\infty \frac{(\Delta)_k^2 \, \Phi_b^{-2k}}{(2\Delta)_k \, k!} \, \langle \tilde{\mathcal{O}}_{\Delta+k}(\tau_1,\tau_2) \rangle_\beta\\
     &=
     \frac{e^{2 \pi \nu}}{Z(\beta)} \int_0^\nu \dd\mu(s_1) \, \dd\mu(s_2) \, e^{-(\beta-\tau_2+\tau_1) E(s_2)} \, e^{-(\tau_2-\tau_1) E(s_1)}
     \times\\
     &\quad \times
     \sum_{k = 0}^\infty \frac{(\Delta)_k^2 \, \Phi_b^{-2k}}{(2\Delta)_k \, k!} \gamma_{\Delta+k}(s_1,s_2).
     % \Gamma(\Delta + k \pm i s_1 \pm i s_2) \, \mathbb{W}(s_1, \nu; \textstyle{\frac{1}{2}} + i s_2, \textstyle{\frac{1}{2}} - i s_2, \Delta + k - i \nu, \Delta + k + i \nu ),
     % \frac{\Phi_b^{2\Delta}}{\displaystyle{\int_0^\nu \dd\mu(s) \, e^{-\beta E(s)}}} \int_0^\nu \dd\mu(s_1) \, \dd\mu(s_2) \, e^{-(\beta-\tau_2+\tau_1) E(s_2)} \, e^{-(\tau_2-\tau_1) E(s_1)} \times\\
     % &\quad \times \Gamma(\Delta \pm i s_1 \pm i s_2) \, \mathbb{W}(s_1, \nu; \textstyle{\frac{1}{2}} + i s_2, \textstyle{\frac{1}{2}} - i s_2, \Delta - i \nu, \Delta + i \nu ),
\end{split}
\end{equation}

We remark that, at large cutoff $\Phi_b \to \infty$, the leading contribution to the two-point function comes from the $k=0$ term of the sum, reproducing the result by \cite{Mertens:2017mtv}.
This can readily be seen by recalling that the operator insertion \eqref{eq:bulk_operator_2F1} reduces to $e^{-\Delta \hat{L}_\text{ren}}$ at large cutoff.

It is now useful to analyze the semiclassical limit of the two-point function \eqref{eq:2pt_function_exact}. In Appendix \ref{app:semiclassical_2pt_function} we discuss how, taking $G_\text{N} \to 0$,\footnote{Once again, one reintroduces $G_\text{N}$ by sending $s \to s/G_\text{N}$ and $\nu \to \nu/G_\text{N}$.} the two-point function \eqref{eq:2pt_function_exact} behaves as
\begin{equation}\label{eq:saddle_L}
	\langle \mathcal{O}_\Delta(\tau_1,\tau_2) \rangle_\beta \overset{G_\text{N} \to 0}{\propto}
    %\left( \frac{\Phi_b}{\cosh(L_\text{E}/2)} \right)^{2 \Delta},
    G_\Delta(L_\text{E}),
    \qquad L_\text{E}(\tau_2-\tau_1,\beta) = 2 \sinh^{-1}\!\left( \frac{\beta}{2 \pi} \abs{ \sin\left( \pi \frac{\tau_2-\tau_1}{\beta} \right) } \right),
\end{equation}
at leading order.
The quantity $L_\text{E}(\tau_2-\tau_1,\beta)$ appearing in \eqref{eq:saddle_L} is precisely the geodesic distance in Euclidean AdS$_2$ between two boundary points separated by time $\abs{\tau_2-\tau_1}$ on a circle of length $\beta$.
To our knowledge, the two-point function \eqref{eq:2pt_function_exact} is the first closed-form proposal in the literature for which the amplitude localizes on the finite cutoff distance $L_\text{E}$ in the $G_\text{N} \to 0$ limit, suggesting that it is a plausible candidate for the two-point function in JT gravity at finite cutoff.\footnote{It is worth emphasizing that the finite cutoff two-point function obtained by naively replacing the energies in the Boltzmann factors with their $T\bar{T}$-deformed counterparts does not localize the amplitude about the geodesic length~\eqref{eq:saddle_L}, as observed in \cite{Griguolo:2025kpi}.}
This fact provides a non-trivial check of the finite cutoff construction and answers one of the open questions formulated in \cite{Griguolo:2025kpi}.

From a path integral point of view, the two-point function is expected to admit a worldline representation of the schematic form
\begin{equation}
   \langle \mathcal{O}_{\Delta} \rangle
   \sim
   \int \mathcal{D}g\,\mathcal{D}\Phi
   \int \mathcal{D}X\,
   \exp\!\left[
   -\frac{\Delta}{\hbar}
   \int d\lambda\,
   \sqrt{
   g_{\mu\nu}(X)
   \dot X^\mu \dot X^\nu
   }
   \right]
   \exp\!\left[
   -\frac{1}{8\pi G_N \hbar}
   S[g,\Phi]
   \right].
\end{equation}
Here, $X^\mu(\lambda)$ denotes the worldline of a probe particle of mass $\Delta$. This expression makes clear that the semiclassical limit involves two distinct steps. First, one takes $G_N\to 0$, which localizes the gravitational path integral on a classical JT geometry. Only after this saddle has been selected does one take the semiclassical limit of the worldline insertion, controlled by $\hbar\to 0$.

In this regime, the worldline path integral is dominated by the shortest trajectory connecting the two insertion points. Therefore the two-point function behaves as
\begin{equation}\label{eq:saddle_PI_worldline}
   \langle \mathcal{O}_{\Delta} \rangle
   \sim
   e^{-\Delta L_\text{E}/\hbar},
    %\left( \frac{\Phi_b}{\cosh(L_\text{E}/2)} \right)^{2 \Delta},
\end{equation}
where $L_\text{E}$ is the length of the geodesic connecting the two points on the thermal circle at which the operators are inserted.
Consistently, the Hamiltonian computation reproduces \eqref{eq:saddle_L} as $G_\text{N} \to 0$, and if one further takes $\hbar \to 0$\footnote{Here $\hbar$ is reintroduced by mapping $\Delta \to \Delta/\hbar$.
To be precise, the identity we are using here is
\begin{equation*}
    G_\Delta(L_\text{E}) \overset{\hbar \to 0}{\sim} \frac{e^{-\Delta (L_\text{E} - 2 \log(2\Phi_b))/\hbar}}{\sqrt{1 - e^{-2 L_\text{E}}}}.
\end{equation*}
Nevertheless, in \eqref{eq:saddle_L_worldline} we are dropping the one-loop factor as we are only interested in comparing our results with the worldline approximation \eqref{eq:saddle_PI_worldline} at exponential order.
}
\begin{equation}\label{eq:saddle_L_worldline}
    \langle \mathcal{O}_\Delta(\tau_1,\tau_2) \rangle_\beta \overset{G_\text{N} \to 0}{\propto}
    G_\Delta(L_\text{E})
    \overset{\hbar \to 0}{\sim} e^{-\Delta L_\text{E}/\hbar}
\end{equation}
at exponential order, yielding the expected semiclassical worldline result.

We must acknowledge that in \eqref{eq:JT_matter_action} no boundary terms or boundary conditions were discussed for the matter fields. As it stands, the results from this section are for matter fields that do not feel the Dirichlet boundary, which instead only affects gravity. We leave a systematic discussion for finite cutoff boundary conditions on matter to future work.

\subsection{Four-point function (time-ordered)}
The machinery developed in the previous section can also be used to compute higher-point correlation functions of bi-local operators in the dual one-dimensional theory. After the two-point function, the simplest instance of correlation function we are able to compute is the four-point function: the latter is obtained as an immediate generalization of the prescription from the previous section
\begin{equation}\label{eq:4pt_function}
\begin{split}
     \langle \mathcal{O}_{\Delta_A}(\tau_1,\tau_2) &\mathcal{O}_{\Delta_B}(\tau_3,\tau_4) \rangle_\beta =\\
     &= \frac{\bra{L=0} e^{-(\tau_4-\tau_3) \hat{H}} \, G_{\Delta_B}(\hat{L}) \, e^{-(\beta-\tau_2+\tau_1-\tau_4+\tau_3) \hat{H}} G_{\Delta_A}(\hat{L}) e^{-(\tau_2 -\tau_1) \hat{H}} \ket{L=0}}{\bra{L=0} e^{-\beta \hat{H}} \ket{L=0}},
\end{split}
\end{equation}
where the insertion times are ordered as $\tau_1 < \tau_2 < \tau_3 < \tau_4$.
Once again, we start preliminarily by considering insertions of $\tilde{G}_\Delta(\hat{L})$ \eqref{eq:bulk_operator_cosh}. Employing the closed form of the integral \eqref{eq:Delta_integral} twice, we find
\begin{equation}\label{eq:4pt_function_preliminary}
    \begin{split}
    \langle \tilde{\mathcal{O}}_{\Delta_A}(\tau_1,\tau_2) \tilde{\mathcal{O}}_{\Delta_B}(\tau_3,\tau_4) \rangle_\beta
    &=
    \frac{e^{2 \pi \nu}}{Z(\beta)} \int_0^\nu \prod_{j=1}^3 \dd\mu(s_j) \,
    \gamma_{\Delta_A}(s_1,s_2) \, \gamma_{\Delta_B}(s_2,s_3) \times\\
    &\quad\times e^{-(\tau_4-\tau_3) E(s_3)} \, e^{-(\beta-\tau_2+\tau_1-\tau_4+\tau_3) E(s_2)} \, e^{-(\tau_2-\tau_1) E(s_1)}
    % \\ &\quad \times \Phi_b^{2\Delta_A} \, \Gamma(\Delta_A \pm is_1 \pm is_2) \, \mathbb{W}(s_1, \nu; \textstyle{\frac{1}{2}} + i s_2, \textstyle{\frac{1}{2}} - i s_2, \Delta_A - i \nu, \Delta_A + i \nu ) \times\\
    % &\quad \times \Phi_b^{2\Delta_B} \, \Gamma(\Delta_B \pm i s_2 \pm i s_3) \, \mathbb{W}(s_3, \nu; \textstyle{\frac{1}{2}} + i s_2, \textstyle{\frac{1}{2}} - i s_2, \Delta_B - i \nu, \Delta_B + i \nu )
    .
\end{split}
\end{equation}
Consequently, using \eqref{eq:Green_function_expansion}, the time-ordered four-point function is readily found to be
\begin{equation}\label{eq:4pt_function_exact}
    \begin{split}
    \vcenter{\hbox{\begin{tikzpicture}[x=0.75pt,y=0.75pt,yscale=-1,xscale=1]
%uncomment if require: \path (0,108); %set diagram left start at 0, and has height of 108

%Shape: Ellipse [id:dp4528008304302231] 
\draw   (30,60) .. controls (30,43.43) and (43.43,30) .. (60,30) .. controls (76.57,30) and (90,43.43) .. (90,60) .. controls (90,76.57) and (76.57,90) .. (60,90) .. controls (43.43,90) and (30,76.57) .. (30,60) -- cycle ;
%Shape: Circle [id:dp635799110068347] 
\draw  [color={rgb, 255:red, 75; green, 19; blue, 254 }  ,draw opacity=1 ][fill={rgb, 255:red, 75; green, 19; blue, 254 }  ,fill opacity=1 ] (81.23,79.72) .. controls (82.07,79.72) and (82.74,80.4) .. (82.74,81.23) .. controls (82.74,82.07) and (82.07,82.74) .. (81.23,82.74) .. controls (80.4,82.74) and (79.72,82.07) .. (79.72,81.23) .. controls (79.72,80.4) and (80.4,79.72) .. (81.23,79.72) -- cycle ;
%Shape: Ellipse [id:dp27156836809545915] 
\draw  [color={rgb, 255:red, 75; green, 19; blue, 254 }  ,draw opacity=1 ][fill={rgb, 255:red, 75; green, 19; blue, 254 }  ,fill opacity=1 ] (38.79,79.73) .. controls (39.62,79.73) and (40.3,80.41) .. (40.3,81.24) .. controls (40.3,82.08) and (39.62,82.76) .. (38.79,82.76) .. controls (37.95,82.76) and (37.28,82.08) .. (37.28,81.24) .. controls (37.28,80.41) and (37.95,79.73) .. (38.79,79.73) -- cycle ;
%Shape: Ellipse [id:dp03554215030394481] 
\draw  [color={rgb, 255:red, 208; green, 2; blue, 27 }  ,draw opacity=1 ][fill={rgb, 255:red, 208; green, 2; blue, 27 }  ,fill opacity=1 ] (81.23,37.29) .. controls (82.07,37.29) and (82.74,37.96) .. (82.74,38.8) .. controls (82.74,39.63) and (82.07,40.31) .. (81.23,40.31) .. controls (80.4,40.31) and (79.72,39.63) .. (79.72,38.8) .. controls (79.72,37.96) and (80.4,37.29) .. (81.23,37.29) -- cycle ;
%Shape: Circle [id:dp7321915588636378] 
\draw  [color={rgb, 255:red, 208; green, 2; blue, 27 }  ,draw opacity=1 ][fill={rgb, 255:red, 208; green, 2; blue, 27 }  ,fill opacity=1 ] (38.79,37.28) .. controls (39.62,37.28) and (40.3,37.95) .. (40.3,38.79) .. controls (40.3,39.62) and (39.62,40.3) .. (38.79,40.3) .. controls (37.95,40.3) and (37.28,39.62) .. (37.28,38.79) .. controls (37.28,37.95) and (37.95,37.28) .. (38.79,37.28) -- cycle ;
%Shape: Arc [id:dp38167497963504504] 
\draw  [draw opacity=0] (38.79,81.24) .. controls (44.22,75.82) and (51.72,72.46) .. (60,72.46) .. controls (68.28,72.46) and (75.78,75.82) .. (81.21,81.24) -- (60,102.46) -- cycle ; \draw  [color={rgb, 255:red, 75; green, 19; blue, 254 }  ,draw opacity=1 ] (38.79,81.24) .. controls (44.22,75.82) and (51.72,72.46) .. (60,72.46) .. controls (68.28,72.46) and (75.78,75.82) .. (81.21,81.24) ;  
%Shape: Arc [id:dp1340120454881597] 
\draw  [draw opacity=0] (38.79,38.79) .. controls (44.22,44.22) and (51.72,47.57) .. (60,47.57) .. controls (68.28,47.57) and (75.78,44.22) .. (81.21,38.79) -- (60,17.57) -- cycle ; \draw  [color={rgb, 255:red, 208; green, 2; blue, 27 }  ,draw opacity=1 ] (38.79,38.79) .. controls (44.22,44.22) and (51.72,47.57) .. (60,47.57) .. controls (68.28,47.57) and (75.78,44.22) .. (81.21,38.79) ;  

% Text Node
\draw (36.79,84.64) node [anchor=north east] [inner sep=0.75pt]  [font=\scriptsize]  {$\tau _{1}$};
% Text Node
\draw (83.21,84.64) node [anchor=north west][inner sep=0.75pt]  [font=\scriptsize]  {$\tau _{2}$};
% Text Node
\draw (83.21,35.39) node [anchor=south west] [inner sep=0.75pt]  [font=\scriptsize]  {$\tau _{3}$};
% Text Node
\draw (36.79,35.39) node [anchor=south east] [inner sep=0.75pt]  [font=\scriptsize]  {$\tau _{4}$};
% Text Node
\draw (60,71.6) node [anchor=south] [inner sep=0.75pt]  [font=\scriptsize,color={rgb, 255:red, 75; green, 19; blue, 254 }  ,opacity=1 ]  {$\Delta _{A}$};
% Text Node
\draw (60,48.4) node [anchor=north] [inner sep=0.75pt]  [font=\scriptsize,color={rgb, 255:red, 208; green, 2; blue, 27 }  ,opacity=1 ]  {$\Delta _{B}$};

\end{tikzpicture}}} &= 
    \langle\mathcal{O}_{\Delta_A}(\tau_1,\tau_2) \mathcal{O}_{\Delta_B}(\tau_3,\tau_4) \rangle_\beta \\
    &=
    \sum_{k,l = 0}^\infty \frac{(\Delta_A)_{k}^2 \, \Phi_b^{-2k}}{(2\Delta_A)_k \, k!}
    \frac{(\Delta_B)_{l}^2 \, \Phi_b^{-2l}}{(2\Delta_B)_l \, l!}\, \langle \tilde{\mathcal{O}}_{\Delta_A+k}(\tau_1,\tau_2) \tilde{\mathcal{O}}_{\Delta_B+l}(\tau_3,\tau_4) \rangle_\beta\\
    &=
    \frac{e^{2 \pi \nu}}{Z(\beta)} \int_0^\nu \prod_{j=1}^3 \dd\mu(s_j) \, e^{-(\tau_4-\tau_3) E(s_3)} \, e^{-(\beta-\tau_2+\tau_1-\tau_4+\tau_3) E(s_2)} \, e^{-(\tau_2-\tau_1) E(s_1)} \times\\
    &\quad \times \sum_{k,l = 0}^\infty \frac{(\Delta_A)_{k}^2 \,\Phi_b^{-2k}}{(2\Delta_A)_k \, k!} 
    % \Gamma(\Delta_A + k \pm is_1 \pm is_2) \, \mathbb{W}(s_1, \nu; \textstyle{\frac{1}{2}} + i s_2, \textstyle{\frac{1}{2}} - i s_2, \Delta_A + k - i \nu, \Delta_A + k + i \nu ) \times\\
    % &\quad \times
    \frac{(\Delta_B)_{l}^2 \,\Phi_b^{-2l}}{(2\Delta_B)_l \, l!} \gamma_{\Delta_A + k}(s_1,s_2) \, \gamma_{\Delta_B + l}(s_2,s_3)
    % \Gamma(\Delta_B + l \pm i s_2 \pm i s_3) \, \mathbb{W}(s_3, \nu; \textstyle{\frac{1}{2}} + i s_2, \textstyle{\frac{1}{2}} - i s_2, \Delta_B + l - i \nu, \Delta_B + l + i \nu )
\end{split}
\end{equation}
in terms of \eqref{eq:4pt_function_preliminary}.

\subsection{Four-point function (OTOC)}
The study of four-point functions with out-of-time-order correlators (OTOCs) allows for the characterization of quantum chaos within the dynamics of systems in a black hole background. In the large cutoff setting, exact knowledge of four-point functions in JT gravity \cite{Mertens:2017mtv} made it possible to confirm the conjecture due to Maldacena, Shenker and Stanford \cite{Maldacena:2015waa} according to which black holes are ``maximally chaotic'' systems, namely they exhibit the largest possible Lyapunov exponent $\lambda_\text{max}=\frac{2\pi}{\hbar \beta}$.

In the current setting, we can mimic the methods first used in \cite{Mertens:2017mtv} to compute a finite cutoff version of the four-point OTOC. The proposal consists of inserting a ``fusion kernel'' within the four-point TOC. The kernel used there is given by the $R$-matrix
\begin{equation}\label{eq:R-matrix}
\begin{split}
    R_{s_4 s_2}&\!\begin{bmatrix}
        s_3 & \Delta_B \\ s_1 & \Delta_A 
    \end{bmatrix}
    = \mathbb{W}(s_4, s_2; \Delta_A + i s_3, \Delta_A - is_3, \Delta_B - is_1, \Delta_B + is_1) \times\\
    &\times \sqrt{\Gamma(\Delta_B \pm is_1 \pm is_2) \Gamma(\Delta_A \pm is_2 \pm is_3) \, \Gamma(\Delta_B \pm is_3 \pm is_4) \Gamma(\Delta_A \pm is_4 \pm is_1)}.
\end{split}
\end{equation}
We conjecture that the prescription to obtain the four-point OTOC in the finite cutoff setting consists in inserting the same kernel \eqref{eq:R-matrix} into the time-ordered four-point TOC, the only difference being that the initial object is now the finite cutoff correlation function \eqref{eq:4pt_function_exact} instead of its large cutoff limit as in \cite{Mertens:2017mtv}. We are motivated in proposing this approach by the fact that the crossing is expected to take place in the bulk and thus the kernel that realizes it should be unrelated to the location of the boundary.
We start from the insertion of $\tilde{G}_\Delta(\hat{L})$ \eqref{eq:bulk_operator_cosh}: for $\tau_1 < \tau_2 < \tau_3 < \tau_4$, we have
\begin{equation}\label{eq:4pt_OTOC_preliminary}
    \begin{split}
        \langle &\tilde{\mathcal{O}}_{\Delta_A}(\tau_1,\tau_3) \tilde{\mathcal{O}}_{\Delta_B}(\tau_2,\tau_4) \rangle_\beta =\\ 
        &= \frac{e^{2 \pi \nu}}{Z(\beta)} \int_0^\nu \prod_{j=1}^4 \dd\mu(s_j) \, e^{-(\beta+\tau_1-\tau_4) E(s_4)} e^{-(\tau_4-\tau_3) E(s_3)} e^{-(\tau_3-\tau_2) E(s_2)} e^{-(\tau_2-\tau_1) E(s_1)} \times\\
        &\quad\times \sqrt{\gamma_{\Delta_B}(s_1,s_2) \, \gamma_{\Delta_A}(s_2,s_3) \, \gamma_{\Delta_B}(s_3,s_4) \, \gamma_{\Delta_A}(s_4,s_1)} \, R_{s_4 s_2}\!\begin{bmatrix}
        s_3 & \Delta_B \\ s_1 & \Delta_A 
        \end{bmatrix}.
        % &\quad \times \Phi_b^{2(\Delta_A + \Delta_B)} \, \Gamma\!\left(\scriptstyle{\Delta_B \pm is_1 \pm is_2}\right) \, \Gamma\!\left(\scriptstyle{\Delta_A \pm is_2 \pm is_3}\right) \, \Gamma\!\left(\scriptstyle{\Delta_B \pm is_3 \pm is_4}\right) \, \Gamma\!\left(\scriptstyle{\Delta_A \pm is_4 \pm is_1}\right) \times\\
        % &\quad \times \sqrt{\mathbb{W}\!\left( \scriptstyle{ s_1, \nu; \frac{1}{2} + i s_2, \frac{1}{2} - i s_2, \Delta_B - i \nu, \Delta_B + i \nu} \right) \, \mathbb{W}\!\left( \scriptstyle{s_2, \nu; \frac{1}{2} + i s_3, \frac{1}{2} - i s_3, \Delta_A - i \nu, \Delta_A + i \nu} \right)} \times\\
        % &\quad \times \sqrt{\mathbb{W}\!\left( \scriptstyle{s_3, \nu; \frac{1}{2} + i s_4, \frac{1}{2} - i s_4, \Delta_B - i \nu, \Delta_B + i \nu} \right) \, \mathbb{W}\!\left( \scriptstyle{s_4, \nu; \frac{1}{2} + i s_1, \frac{1}{2} - i s_1, \Delta_A - i \nu, \Delta_A + i \nu} \right)} \times\\
        % &\quad \times \mathbb{W}\!\left( \scriptstyle{s_4, s_2; \Delta_A + i s_3, \Delta_A - i s_3, \Delta_B - is_1, \Delta_B + is_1} \right).
    \end{split}
\end{equation}
Finally, as a result of our conjecture, we propose the four-point OTOC to be given by
\begin{equation}\label{eq:4pt_OTOC}
    \begin{split}
    &\vcenter{\hbox{\begin{tikzpicture}[x=0.75pt,y=0.75pt,yscale=-1,xscale=1]
%uncomment if require: \path (0,108); %set diagram left start at 0, and has height of 108

%Shape: Ellipse [id:dp07233853854841343] 
\draw   (30,60) .. controls (30,43.43) and (43.43,30) .. (60,30) .. controls (76.57,30) and (90,43.43) .. (90,60) .. controls (90,76.57) and (76.57,90) .. (60,90) .. controls (43.43,90) and (30,76.57) .. (30,60) -- cycle ;
%Shape: Circle [id:dp7547853600679387] 
\draw  [color={rgb, 255:red, 208; green, 2; blue, 27 }  ,draw opacity=1 ][fill={rgb, 255:red, 208; green, 2; blue, 27 }  ,fill opacity=1 ] (81.23,79.72) .. controls (82.07,79.72) and (82.74,80.4) .. (82.74,81.23) .. controls (82.74,82.07) and (82.07,82.74) .. (81.23,82.74) .. controls (80.4,82.74) and (79.72,82.07) .. (79.72,81.23) .. controls (79.72,80.4) and (80.4,79.72) .. (81.23,79.72) -- cycle ;
%Shape: Ellipse [id:dp7556281020879838] 
\draw  [color={rgb, 255:red, 75; green, 19; blue, 254 }  ,draw opacity=1 ][fill={rgb, 255:red, 75; green, 19; blue, 254 }  ,fill opacity=1 ] (38.79,79.73) .. controls (39.62,79.73) and (40.3,80.41) .. (40.3,81.24) .. controls (40.3,82.08) and (39.62,82.76) .. (38.79,82.76) .. controls (37.95,82.76) and (37.28,82.08) .. (37.28,81.24) .. controls (37.28,80.41) and (37.95,79.73) .. (38.79,79.73) -- cycle ;
%Shape: Ellipse [id:dp7132781470967113] 
\draw  [color={rgb, 255:red, 75; green, 19; blue, 254 }  ,draw opacity=1 ][fill={rgb, 255:red, 75; green, 19; blue, 254 }  ,fill opacity=1 ] (81.23,37.29) .. controls (82.07,37.29) and (82.74,37.96) .. (82.74,38.8) .. controls (82.74,39.63) and (82.07,40.31) .. (81.23,40.31) .. controls (80.4,40.31) and (79.72,39.63) .. (79.72,38.8) .. controls (79.72,37.96) and (80.4,37.29) .. (81.23,37.29) -- cycle ;
%Shape: Circle [id:dp9865360301572264] 
\draw  [color={rgb, 255:red, 208; green, 2; blue, 27 }  ,draw opacity=1 ][fill={rgb, 255:red, 208; green, 2; blue, 27 }  ,fill opacity=1 ] (38.79,37.28) .. controls (39.62,37.28) and (40.3,37.95) .. (40.3,38.79) .. controls (40.3,39.62) and (39.62,40.3) .. (38.79,40.3) .. controls (37.95,40.3) and (37.28,39.62) .. (37.28,38.79) .. controls (37.28,37.95) and (37.95,37.28) .. (38.79,37.28) -- cycle ;
%Straight Lines [id:da6775234447817879] 
\draw [color={rgb, 255:red, 208; green, 2; blue, 27 }  ,draw opacity=1 ]   (38.79,38.79) -- (81.23,81.23) ;
%Straight Lines [id:da8917446091160319] 
\draw [color={rgb, 255:red, 255; green, 255; blue, 255 }  ,draw opacity=1 ][line width=3.75]    (50,70) -- (70,50) ;
%Straight Lines [id:da8020067315693074] 
\draw [color={rgb, 255:red, 75; green, 19; blue, 254 }  ,draw opacity=1 ]   (81.23,38.8) -- (38.79,81.24) ;

% Text Node
\draw (36.79,84.64) node [anchor=north east] [inner sep=0.75pt]  [font=\scriptsize]  {$\tau _{1}$};
% Text Node
\draw (83.21,84.64) node [anchor=north west][inner sep=0.75pt]  [font=\scriptsize]  {$\tau _{2}$};
% Text Node
\draw (83.23,35.4) node [anchor=south west] [inner sep=0.75pt]  [font=\scriptsize]  {$\tau _{3}$};
% Text Node
\draw (36.79,35.39) node [anchor=south east] [inner sep=0.75pt]  [font=\scriptsize]  {$\tau _{4}$};
% Text Node
\draw (43.79,81.24) node [anchor=west] [inner sep=0.75pt]  [font=\scriptsize,color={rgb, 255:red, 75; green, 19; blue, 254 }  ,opacity=1 ]  {$\Delta _{A}$};
% Text Node
\draw (43.79,38.79) node [anchor=west] [inner sep=0.75pt]  [font=\scriptsize,color={rgb, 255:red, 208; green, 2; blue, 27 }  ,opacity=1 ]  {$\Delta _{B}$};

\end{tikzpicture}}} = 
    \langle \mathcal{O}_{\Delta_A}(\tau_1,\tau_3) \mathcal{O}_{\Delta_B}(\tau_2,\tau_4) \rangle_\beta \\
    &= \sum_{k,l = 0}^\infty \frac{(\Delta_A)_{k}^2 \, \Phi_b^{-2k}}{(2\Delta_A)_k \, k!}
    \frac{(\Delta_B)_{l}^2 \, \Phi_b^{-2l}}{(2\Delta_B)_l \, l!}\, \langle \tilde{\mathcal{O}}_{\Delta_A+k}(\tau_1,\tau_3) \tilde{\mathcal{O}}_{\Delta_B+l}(\tau_2,\tau_4) \rangle_\beta\\
    &= \frac{e^{2 \pi \nu}}{Z(\beta)} \int_0^\nu \prod_{j=1}^4 \dd\mu(s_j) \, e^{-(\beta+\tau_1-\tau_4) E(s_4)} e^{-(\tau_4-\tau_3) E(s_3)} e^{-(\tau_3-\tau_2) E(s_2)} e^{-(\tau_2-\tau_1) E(s_1)} \times\\
    &\quad \times \sum_{k,l = 0}^\infty \frac{(\Delta_A)_{k}^2 \, \Phi_b^{-2k}}{(2\Delta_A)_k \, k!}
    \frac{(\Delta_B)_{l}^2 \, \Phi_b^{-2l}}{(2\Delta_B)_l \, l!} \, \sqrt{\gamma_{\Delta_B + l}(s_1,s_2) \, \gamma_{\Delta_A + k}(s_2,s_3) \, \gamma_{\Delta_B + l}(s_3,s_4) \, \gamma_{\Delta_A + k}(s_4,s_1)} \times\\
    &\quad\times R_{s_4 s_2}\!\begin{bmatrix}
    s_3 & \Delta_B + l \\ s_1 & \Delta_A + k 
    \end{bmatrix}.
    % &\quad \times \Gamma\!\left(\scriptstyle{\Delta_B + l \pm is_1 \pm is_2}\right) \, \Gamma\!\left(\scriptstyle{\Delta_A + k \pm is_2 \pm is_3}\right) \, \Gamma\!\left(\scriptstyle{\Delta_B + l \pm is_3 \pm is_4}\right) \, \Gamma\!\left(\scriptstyle{\Delta_A + k \pm is_4 \pm is_1}\right) \times\\
    % &\quad \times \sqrt{\mathbb{W}\!\left( \scriptstyle{ s_1, \nu; \frac{1}{2} + i s_2, \frac{1}{2} - i s_2, \Delta_B + l - i \nu, \Delta_B + l + i \nu} \right) \, \mathbb{W}\!\left( \scriptstyle{s_2, \nu; \frac{1}{2} + i s_3, \frac{1}{2} - i s_3, \Delta_A + k - i \nu, \Delta_A + k + i \nu} \right)} \times\\
    % &\quad \times \sqrt{\mathbb{W}\!\left( \scriptstyle{s_3, \nu; \frac{1}{2} + i s_4, \frac{1}{2} - i s_4, \Delta_B + l - i \nu, \Delta_B + l + i \nu} \right) \, \mathbb{W}\!\left( \scriptstyle{s_4, \nu; \frac{1}{2} + i s_1, \frac{1}{2} - i s_1, \Delta_A + k - i \nu, \Delta_A + k + i \nu} \right)} \times\\
    %  &\quad \times \mathbb{W}\!\left( \scriptstyle{s_4, s_2; \Delta_A + k + i s_3, \Delta_A + k - i s_3, \Delta_B + l - is_1, \Delta_B + l + is_1} \right).
\end{split}
\end{equation}
It would be interesting to investigate whether the Lyapunov exponent corresponding to \eqref{eq:4pt_OTOC} is maximal also in the finite cutoff framework, comparing the exact formulas presented here with the analysis of chaos in the deformed Schwarzian theory \cite{Gross:2019ach}.

\subsection{Diagrammatic rules}
The results from the previous sections can be summarized in terms of diagrammatic rules for the computation of $n$-point correlation functions in JT gravity at finite cutoff, and can be seen as a generalization of the ones from ordinary JT gravity \cite{Mertens:2017mtv}.
A generic correlation function at finite cutoff is found as follows: first, assign a momentum/energy label $s_j$ to each bulk sector in the diagram so that each contributes an integration measure
\begin{equation}
    \int_0^\nu \dd \mu(s_j) = \int_0^\nu \rho(s_j) \, \dd s_j = \int_0^\nu  \frac{2 s_j \, \sinh(2 \pi s_j)}{\cosh(2 \pi s_j) + \cosh(2 \pi \nu)} \dd s_j.
\end{equation}
Then, every boundary arc corresponds to an exponential factor
\begin{equation}
    \vcenter{\hbox{\begin{tikzpicture}[x=0.75pt,y=0.75pt,yscale=-1,xscale=1]
%uncomment if require: \path (0,92); %set diagram left start at 0, and has height of 92

%Shape: Arc [id:dp45146213919360634] 
\draw  [draw opacity=0] (31.8,49.74) .. controls (35.99,38.22) and (47.04,30) .. (60,30) .. controls (72.96,30) and (84.01,38.22) .. (88.2,49.74) -- (60,60) -- cycle ; \draw   (31.8,49.74) .. controls (35.99,38.22) and (47.04,30) .. (60,30) .. controls (72.96,30) and (84.01,38.22) .. (88.2,49.74) ;  
%Shape: Ellipse [id:dp9489798046138754] 
\draw  [color={rgb, 255:red, 0; green, 0; blue, 0 }  ,draw opacity=1 ][fill={rgb, 255:red, 0; green, 0; blue, 0 }  ,fill opacity=1 ] (88.2,48.23) .. controls (89.03,48.23) and (89.71,48.9) .. (89.71,49.74) .. controls (89.71,50.57) and (89.03,51.25) .. (88.2,51.25) .. controls (87.36,51.25) and (86.69,50.57) .. (86.69,49.74) .. controls (86.69,48.9) and (87.36,48.23) .. (88.2,48.23) -- cycle ;
%Shape: Ellipse [id:dp8758898354596837] 
\draw  [color={rgb, 255:red, 0; green, 0; blue, 0 }  ,draw opacity=1 ][fill={rgb, 255:red, 0; green, 0; blue, 0 }  ,fill opacity=1 ] (31.8,48.23) .. controls (32.64,48.23) and (33.31,48.9) .. (33.31,49.74) .. controls (33.31,50.57) and (32.64,51.25) .. (31.8,51.25) .. controls (30.97,51.25) and (30.29,50.57) .. (30.29,49.74) .. controls (30.29,48.9) and (30.97,48.23) .. (31.8,48.23) -- cycle ;

% Text Node
\draw (60,42) node  [font=\scriptsize]  {$s_{j}$};
% Text Node
\draw (29.8,53.14) node [anchor=north east] [inner sep=0.75pt]  [font=\scriptsize]  {$\tau _{2}$};
% Text Node
\draw (90.2,53.14) node [anchor=north west][inner sep=0.75pt]  [font=\scriptsize]  {$\tau _{1}$};

\end{tikzpicture}}} = e^{-(\tau_2 - \tau_1) E(s_j)},
\end{equation}
where the label $s_j$ is the same as
that of the bulk sector bounded by the arc.
Each vertex contributes
\begin{equation}
    \vcenter{\hbox{\begin{tikzpicture}[x=0.75pt,y=0.75pt,yscale=-1,xscale=1]
%uncomment if require: \path (0,92); %set diagram left start at 0, and has height of 92

%Shape: Arc [id:dp5418980408769825] 
\draw  [draw opacity=0] (70.26,11.8) .. controls (81.78,15.99) and (90,27.04) .. (90,40) .. controls (90,52.96) and (81.78,64.01) .. (70.26,68.2) -- (60,40) -- cycle ; \draw   (70.26,11.8) .. controls (81.78,15.99) and (90,27.04) .. (90,40) .. controls (90,52.96) and (81.78,64.01) .. (70.26,68.2) ;  
%Straight Lines [id:da13261003377031944] 
\draw [color={rgb, 255:red, 75; green, 19; blue, 254 }  ,draw opacity=1 ]   (90,40) -- (60,40) ;
%Shape: Ellipse [id:dp6413856313609425] 
\draw  [color={rgb, 255:red, 75; green, 19; blue, 254 }  ,draw opacity=1 ][fill={rgb, 255:red, 75; green, 19; blue, 254 }  ,fill opacity=1 ] (90,38.49) .. controls (90.83,38.49) and (91.51,39.17) .. (91.51,40) .. controls (91.51,40.83) and (90.83,41.51) .. (90,41.51) .. controls (89.17,41.51) and (88.49,40.83) .. (88.49,40) .. controls (88.49,39.17) and (89.17,38.49) .. (90,38.49) -- cycle ;

% Text Node
\draw (75,26.5) node  [font=\scriptsize]  {$s_{1}$};
% Text Node
\draw (75,53.5) node  [font=\scriptsize]  {$s_{2}$};
% Text Node
\draw (59,40) node [anchor=east] [inner sep=0.75pt]  [font=\scriptsize,color={rgb, 255:red, 75; green, 19; blue, 254 }  ,opacity=1 ]  {$\Delta $};

\end{tikzpicture}}} =
    \sqrt{\gamma_\Delta(s_1,s_2)}
    % \Phi_b^\Delta \sqrt{ \sum_{k = 0}^\infty \frac{(\Delta)_{k}^2}{(2\Delta)_k \, k!} \Gamma(\Delta + k \pm is_1 \pm is_2) \, \mathbb{W}(s_1,\nu; 1/2 + is_2, 1/2 - is_2, \Delta + k - i\nu, \Delta + k + i\nu)}.
\end{equation}
Each crossing between two matter lines in the bulk corresponds to
\begin{equation}
    \begin{split}
        \vcenter{\hbox{\begin{tikzpicture}[x=0.75pt,y=0.75pt,yscale=-1,xscale=1]
%uncomment if require: \path (0,108); %set diagram left start at 0, and has height of 108

%Straight Lines [id:da4068651937310872] 
\draw [color={rgb, 255:red, 208; green, 2; blue, 27 }  ,draw opacity=1 ]   (38.79,38.79) -- (81.23,81.23) ;
%Straight Lines [id:da9423492681284427] 
\draw [color={rgb, 255:red, 255; green, 255; blue, 255 }  ,draw opacity=1 ][line width=3.75]    (50,70) -- (70,50) ;
%Straight Lines [id:da9302289918664595] 
\draw [color={rgb, 255:red, 75; green, 19; blue, 254 }  ,draw opacity=1 ]   (81.23,38.8) -- (38.79,81.24) ;

% Text Node
\draw (83.23,44.2) node [anchor=north] [inner sep=0.75pt]  [font=\scriptsize,color={rgb, 255:red, 75; green, 19; blue, 254 }  ,opacity=1 ]  {$\Delta _{A}$};
% Text Node
\draw (36.79,44.19) node [anchor=north] [inner sep=0.75pt]  [font=\scriptsize,color={rgb, 255:red, 208; green, 2; blue, 27 }  ,opacity=1 ]  {$\Delta _{B}$};
% Text Node
\draw (60,71.4) node [anchor=north] [inner sep=0.75pt]  [font=\scriptsize]  {$s_{1}$};
% Text Node
\draw (70,60) node [anchor=west] [inner sep=0.75pt]  [font=\scriptsize]  {$s_{2}$};
% Text Node
\draw (60,48.6) node [anchor=south] [inner sep=0.75pt]  [font=\scriptsize]  {$s_{3}$};
% Text Node
\draw (50,60) node [anchor=east] [inner sep=0.75pt]  [font=\scriptsize]  {$s_{4}$};

\end{tikzpicture}}} &= R_{s_4 s_2}\!\begin{bmatrix}
        s_3 & \Delta_B \\ s_1 & \Delta_A 
        \end{bmatrix}
        % \\
        % &= \sqrt{\Gamma(\Delta_B \pm is_1 \pm is_2) \Gamma(\Delta_A \pm is_2 \pm is_3) \Gamma(\Delta_B \pm is_3 \pm is_4) \Gamma(\Delta_A \pm is_4 \pm is_1)} \times\\
        % &\quad \times \mathbb{W}(s_4, s_2; \Delta_A + i s_3, \Delta_A - is_3, \Delta_B - is_1, \Delta_B + is_1).
    \end{split}
\end{equation}
Finally, divide by the rescaled partition function
\begin{equation}
    e^{-2 \pi \nu} \, Z(\beta) = \int_0^\nu \dd\mu(s) \, e^{-\beta E(s)}
\end{equation}
and sum over all matter line labels as
\begin{equation}
    \sum_{k=0}^\infty \frac{(\Delta)_k^2 \, \Phi_b^{-2k}}{(2 \Delta)_k \, k!} (\Delta \to \Delta + k).
\end{equation}
For diagrams without crossings, both endpoints of every line border the same pair of sectors,
so the sum over $k$ can be performed for each line individually, allowing us to recover \eqref{eq:2pt_function_exact} and
\eqref{eq:4pt_function_exact}.

\section{The scattering problem and $\mathfrak{sl}(2,\mathbb{R})$ representation theory}\label{sec:grouptheory}
In this section we begin to connect the scattering problem discussed above with the representation theory of $\mathfrak{sl}(2,\mathbb{R})$. The analysis presented here should be regarded as preliminary. A more complete treatment, including a first-order formulation of the construction in terms of $BF$ theory, will be presented in future work \cite{Belaey:2026xvh}.

We begin by recalling some basic facts about $\mathfrak{sl}(2,\mathbb{R})$. The algebra is generated by three elements $H,E,F$ satisfying
\begin{equation}
    [H,E]=E,
    \qquad
    [H,F]=-F,
    \qquad
    [E,F]=2H.
\end{equation}
A convenient matrix realization is
\begin{equation}\label{eq:sl2R_algebra}
    E=
    \begin{pmatrix}
    0&1\\
    0&0
    \end{pmatrix},
    \qquad
    F=
    \begin{pmatrix}
    0&0\\
    1&0
    \end{pmatrix},
    \qquad
    H= \frac{1}{2}
    \begin{pmatrix}
    1&0\\
    0&-1
    \end{pmatrix}.
\end{equation}
With these conventions, the quadratic Casimir is
\begin{equation}
    \Omega
    =
    H^2+\frac{1}{2} \left( EF + FE \right),
\end{equation}
or, introducing the useful combinations $X=\frac{E+F}{2}$ and $Y=\frac{E-F}{2}$,
\begin{equation}
    [H,X] = Y, \qquad [H,Y] = X, \qquad [X,Y] = -H, \qquad\text{and}\qquad \Omega = H^2 + X^2 - Y^2.
\end{equation}
We fix the constant value of the Casimir on a given unitary irreducible representation to be
\begin{equation}\label{eq:Casimir_value}
    \Omega = j(j+1) = - \frac{1}{4} - s^2,
\end{equation}
where the condition $j = -\frac{1}{2} + is$ follows from unitarity for the principal continuous series representations. This means that we label each $\mathfrak{sl}(2,\mathbb{R})$ unitary irrep with a positive real number $s \geq 0$.

We will be interested in the regular representation of the corresponding group on functions defined on a suitable coordinate patch. More precisely we consider a class of elements
\begin{equation}\label{eq:decomposition}
    g = e^{aX} e^{LH} e^{bX},
    \qquad a>0,\quad L>0,\quad b>0.
\end{equation}
This decomposition covers an open subset of hyperbolic elements of $\text{SL}(2,\mathbb{R})$, that is elements with trace larger than $2$, and it provides a useful coordinate chart for the present discussion. We have chosen the radial coordinate to be directly the JT geodesic length $L$.

The Lie algebra acts on functions $f(g)$ by infinitesimal left and right multiplication according to
\begin{equation}\label{eq:left_right_regular_action}
    L_T f(g) = \frac{\dd}{\dd t}\bigg\lvert_{t=0} f\!\left( e^{-tT} g \right),
    \qquad
    R_T f(g) = \frac{\dd}{\dd t}\bigg\lvert_{t=0} 
    f\!\left( g e^{tT} \right),
    \qquad
    T\in\mathfrak{sl}(2,\mathbb R).
\end{equation}
In the coordinates introduced above, the right-invariant vector fields are
\begin{equation}
    \begin{split}
        L_X &= -\partial_a,\\
        L_H &= \frac{\sinh{a}}{\tanh{L}}\,\partial_a - \cosh{a}\,\partial_{L} - \frac{\sinh{a}}{\sinh{L}}\,\partial_b,\\
        L_Y &= \frac{\cosh{a}}{\tanh{L}}\,\partial_a - \sinh{a}\,\partial_{L} -  \frac{\cosh{a}}{\sinh{L}}\,\partial_b.
    \end{split}
\end{equation}
Similarly, the left-invariant vector fields are
\begin{equation}
    \begin{split}
        R_X &= \partial_b,\\
        R_H &= \frac{\sinh{b}}{\sinh{L}}\,\partial_a + \cosh{b}\,\partial_{L} - \frac{\sinh{b}}{\tanh{L}}\,\partial_b,\\
        R_Y &= - \frac{\cosh{b}}{\sinh L}\,\partial_a - \sinh{b}\,\partial_{L} + \frac{\cosh{b}}{\tanh{L}}\,\partial_b.
    \end{split}
\end{equation}
Substituting these vector fields into the quadratic Casimir gives the differential operator
\begin{equation}\label{eq:casimir_differential}
    \Omega = \partial_{L}^2 + \frac{1}{\tanh{L}} \, \partial_{L} + \frac{\partial_a \, \partial_b}{\cosh^2(L/2)} - \frac{(\partial_a - \partial_b)^2}{\sinh^2\!{L}}.
\end{equation}
Since we can rewrite
\begin{equation}\label{eq:Casimir_kinetic}
    \partial_L^2 + \frac{1}{\tanh{L}} \, \partial_L = \frac{1}{\sinh{L}} \, \partial_L \left( \sinh{L} \, \partial_L \right),
\end{equation}
it is natural to equip this space of functions with the non-flat measure $\sinh{L} \, \dd L \, \dd a \, \dd b$. Indeed, this measure matches the Haar measure with the parameterization \eqref{eq:decomposition}: we find
\begin{equation}\label{eq:Haar_metric}
    \dd s^2 = 2 \text{Tr}\left[ (g^{-1} \dd g)^2 \right] = \dd L^2 + \dd a^2 + \dd b^2 + \cosh{L} \, (\dd a \, \dd b + \dd b \, \dd a),
\end{equation}
leading to
\begin{equation}
    \dd \mu_\text{Haar}(L,a,b) = \sqrt{-\det(\dd s^2)} = \sinh{L} \, \dd L \, \dd a \, \dd b.
\end{equation}
Consistently, \eqref{eq:casimir_differential} is precisely the Laplace--Beltrami operator for the metric \eqref{eq:Haar_metric}.

The connection with the P\"oschl--Teller problem \eqref{eq:Schroedinger_equation}
becomes apparent once we consider the action of $\mathfrak{sl}(2,\mathbb{R})$ on
matrix elements of the form\footnote{We write the eigenvalues of the generators
as $i\phi$, with $\phi\in\mathbb{R}$, since the generators are anti-Hermitian in any unitary representation.}
\begin{equation}\label{eq:matrix_element}
    R^s_{\phi_\text{L},\phi_\text{R}}(g) \equiv \bra{s, X = i \phi_\text{L}} g \ket{s, X = i \phi_\text{R}},
\end{equation}
where the bra and the ket belong to the irreducible representation labeled by
$s$ and are eigenvectors of $X$, i.e.\ $X \ket{s, X = i \phi} = i \phi \ket{s, X = i \phi}$.\footnote{It must be mentioned that the spectrum of the hyperbolic generator $X$ has multiplicity two. This means that the notation $\ket{s,X=i\phi}$ is ambiguous and by itself does not define a specific eigenvector. Nevertheless, for the present discussion it is sufficient to know that it solves $X \ket{s,X=i\phi} = i\phi \ket{s,X=i\phi}$.}
Since the matrix elements \eqref{eq:matrix_element} are functions of the group
element $g$, they carry a representation of $\mathfrak{sl}(2,\mathbb{R})$, on
which the generators act according to \eqref{eq:left_right_regular_action}. It
follows that, by construction, each matrix element of this form simultaneously
solves the three differential equations
\begin{equation}\label{eq:matrix_element_equations}
    \begin{split}
        \Omega \, R^s_{\phi_\text{L},\phi_\text{R}}(g) &= \left( -\frac{1}{4} - s^2 \right) \, R^s_{\phi_\text{L},\phi_\text{R}}(g),\\
        L_X \, R^s_{\phi_\text{L},\phi_\text{R}}(g) &= - i \phi_\text{L} \,R^s_{\phi_\text{L},\phi_\text{R}}(g),\\
        R_X \, R^s_{\phi_\text{L},\phi_\text{R}}(g) &= + i \phi_\text{R} \,R^s_{\phi_\text{L},\phi_\text{R}}(g).
    \end{split}
\end{equation}
We now combine the differential form of the Casimir \eqref{eq:casimir_differential}
with the last two conditions, in which $L_X=-\partial_a$ and $R_X=\partial_b$ act
as multiplication by $-i\phi_\text{L}$ and $i\phi_\text{R}$, respectively. Using
the ansatz
\begin{equation}
    R^s_{\phi_\text{L},\phi_\text{R}}(g) = \frac{e^{i (\phi_\text{L} a + \phi_\text{R} b)}}{\sqrt{\sinh{L}} }\, \psi_s(L),
\end{equation}
we find that the reduced matrix element $\psi_s(L)$ solves
\begin{equation}\label{eq:reduced_casimir}
    \left( - \partial_L^2 + \frac{\phi_\text{L} \, \phi_\text{R}}{\cosh^2(L/2)} - \frac{\frac{1}{4} + (\phi_\text{L} - \phi_\text{R})^2}{\sinh^2\!{L}} \right) \psi_s(L) = s^2 \, \psi_s(L).
\end{equation}
For $\phi_\text{L} = \phi_\text{R} = \nu$, the reduced Casimir equation
\eqref{eq:reduced_casimir} coincides with the Schr\"odinger equation
\eqref{eq:Schroedinger_equation} obtained from the non-trivial ordering of
Section~\ref{sec:canonical_quantization}, which therefore acquires a
representation-theoretic origin.

Some comments are in order. First, the construction above relies primarily on the local $\mathfrak{sl}(2,\mathbb{R})$ algebraic structure, rather than on a specific global choice of group. Also, the conditions
\begin{equation}
    a>0,
    \qquad
    b>0,
    \qquad
    L>0
\end{equation}
in the decomposition \eqref{eq:decomposition} do not define a subgroup: in general, the inverse of an element in this patch does not belong to the same patch. This is not a problem for the discussion presented, since the parameterization was only used as a local coordinate chart on an ambient space containing these elements. The local differential operators, and in particular the radial Casimir, are determined by the Lie algebra and by this choice of coordinates.

Consequently, the construction does not by itself select a unique global object. It is compatible, for example, with working on $\text{SL}(2,\mathbb{R})$, on its universal cover, or on a positive subsemigroup such as $SL^+(2,\mathbb{R})$. The latter possibility is particularly interesting, since positive semigroup structures have been argued to underlie the standard JT gravity construction \cite{Blommaert:2018oro,Blommaert:2018iqz}.

In the ordinary JT case, however, one usually works in a different polarization. Rather than fixing the left and right actions of the hyperbolic generator $X=\frac{E+F}{2}$, one fixes parabolic generators, typically $L_F$ and $R_E$. Adopting a Gauss parameterization
\begin{equation}
    g
    =
    e^{xF}e^{L_\text{ren} H}e^{yE},
\end{equation}
the Casimir takes the differential form
\begin{equation}
    \Omega
    = \partial_{L_\text{ren}}^2 + \partial_{L_\text{ren}} + e^{-L_\text{ren}} \partial_x \partial_y
    = e^{-L_\text{ren}} \partial_{L_\text{ren}} e^{L_\text{ren}} \partial_{L_\text{ren}} + e^{-L_\text{ren}} \partial_x \partial_y.
\end{equation}
Then, upon enforcing the constraints
\begin{equation}
    L_F f = -i\lambda f,
    \qquad
    R_E f = i\mu f,
\end{equation}
and making an ansatz $R^s_{\lambda,\mu} = e^{i(\lambda x+ \mu y)} e^{-L_\text{ren}/2} \psi_s(L_\text{ren})$ for the matrix element, the reduced Casimir equation becomes the Liouville equation
\begin{equation}
    \left( -\partial_{L_\text{ren}}^2 + \lambda \mu \, e^{-L_\text{ren}} \right) \psi_s(L_\text{ren}) = s^2 \, \psi_s(L_\text{ren}).
\end{equation}

Our construction should be viewed as the finite cutoff analogue of this story, and therefore it should contain the ordinary JT description in an appropriate scaling limit. To see this explicitly, define
\begin{equation}
    L
    =
    2\log(2\Phi_b) + L_\text{ren}.
\end{equation}
Then the finite cutoff decomposition becomes
\begin{equation}
    g
    =
    e^{aX}
    e^{LH}
    e^{bX}
    =
    e^{aX}
    e^{2\log(2\Phi_b)H}
    e^{L_\text{ren} H}
    e^{bX}.
\end{equation}
Splitting the middle factor symmetrically, and commuting the two factors of $e^{\log(2\Phi_b)H}$ to the left and to the right, gives
\begin{equation}
\begin{aligned}
    e^{aX}e^{LH}e^{bX}
    &=
    e^{\log(2\Phi_b)H}
    e^{\frac{a}{2}((2\Phi_b)^{-1}E+2\Phi_b F)}
    e^{L_\text{ren} H}
    e^{\frac{b}{2}(2\Phi_b E+(2\Phi_b)^{-1}F)}
    e^{\log(2\Phi_b)H}.
\end{aligned}
\end{equation}
Here we used
\begin{equation}
    e^{-\log(2\Phi_b)H} E e^{\log(2\Phi_b)H}
    =
    (2\Phi_b)^{-1}E,
    \qquad
    e^{-\log(2\Phi_b)H} F e^{\log(2\Phi_b)H}
    =
    2\Phi_b F.
\end{equation}
Thus, in order to obtain a finite large-\(\Phi_b\) limit, we keep fixed the rescaled variables
\begin{equation}
    x=a\Phi_b,
    \qquad
    y=b\Phi_b.
\end{equation}
Taking \(\Phi_b\to\infty\) with \(x,y,L_\text{ren}\) fixed, we find
\begin{equation}\label{eq:parabolic_limit}
    e^{aX}e^{LH}e^{bX}
    \sim
    e^{\log(2\Phi_b)H}
    e^{xF}
    e^{L_\text{ren} H}
    e^{yE}
    e^{\log(2\Phi_b)H}.
\end{equation}
The two outer factors are constant group elements with respect to the coordinates \(x,y,L_\text{ren}\). They therefore do not affect the local form of the reduced Casimir. Consequently, in the scaling limit, the finite cutoff hyperbolic parameterization degenerates into the parabolic Gauss parameterization relevant for ordinary JT gravity \cite{Blommaert:2018oro,Blommaert:2018iqz}.

In this limit the left \(X\)-translation retains only its \(F\)-component, while the right \(X\)-translation retains only its \(E\)-component. Thus fixing \(L_X\) and \(R_X\) at finite cutoff becomes, in the large cutoff limit, the familiar parabolic reduction in which one fixes \(L_F\) and \(R_E\). In this precise sense, the P\"oschl--Teller scattering problem obtained above contains the Liouville scattering problem of ordinary JT gravity as its large cutoff limit.

There is a mild arbitrariness in the scaling limit described above. Indeed, one could more generally take
\begin{equation}
    L
    =
    2\log(C\Phi_b)+L_\text{ren},
    \qquad
    C>0 .
\end{equation}
This modification does not affect the final result. Its only effect is to induce a constant rescaling of the coordinates \(x\) and \(y\) that appear in the parabolic parameterization.

This freedom under constant rescalings of the parabolic coordinates was already observed in the first-order formulation of JT gravity \cite{Oliver_Coussaert_1995,Blommaert:2023opb}. From the present perspective, it arises naturally from the large-\(\Phi_b\) limit, which obscures the finite cutoff origin of the length variable \(L\). This is directly analogous to the gravitational relation between the finite cutoff geodesic length and the renormalized length in the asymptotic theory.

When discussing the two-point function in Section \ref{sec:2_point_function}, we anticipated that the choice of insertion \eqref{eq:bulk_operator_2F1}
\begin{equation}
    G_\Delta(\hat L)
    =
    \left( \frac{\Phi_b}{\cosh\!{(\hat{L}/2)}}\right)^{2\Delta} {}_2F_1\!\left( \Delta, \Delta, 2\Delta; \frac{1}{\cosh^2\!{(\hat{L}/2)}} \right)
\end{equation}
is natural from the viewpoint of $\mathfrak{sl}(2,\mathbb{R})$ representation theory. We now explain this statement more explicitly.
First, let us recall what happens in ordinary JT gravity at large cutoff: in Section 4.3.2 of \cite{Blommaert:2018oro}, it is argued that matter insertions correspond to boundary-anchored Wilson lines in a discrete series representation labeled by $j = -\Delta$ (here $j$ is the Casimir eigenvalue label in \eqref{eq:Casimir_value}, $\Omega = \Delta(\Delta-1)$) and $F = 0 = E$, i.e. the matrix element
\begin{equation}
    R^\Delta_{0,0}(g) = \bra{\Delta,F=0} e^{xF}e^{L_\text{ren}H}e^{yE} \ket{\Delta,E=0} = e^{-\Delta L_\text{ren}}.
\end{equation}
This explains the operator insertion \eqref{eq:bilocal_operator_schwarzian_thermal} from a first-order formulation perspective.

This construction can be readily generalized to the finite cutoff setup. Indeed, substituting $-\frac{1}{4}-s^2 \to \Delta(\Delta-1)$ and $\phi_\text{L} = 0 = \phi_R$ into the matrix element equations \eqref{eq:matrix_element_equations}, the reduced Casimir equation becomes
\begin{equation}
    \left( \frac{1}{\sinh L} \partial_L \, \sinh L \, \partial_L - \Delta(\Delta-1) \right) R^\Delta_{0,0}(L) = 0.
\end{equation}
Solutions to this equation are precisely given by the AdS$_2$ propagators \eqref{eq:AdS_propagator}, specifically they are combinations of $G_\Delta(L)$ and $G_{1-\Delta}(L)$ for generic $\Delta$. Finally, requiring the solution to decay at large distances as $e^{-\Delta L}$ selects $R^\Delta_{0,0}(L) = G_\Delta(L)$.

\section{What if the boundary can move beyond the horizon?}
\label{subsec:boundary_behind_horizon}
The aim of this section is to clarify a point that remained implicit in the previous discussion. In the sections above, when computing observables, we effectively restricted the spectral decomposition to the exterior sector. More precisely, we used the operator
\begin{equation}
    \hat{\Pi}
    =
    \int_0^{\nu}
    \diff{s}\, \rho(s) 
    \ket{s}\bra{s},
\end{equation}
as the relevant resolution of the identity, but this reasoning is consistent only if the physical Hilbert space is defined from the start as the exterior Hilbert space
\begin{equation}
     \mathcal{H}_{\text{ext}}
    =
    \operatorname{span}
    \left\{
    \ket{s}:
    \braket{L=0}{s}=0,
    \quad
    0<s<\nu
    \right\}.
\end{equation}
The restriction \(s<\nu\), or equivalently \(\Phi_h<\Phi_b\), was suggested by the expression for the Hamiltonian
\begin{equation}
     H
     =
     \Phi_b
     -
     \sqrt{\Phi_b^2-\Phi_h^2},
\end{equation}
which was derived under the assumption that the cutoff boundary lies outside the event horizon. If one naively extends it to \(\Phi_h>\Phi_b\), the energy becomes complex. This signals that the exterior expression for the Hamiltonian cannot simply be analytically continued beyond the horizon without further input. Indeed, when the curve \(\Phi=\Phi_b\) crosses the horizon, it goes from being timelike to being spacelike and it is therefore not immediately obvious how to define the corresponding Hamiltonian evolution.

Before proposing a possible resolution, let us explain why this restriction may be problematic. The issue is related to the fact that removing part of the energy spectrum changes the meaning of the $\{\ket{L}\}$-basis as well. The point can be understood from an elementary example in ordinary quantum mechanics. On the full line, momentum eigenstates form a complete basis,
\begin{equation}
    1
    =
    \int_{-\infty}^{+\infty}
    \diff{p} 
    \ket{p}\bra{p}.
\end{equation}
Because of this completeness, one can construct sharply localized position eigenstates satisfying
\begin{equation}
    \braket{x}{x'}
    =
    \int_{-\infty}^{+\infty}
    \frac{\diff{p}}{2\pi}
    e^{ip(x-x')}
    =
    \delta(x-x') .
\end{equation}
Suppose, however, that we keep only momenta in a finite interval, say \(\abs{p} < \Lambda\). Then the identity is replaced by the restricted operator
\begin{equation}
    \hat{\Pi}_\Lambda
    =
    \int_{-\Lambda}^{+\Lambda}
    \diff{p}
    \ket{p}\bra{p},
\end{equation}
and the corresponding projected position states are
\begin{equation}
    \ket{x}_\Lambda
    \equiv
    \hat{\Pi}_\Lambda \ket{x}.
\end{equation}
Crucially, these states are no longer orthonormal as their overlap is
\begin{equation}
    _{\Lambda}\braket{x}{x'}_\Lambda
    =
    \bra{x}\Pi_\Lambda\ket{x'}
    =
    \int_{-\Lambda}^{+\Lambda}
    \frac{\diff{p}}{2\pi}
    e^{ip(x-x')}
    =
    \frac{\sin\left[\Lambda(x-x')\right]}
    {\pi(x-x')} .
\end{equation}
The absence of a delta function on the right-hand side teaches us that once only a portion of the momentum spectrum is retained, the restricted theory no longer admits states with sharply defined position.

The same phenomenon occurs in our gravitational setting. In the full finite cutoff theory, the states $\{\ket{s}\}$ form a complete basis on the half-line of geodesic lengths. This completeness is what allows one to reconstruct the length eigenstates \(\ket{L}\) and to impose the usual orthogonality relation $\bra{L}\ket{L'}=\delta(L-L')$.
By contrast, if we keep only the exterior sector \(s<\nu\), the completeness relation is lost. The kernel obtained from the restricted set of modes is
\begin{equation}
    \bra{L}\hat{\Pi}\ket{L'}
    =
    \int_0^{\nu}
    \diff{s}\, \rho(s)\,
    \psi_{s}(L)
    \psi_{s}(L')^* ,
\end{equation}
and this is not equal to \(\delta(L-L')\). Therefore the projected states \(\hat{\Pi}\ket{L}\) do not define an orthonormal length basis. Equivalently, after the restriction to \(\Phi_h<\Phi_b\), one can no longer assign a sharply definite value to the geodesic length \(L\).

This is especially relevant for the prescriptions used above to compute observables. Those prescriptions rely on the existence of the state \(\ket{L=0}\), but in the restricted theory we replace it with $\hat{\Pi}\ket{L=0}$, which is not a state of definite length. As a result, the geometric interpretation of the open-channel amplitude as a transition between zero-length states becomes less transparent.

There is also a second issue concerning the dynamics of the theory. In a Hamiltonian formulation, restricting the spectrum is not automatic and typically requires justification, unless the restriction follows from a superselection rule or is built into the definition of the theory itself. The fact that the exterior Hamiltonian \eqref{eq:hamiltonian_quantum} becomes complex for $\Phi_h > \Phi_b$ does not by itself directly mean that the corresponding states should be discarded. Rather, it might indicate that the exterior Hamiltonian \eqref{eq:hamiltonian_quantum} is not the correct self-adjoint operator on the entire Hilbert space. A closely related issue appears in other contexts, such as in the $T\bar{T}$ deformation of two-dimensional CFTs, where a naive continuation can produce complex energies unless the correct branch or completion of the theory is specified.  

Thus, it remains an open question whether the exterior-only description we have discussed so far is dynamically complete. A resolution to this problem could be to extend the theory in order to include the sector $\Phi_h > \Phi_b$, corresponding to geometries in which the boundary lies behind the horizon.

\subsection{Foliating the black hole interior}
In this subsection we propose a possible resolution of the problems discussed above. The current discussion should be regarded as speculative, and its purpose is to indicate a natural way of extending the exterior Hamiltonian description \eqref{eq:hamiltonian_quantum} to the regime in which the cutoff boundary crosses the horizon.
\begin{figure}
    \centering
    \includegraphics[width=0.655\linewidth]{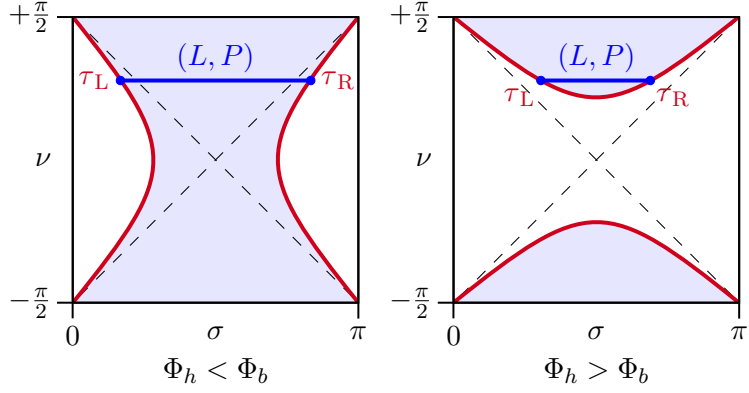}
    \caption{We foliate the Penrose diagram with horizontal boundary-to-boundary geodesics both in the case $\Phi_h < \Phi_b$ and $\Phi_h > \Phi_b$. In the latter setting, the parameters $\tau_\text{L}$ and $\tau_\text{R}$ labeling the endpoints actually have the interpretation of spatial coordinates.}
    \label{fig:Penrose_diagram_interior}
\end{figure}
As reviewed in Section \ref{sec:classical_JT_gravity}, the boundary of the spacetime is defined by the condition
\begin{equation}
    \Phi\big|_{\partial\mathcal{M}} = \Phi_b .
\end{equation}
In Schwarzschild gauge, the cutoff curve is located at \(r=\Phi_b\), while the horizon is at \(r=\Phi_h\). The induced metric on the cutoff curve is therefore
\begin{equation}
    \dd s^2_{\partial\mathcal{M}}
    =
    \begin{cases}
        -\left(\Phi_b^2-\Phi_h^2\right)\dd t^2,
        &\qquad \Phi_h<\Phi_b,\\[0.4em]
        +\left(\Phi_h^2-\Phi_b^2\right)\dd t^2,
        &\qquad \Phi_h>\Phi_b.
    \end{cases}
\end{equation}
Thus, in the exterior branch \(\Phi_h<\Phi_b\), the cutoff curve is timelike and the parameter \(\tau\) introduced in \eqref{eq:proper_time} is the proper time along the boundary. In the interior branch \(\Phi_h>\Phi_b\), instead, the cutoff curve is spacelike and the interpretation of the corresponding conserved charge is therefore more subtle.

Following the logic of \cite{AliAhmad:2025kki}, we propose that the boundary evolution associated with a spacelike cutoff curve should be interpreted as Euclidean evolution generated by the corresponding Brown--York charge. 
This mechanism was proposed in \cite{AliAhmad:2025kki}, where the gravitational bulk dual of a two-dimensional CFT deformed by the \(T\bar T+\Lambda\) operator involves a change in the signature of the dual three-dimensional metric configurations \cite{Gorbenko:2018oov,Torroba:2022jrk, Chang:2025ays, Shyam:2021ciy}. The present setting is analogous in spirit, with the main difference that the bulk theory is two-dimensional.

More explicitly, we define the conserved charge behind the horizon by applying the Brown--York prescription to a cutoff surface with timelike normal. Equivalently, the charge is computed by contracting the boundary stress tensor \(T_{\alpha\beta}\) with the appropriate Killing vector field tangent to the boundary curve, with the sign dictated by the spacelike nature of the cutoff curve.\footnote{
For a spacelike boundary, $n^2= - 1$. Thus, the appropriate definition of the Brown--York tensor is
\[
    T^{\text{L}/\text{R}}_{\alpha\beta}
    \equiv
    \frac{2}{\sqrt{\abs{h}}}
    \frac{\delta S_{\text{JT}}}{\delta h^{\alpha\beta}} \bigg\lvert_{\partial\mathcal{M}_{\text{L}/\text{R}}}
    =
    - \left( n^\mu \nabla_\mu \Phi - \Phi \right)\bigg\lvert_{\partial\mathcal{M}_{\text{L}/\text{R}}}
    h_{\alpha\beta}.
\]
}
This gives
\begin{equation}
    H_{\rm int}(\Phi_h)
    =
    \Phi_b+\sqrt{\Phi_h^2-\Phi_b^2},
\end{equation}
and the complete Hamiltonian is then taken to be
\begin{equation}\label{eq:hamiltonian_piecewise}
    H(\Phi_h)
    =
    \begin{cases}
        \Phi_b-\sqrt{\Phi_b^2-\Phi_h^2},
        &\qquad \Phi_h<\Phi_b,\\[0.4em]
        \Phi_b+\sqrt{\Phi_h^2-\Phi_b^2},
        &\qquad \Phi_h>\Phi_b.
    \end{cases}
\end{equation}
Following the same steps as in Section \ref{sec:JT_scattering}, we now use as a phase space variable the length of the horizontal boundary-to-boundary geodesic that connects two boundary points as in the right panel of Figure \ref{fig:Penrose_diagram_interior}.
An explicit computation yields
\begin{equation}
   L(\tau,\Phi_h)
   =
   2\sinh^{-1}\!\left[
   \frac{\sqrt{\Phi_h^2-\Phi_b^2}}{\Phi_h}
   \left|
   \sinh\left(
   \frac{\Phi_h}{\sqrt{\Phi_h^2-\Phi_b^2}}
   \frac{\tau}{2}
   \right)
   \right|
   \right],
\end{equation}
where \(\tau\) is now the proper distance separating the endpoints of the geodesic along the spacelike cutoff boundary. We propose to interpret \(\tau\) as the parameter along which the Euclidean Hamiltonian evolution takes place.

As in Section \ref{sec:JT_scattering}, one can determine the momentum \(P\) conjugate to the geodesic length \(L\). Then, the interior Hamiltonian expressed in terms of the canonical pair \((L,P)\) becomes
\begin{equation}
     H_{\rm int}(L,P)
     =
     \Phi_b
     +
     \sqrt{
     \left(P^2
     +
     \frac{\Phi_b^2}{\cosh^2(L/2)} \right)
     -
     \Phi_b^2
     }.
\end{equation}
Thus the quantity \(\Phi_h^2(L,P)\) is again the P\"oschl--Teller Hamiltonian \eqref{eq:Phi_h^2}.
Consequently, both the phase space trajectories and the quantization of the theory can be discussed using the same scattering problem analyzed in the previous sections.

This observation is important because it addresses, at the level of the spectral
problem, the issues described above. With this prescription, the energies do not become complex after crossing the horizon. Moreover, the Hamiltonian can be promoted to a self-adjoint operator on the full scattering Hilbert space
\begin{equation}\label{eq:Hilbert_space_full}
    \mathcal{H}
    =
    \text{span}
    \left\{
    \ket{s}:
    \braket{L=0}{s}=0
    \right\},
\end{equation}
with no restriction to \(s<\nu\). The completeness of the \(\ket{s}\) basis is then restored, and therefore the length states obey $\bra{L}\ket{L'}=\delta(L-L')$.
This allows us to recover the interpretation of \(\ket{L}\) as sharply localized length states, and in particular gives a cleaner meaning to the formal state \(\ket{L=0}\) appearing in the open-channel prescription \eqref{eq:partition_function_prescription}.
% A real and self-adjoint spectral evolution on the full scattering Hilbert space
% does not by itself establish a complete gravitational UV completion: deriving the
% interior evolution, and its matching to the exterior, from a variational
% principle remains an open problem.

In this extended setup, the main modification in the computation of observables is that the resolution of the identity is no longer truncated at \(s=\nu\) and actually takes the form of Equation \eqref{eq:identity}.
For example, with the same prescription \eqref{pippo} and the same topological
factor as in Section~\ref{sec:partition_function}, but without the projector, the
disk partition function becomes
\begin{equation}
\label{eq:partition_function_full}
Z_{\rm full}(\beta,\nu)
=e^{2\pi\nu}\int_0^\infty \dd s\,\rho(s)\,e^{-\beta E(s)} ,
\end{equation}
where the energies are
\begin{equation}\label{eq:energy_piecewise}
    E(s) =
    \begin{cases}
    \nu - \sqrt{\nu^2 - s^2},
    &\qquad s<\nu,\\[0.4em]
    \nu + \sqrt{s^2 - \nu^2},
    &\qquad s>\nu.
    \end{cases}
\end{equation}
Equivalently, after changing variables from \(s\) to \(E\), the partition function can be written in spectral form and the corresponding spectral density is
\begin{equation}\label{eq:spectral_densityy}
    \rho(E) =
    \begin{cases}
        \displaystyle e^{2\pi\nu}
        \frac{
        2(\nu-E)
        \sinh\left(
        2\pi\sqrt{\nu^2-(\nu-E)^2}
        \right)}
        {
        \cosh\left(
        2\pi\sqrt{\nu^2-(\nu-E)^2}
        \right)
        +
        \cosh(2\pi\nu)
        },
        &\qquad E<\nu,
        \\[1.2em]
        \displaystyle e^{2\pi\nu}
        \frac{
        2(E-\nu)\,
        \sinh\left(
        2\pi\sqrt{\nu^2+(E-\nu)^2}
        \right)}
        {
        \cosh\left(
        2\pi\sqrt{\nu^2+(E-\nu)^2}
        \right)
        +
        \cosh(2\pi\nu)
        },
        &\qquad E>\nu .
    \end{cases}
\end{equation}
\begin{figure}
    \centering
    \includegraphics{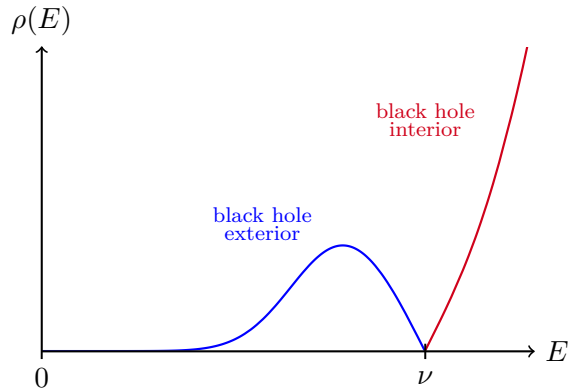}
    \caption{The piecewise spectral density of the extended theory \eqref{eq:spectral_densityy}. The blue curve takes contributions from states such that $\Phi_h < \Phi_b$ and was present already in \eqref{eq:spectral_density}. The red curve is a novel contribution from states corresponding to geometries where the entire spacetime lies behind the black hole horizon, i.e. $\Phi_h > \Phi_b$.}
    \label{fig:spectral_density_piecewise}
\end{figure}
As shown in Figure \ref{fig:spectral_density_piecewise}, the density \eqref{eq:spectral_densityy} is defined for the full range of energies. It vanishes at \(E=0\) and at the transition point \(E=\nu\), while at large energies it grows linearly as
\begin{equation}
\label{eq:spectral_density_full_asymptotic}
    \rho(E)
    \overset{E\to\infty}{\sim}
    2e^{2\pi\nu}E .
\end{equation}
Finally, as discussed in Appendix \ref{app:semiclassical_partition_function}, the branch behind the horizon does not contribute to the leading semiclassical saddle. Therefore, the semiclassical results \eqref{eq:partition_function_semiclassical} obtained in the previous sections are unchanged by its inclusion.

\section{Concluding remarks}\label{sec:conclusions}
\subsection{Comparison with $T\bar{T}$ and Wheeler--DeWitt}\label{sec:TT}
Ordinary JT gravity at large cutoff has become one of the most fruitful toy models of quantum gravity thanks to the analytic control it exhibits and the detailed knowledge of its holographic dual, the latter being the Schwarzian quantum mechanics emerging in the low-energy regime of SYK. It would be very interesting to find a one-dimensional theory dual to JT gravity at finite cutoff. We remark that, while the exact form of the action of this putative dual theory is not known, it is determined by an exact Riccati-type equation \cite{Griguolo:2025kpi} whose solution, in principle, can be constructed perturbatively to arbitrary order around the large cutoff limit.

In \cite{McGough:2016lol}, it was originally proposed to relate three-dimensional AdS gravity at finite cutoff to a $T\bar{T}$-deformation of its dual two-dimensional CFT.
The authors of \cite{Gross:2019ach,Gross:2019uxi,Chakraborty_2020} proposed to consider a one-dimensional analogue of the $T\bar{T}$-deformation of the Schwarzian theory as the holographic dual to JT gravity at finite cutoff. We now briefly recall their proposal. Consider a quantum Hamiltonian $\hat{H}_0$ and the corresponding eigenvalue problem
\begin{equation}
    \hat{H}_0 \ket{E_0} = E_0 \ket{E_0}.
\end{equation}
The deformation proposed in \cite{Gross:2019ach,Gross:2019uxi,Chakraborty_2020} is realized by a flow equation for the energy levels of the theory
\begin{equation}\label{eq:TT_QM}
    \pdv{E_\lambda}{\lambda} = \frac{E_\lambda^2}{1/2 - 2 \lambda E_\lambda}, \qquad E_{\lambda=0} = E_0,
\end{equation}
where $\lambda$ is the deformation parameter. The initial condition on the right of Equation \eqref{eq:TT_QM} requires the deformed energies $E_\lambda$ to reduce to the ones of the undeformed theory $E_0$ when the deformation parameter is turned off, yielding the unique solution
\begin{equation}\label{eq:TT_energy}
    E_\lambda = \frac{1}{4 \lambda} \left( 1 - \sqrt{1 - 8 \lambda E_0} \right).
\end{equation}
Promoting the deformation \eqref{eq:TT_energy} at the operator level, the Hamiltonian $\hat{H}_0$ flows to
\begin{equation}\label{eq:TT_operator}
    \hat{H}_\lambda = \frac{1}{4 \lambda} \left( 1 - \sqrt{1 - 8 \lambda \hat{H}_0} \right).
\end{equation}
Since $\hat{H}_\lambda$ is purely a function of $\hat{H}_0$, the eigenvectors of the latter are also eigenvectors of the former, $\ket{E_\lambda} = \ket{E_0}$, and their eigenvalues are related by \eqref{eq:TT_energy}.

So far the discussion was completely general. Specializing now to ordinary JT gravity at large cutoff, it is known that the dual QM theory is described by the Schwarzian action.
% \begin{equation}
%     S_\text{Sch}[t_\text{E}] = \int_0^{\beta_\text{JT}} \dd u \, \bigg\{ \tan\left( \frac{\pi}{\beta_\text{JT}} t_\text{E}(u) \right), u \bigg\}.
% \end{equation}
In \cite{Harlow:2018tqv} it was shown that the partition function of JT gravity (and of Schwarzian theory as well) can be computed as an amplitude within a theory described by the Liouville Hamiltonian \eqref{eq:hamiltonian_Liouville}
\begin{equation}\label{eq:hamiltonian_Liouville_TT}
    \hat{H}_\text{JT} = \frac{1}{2 \Phi_b} \left( \hat{P}^2 + e^{-\hat{L}_\text{ren}} \right),
\end{equation}
yielding\footnote{Here we are omitting a factor of $\left( \lim_{L_\text{ren} \to -\infty} \frac{1}{\pi} K_{2is}\!\left( 2 e^{-L_\text{ren}/2} \right) \right)^2$ in front, which is an $s$-independent shift to the entropy coming from the $L_\text{ren} \to -\infty$ behavior of the wavefunctions \eqref{eq:eigenfunction_large_cutoff}.}
\begin{equation}
    \begin{split}
        Z(\beta_\text{JT})
        &= \bra{L_\text{ren}=-\infty} e^{-\beta \hat{H}_\text{JT}} \ket{L_\text{ren}=-\infty}
        \propto \int_0^\infty \dd s \, 4 s \sinh(2 \pi s) \, e^{-\beta_\text{JT} s^2/2},
    \end{split}
\end{equation}
where we used the renormalized inverse temperature \eqref{eq:beta_renormalized}.
According to the proposal \eqref{eq:TT_operator}, the corresponding finite cutoff theory would be obtained by flowing the Liouville Hamiltonian $\hat{H}_0 = \hat{H}_\text{JT}$ from $\lambda = 0$ to $\lambda = \frac{1}{4 \Phi_b}$:
\begin{equation}\label{eq:Hlambda}
    \hat{H}_{\lambda} = \Phi_b - \sqrt{\Phi_b^2 - \left( \hat{P}^2 + e^{-\hat{L}_\text{ren}} \right)},
\end{equation}
while the states are left invariant by the flow. Hence, the partition function of the theory should be obtained as the amplitude\footnote{Here we neglect the issue of the energies becoming complex when $s>\nu$, as it does not affect the current conclusions. See \cite{Iliesiu:2020zld,Griguolo:2021wgy} for possible resolutions of the issue in the context of $T\bar{T}$ deformation.}
\begin{equation}\label{eq:TT_partition_function}
    \begin{split}
        Z(\beta)
        &= \bra{L_\text{ren}=-\infty} e^{-\beta \hat{H}_{\lambda}} \ket{L_\text{ren}=-\infty}
        \propto \int_0^\infty \dd s \, 4 s \sinh(2 \pi s) \, e^{-\beta (\nu - \sqrt{\nu^2 - s^2})}.
    \end{split}
\end{equation}
Neglecting the unphysical prefactors coming from asymptotics of the eigenfunctions, the partition function \eqref{eq:TT_partition_function} differs from our result \eqref{eq:partition_function} due to the presence of a different spectral density. We propose to trace back the cause of the discrepancy to the introduction of a renormalized length
\begin{equation}\label{eq:renorm}
    L_\text{ren} = L - 2 \log(2 \Phi_b)
\end{equation}
in the large cutoff setting. Indeed, the state of definite renormalized length $\ket{L_\text{ren} = -\infty}$ that is used to compute the partition function \eqref{eq:TT_partition_function} is part of the Hilbert space of the theory only in the strict large cutoff regime, as it does not correspond to any physical configuration of the system in the finite cutoff case since $L \geq 0$ and $\Phi_b < \infty$ there.
A second important difference is that the Hamiltonian \eqref{eq:hamiltonian_quantum} we obtained from gravity is actually found by deforming the P\"oschl--Teller Hamiltonian
\begin{equation}
    P^2 + \frac{\Phi_b^2}{\cosh^2(L/2)}
\end{equation}
according to \eqref{eq:TT_operator}, instead of taking the Liouville Hamiltonian \eqref{eq:hamiltonian_Liouville_TT} as the starting point.

In \cite{Iliesiu:2020zld}, the $T\bar{T}$-deformed Schwarzian partition function \eqref{eq:TT_partition_function} was obtained from the gravitational side of the duality as a solution of the corresponding Wheeler--DeWitt equation\footnote{We adapt their notation to ours as $M = s^2$, $L = \beta$, $\phi_b = \nu$.}
\begin{equation}\label{eq:WdW_equation}
    \left( (\beta \partial_\beta - 1) \partial_\nu - \beta^2 \nu \right) \Psi(\beta,\nu) = 0.
\end{equation}
Solutions of \eqref{eq:WdW_equation} are linear combinations of $e^{\pm\beta \sqrt{\nu^2-s^2}}$, and the $s$-dependent coefficients in \cite{Iliesiu:2020zld} are uniquely fixed in order to have consistency with the known large cutoff result:
\begin{equation}\label{eq:WdW_wavefunction}
    \Psi(\beta,\nu) = \int \text{d}s \, 4 s \sinh(2 \pi s) \, e^{\beta \sqrt{\nu^2-s^2}}.
\end{equation}
Then, the partition function of the theory is identified with the wavefunction \eqref{eq:WdW_wavefunction} with the counterterm removed, $\Psi(\beta,\nu) = e^{\beta\nu} Z(\beta,\nu)$.

Instead, our proposal for the partition function \eqref{eq:partition_function} (with the counterterm $e^{-\beta \nu}$ removed) cannot be obtained as a combination of $e^{\pm \beta \sqrt{\nu^2-s^2}}$ with $\nu$-independent coefficients and therefore it is not a solution of the Wheeler--DeWitt equation.
This fact should not be surprising since Equation \eqref{eq:WdW_equation} was derived in \cite{Iliesiu:2020zld} by performing a foliation of the Euclidean disk, whereas in Section \ref{sec:Poschl_Teller_instantons} we discussed in detail why our result \eqref{eq:partition_function} is actually picking up contributions from topologies other than the smooth disk.
We conclude by pointing out that Equation \eqref{eq:spectralexpansion} allows us to write
\begin{equation}
    Z(\beta) = e^{-\beta\nu} \sum_{n=1}^\infty (-1)^{n-1} Z_n(\beta,\nu),
\end{equation}
where we introduced the partition functions for disks with $n-1$ pinches
\begin{equation}\label{eq:partition_function_pinched_disk}
    Z_n(\beta,\nu) = \int \text{d}s \, 4 s \, \sinh(2 \pi n s) \, e^{- 2 \pi (n-1) \nu} \, e^{+ \beta \sqrt{\nu^2 - s^2}}.
\end{equation}
As expected, $Z_1(\beta,\nu) = \Psi(\beta,\nu)$.
An explicit computation then shows that each term \eqref{eq:partition_function_pinched_disk} solves a modified version of the Wheeler--DeWitt equation \eqref{eq:WdW_equation} that is obtained by shifting $\partial_\nu \to \partial_\nu + 2 \pi (n-1)$, i.e.
\begin{equation}\label{eq:Wdw_pinched}
    \left( (\beta \partial_\beta - 1) (\partial_\nu + 2 \pi (n-1)) - \beta^2 \nu \right) Z_n(\beta,\nu) = 0.
\end{equation}
This suggests that an analogue of the Wheeler--DeWitt equation holds once the topology of the manifold is fixed, even if we lack a geometrical derivation for it.
Moreover, we must mention that choosing the integration contour in \eqref{eq:partition_function_pinched_disk} to range from $0$ to $\nu$ would produce a boundary term in \eqref{eq:Wdw_pinched}, but this issue would also be present in \eqref{eq:WdW_wavefunction}.

With this knowledge, it is interesting to ask whether it is possible to find a generalization of the Wheeler--DeWitt equation that takes into account all contributions from the pinched disks topologies at once.

\subsection{Outlook}\label{sec:outlook}
We end this work with some more speculative remarks and related future directions.

\paragraph{\sffamily\bfseries Further understanding of pinched disks and the black hole interior.}
In Section \ref{sec:Poschl_Teller_instantons} we have identified the actions of the additional saddles as contributions from geometries with pinched disks topology.
It would be interesting to derive their complete contributions from the gravitational path integral \cite{Buchmuller:2024ksd}, including the fluctuation determinants and the phases associated with the pinches. This is expected to require a path integration measure defined consistently at finite cutoff \cite{Moitra:2021uiv}.
Such a derivation should also produce a generalized Wheeler--DeWitt equation for the sum over pinched configurations as a Ward identity.

A second problem we would like to address is to obtain the proposed black hole interior evolution from a variational principle that treats timelike, null, and spacelike cutoff curves consistently.
This could clarify the gravitational meaning of the self-adjoint evolution on the full Hilbert space \eqref{eq:Hilbert_space_full}.
An analysis near $s=\nu$ would also determine how the large cutoff and high-temperature limits interact.

\paragraph{\sffamily\bfseries Finite cutoff amplitudes in terms of large cutoff Feynman rules.}
As we observed in Equation~\eqref{eq:partastwopoint}, the finite cutoff partition function appears to behave as a two-point function in the large cutoff limit of JT gravity. Moreover, in deriving the exact two-point function~\eqref{eq:2pt_function_exact}, we found the Wilson function to emerge naturally. In the large cutoff theory, this function is known to encode either particle crossings in the bulk \cite{Lam_2018} or, equivalently, the crossing of Wilson lines in the first order formulation \cite{Blommaert:2018oro,Iliesiu:2019xuh}. Together, these observations suggest that finite cutoff amplitudes may admit a reinterpretation in terms of their large cutoff counterparts. From the group-theoretic perspective, this interpretation is especially compelling, as the finite  and large cutoff descriptions can be understood as different coordinate parameterizations of the same $\text{SL}(2,\mathbb{R})$ group manifold, as discussed in Section~\ref{sec:grouptheory}.

More specifically, we are led to conjecture that the finite cutoff boundary can be viewed as an additional particle of weight $\Delta=\frac{1}{2}$ propagating on the large cutoff disk. Under this interpretation, the finite cutoff partition function would correspond to a two-point function involving this boundary particle, while the finite cutoff two-point amplitude would arise from the crossing of the standard matter particle of weight $\Delta$ with the additional $\Delta=\frac{1}{2}$ particle. Establishing this picture rigorously within the framework of the large cutoff Feynman rules~\cite{Blommaert:2018oro,Mertens:2017mtv} would be highly interesting and could shed light on the origin of the distinguished weight $\Delta=\frac{1}{2}$ appearing throughout our formulas.

A possible explanation in terms of EOW branes should be explored in future work. This proposal is motivated by the observation that the spectral density seems to be a doubled version of the EOW brane spectral density~\cite{Gao:2021uro}.

\paragraph{\sffamily\bfseries First order BF theory perspective and one-sided wavefunctions.}
Another very interesting future direction is the development of a complete first order formulation of JT gravity at finite cutoff. Such a framework would complete the preliminary steps initiated in Section~\ref{sec:grouptheory} and provide a full finite cutoff embedding of the standard BF formulation of JT gravity~\cite{Blommaert:2018oro,Iliesiu:2019xuh,Blommaert:2018iqz}. This program will be pursued in future work~\cite{Belaey:2026xvh}.

A first application of this framework would be the construction of one-sided wavefunctions of the universe, in which one end of the Cauchy slice is placed at finite cutoff while the other remains at asymptotic infinity. Such states could provide a bridge between the finite and large cutoff descriptions and may clarify whether the finite cutoff partition function admits an interpretation as a genuine trace, rather than as the transition amplitude appearing in~\eqref{eq:partition_function_prescription}.

A second application concerns the computation of the finite cutoff trumpet amplitude, which appears as a particularly natural observable in the BF formulation and would provide the key ingredient for extending the framework developed in this paper to higher genus topologies \cite{Bhattacharyya_2026}. Given the nonperturbative effects identified here, together with their natural group-theoretic interpretation, we anticipate that both the trumpet amplitude and its higher genus completion may differ from existing proposals in the literature~\cite{Iliesiu:2020zld,Griguolo:2021wgy}, receiving additional contributions. If so, this would in turn require a modification of the spectral curve introduced in~\cite{Griguolo:2021wgy}.

\paragraph{\sffamily\bfseries Applications and extensions.}
A natural extension of the framework presented in this paper is the study of more general dilaton potentials \cite{Turiaci:2020fjj,Kruthoff:2024gxc}, an investigation that has already been initiated in \cite{Griguolo:2025kpi}. In a companion paper \cite{futurepaper}, we will develop an exact canonical quantization of the paradigmatic cases of sinh and sine dilaton potentials in the presence of finite Dirichlet walls. Sine dilaton gravity \cite{Blommaert:2024whf,Blommaert:2025avl,Blommaert:2025eps,Bossi:2024tvh}  is particularly appealing because of its duality with the double-scaling limit of the SYK model \cite{Berkooz:2018jqr,Berkooz:2018qkz}, which has attracted considerable attention in recent years \cite{Berkooz:2022mfk,Lin:2022rbf,Goel:2023svz,Lin:2023trc,Almheiri:2024ayc,Berkooz:2024evs,Belaey:2025ijg,vanderHeijden:2025zkr,Blommaert:2025eps,Schouten:2025tvn, Alfinito:2026cky,Aguilar-Gutierrez:2026jjv}.
Sinh dilaton gravity, instead, is a reformulation of Liouville gravity \cite{Mertens:2020hbs}.

We also expect that the canonical quantization strategy developed in this paper can be successfully extended to dS$_2$ \cite{Maldacena:2019cbz,Cotler:2024xzz,Fanaras:2021awm}, by foliating the spacetime with timelike slices anchored at future and past infinity \cite{Heller:2025ddj}. Such a construction could pave the way toward a finite-distance description of cosmological correlators \cite{Fumagalli:2024msi}. More broadly, it would establish a connection between the finite cutoff quantization presented here and recent developments on closed universes \cite{Blommaert:2025rgw,Iliesiu:2024cnh}, quantum diamonds \cite{Blommaert:2026lvp}, and the incorporation of observers into quantum gravity \cite{Blommaert:2026ofx,Tietto:2025oxn, Abdalla:2025gzn}.

Finally, it would be interesting to apply the results of this analysis to the dynamics of near-extremal black holes, where a Schwarzian mode emerges at the boundary of the infinitely long near-horizon throat \cite{Iliesiu:2020qvm,Nayak:2018qej,Castro:2022cuo}. The spectral density and two-point functions computed here could provide valuable probes of the low-energy dynamics of four-dimensional black holes \cite{Emparan:2025sao,Emparan:2025qqf, Betzios:2025sct}, and may also offer insight into the evaporation process \cite{Brown:2024ajk} as one probes progressively deeper into the quantum throat.

\section*{Acknowledgments}
We thank Sergio E. Aguilar-Gutierrez, Andreas Belaey, Andreas Blommaert, Wilfried Buchmüller, Thomas Mertens, Alexander Westphal and Qi-Feng Wu for useful discussions. LG, LR, DS, AT have been supported in part by the Italian Ministero
dell’Università e della Ricerca (MUR), and by Istituto Nazionale di Fisica Nucleare (INFN)
through the ``Gauge and String Theory'' (GAST) research project. JP acknowledges financial support from the European Research Council (grant BHHQG-101040024). Funded
by the European Union. Views and opinions expressed are however those of the author(s)
only and do not necessarily reflect those of the European Union or the European Research
Council. Neither the European Union nor the granting authority can be held responsible
for them.
JP acknowledges fruitful discussions during the workshop ``Observers, wormholes and complex saddles in cosmology'', organized at the Bernoulli Center for Fundamental Studies (EPFL, Lausanne) from 18--22 May 2026.

\newpage
\appendix
\addcontentsline{toc}{section}{Appendices}
\begin{flushleft}
\huge\bfseries\sffamily\color{black}   Appendices
\end{flushleft}

\section{Quantum P\"oschl--Teller with trivial ordering}\label{appendix:flat}
In this Appendix, we quantize the P\"oschl--Teller Hamiltonian \eqref{eq:Phi_h^2} adopting a trivial operator ordering of the canonical variables. We obtain the corresponding eigenfunctions and use them to compute the partition function of the model, then we discuss why these results are less satisfactory than those obtained with the operator ordering of Section
\ref{sec:canonical_quantization} of the main text. 

With this understanding, we begin by considering the straightforward quantization of the classical Hamiltonian \eqref{eq:Phi_h^2},
\begin{equation}\label{eq:Poschl-Teller2}
    \hat{\Phi}_h^2(\hat{L},\hat{P}) = \hat{P}^2 + \frac{\Phi_b^2}{\cosh^2(\hat{L}/2)} ,
\end{equation}
which preserves the form of its classical counterpart.

For notational convenience, we rescale $\Phi_h$ and $\Phi_b$ by Planck's constant and introduce
\begin{equation}\label{eq:parameters_s_delta}
    s \equiv \frac{\Phi_h}{\hbar},
    \qquad
    \frac{\Phi_b^2}{\hbar^2}
    \equiv
    \lambda^2 + \frac{1}{16}.
\end{equation}
Furthermore, we set $\hbar=1$ for the remainder of this appendix. With these choices, the time-independent Schr\"odinger equation associated with \eqref{eq:Poschl-Teller2} becomes
\begin{equation}\label{eq:Schroedinger_equation2}
    - \dv[2]{\psi_s(L)}{L}
    + \frac{\lambda^2 + \frac{1}{16}}
           {\cosh^2(L/2)}
      \psi_s(L)
    =
    s^2 \, \psi_s(L),
\end{equation}
where $\psi_s(L)=\braket{L}{s}$ denotes the energy eigenfunction.

For fixed values of the parameters, two independent solutions of \eqref{eq:Schroedinger_equation2} are given by the associated Legendre functions
\begin{equation}\label{eq:Legendre_function}
    \begin{split}
    P^{\pm 2is}_{-\frac{1}{2}+2i\lambda}\!\left(\tanh\textstyle{\frac{L}{2}}\right)
    &=
    \frac{e^{\pm i s L}}{\Gamma(1 \mp 2is)} \,
    {}_2F_1\!\left(
        \frac{1}{2} - 2i\lambda,
        \frac{1}{2} + 2i\lambda,
        1 \mp 2is;
        \frac{1-\tanh\textstyle{\frac{L}{2}}}{2}
    \right).
    \end{split}
\end{equation}
For a compendium of properties of these special functions see, for instance, Section 3.4 of \cite{bateman_1953_cnd32-h9x80} and Section 4.3 of \cite{doi:10.1137/1009129}.
In the present case, the spectral problem \eqref{eq:Schroedinger_equation2} is supplemented by the physical requirement that $\psi_s(L \leq 0)=0$, following from the interpretation of $L$ as a geodesic length. This condition forces us to pick the combination of Legendre functions \eqref{eq:Legendre_function} that vanishes at the origin:
\begin{equation}\label{eq:eigenfunction_Legendre}
    \psi_s(L)
    =
    P^{2is}_{-\frac{1}{2}+2 i \lambda}\!\left( \tanh\textstyle{\frac{L}{2}} \right) - e^{i \eta(s,\lambda)} \, P^{-2is}_{-\frac{1}{2}+2 i \lambda}\!\left( \tanh\textstyle{\frac{L}{2}} \right),
\end{equation}
where the phase is defined by
\begin{equation}\label{eq:phase}
    e^{i \eta(s,\lambda)}
    \equiv
    \frac{P^{2is}_{-\frac{1}{2}+2 i \lambda}(0)}{P^{-2is}_{-\frac{1}{2}+2 i \lambda}(0)}
    =
    2^{4is} \frac{\Gamma\!\left(\textstyle{\frac{3}{4}} + i \lambda + is\right) \Gamma\!\left(\textstyle{\frac{3}{4}} - i \lambda + is\right)}{\Gamma\!\left(\textstyle{\frac{3}{4}} - i \lambda - is\right) \Gamma\!\left(\textstyle{\frac{3}{4}} + i \lambda - is\right)}.
    %=
    %\frac{\cosh(2 \pi \lambda) + i \sinh(2 \pi s)}{\pi} \, \Gamma\!\left( \textstyle{\frac{1}{2}} + 2i(s + \lambda) \right) \Gamma\!\left( \textstyle{\frac{1}{2}} + 2i(s - \lambda) \right).
\end{equation}

Since the Legendre functions we are considering are of purely imaginary order and the real part of their degree is $-\frac{1}{2}$, the wavefunctions \eqref{eq:eigenfunction_Legendre} are odd functions of $L$ and obey the orthogonality condition \cite{sbielski2013,feldbrugge2023orthogonalityrelationsconicalfunctions}
% \begin{equation}\label{eq:orthogonality_Legendre}
%     \begin{split}
%         \int_{0}^{\infty} \dd L \, P^{2is}_{-\frac{1}{2}+2i\lambda}\!\left(\tanh\textstyle{\frac{L}{2}}\right) \, &P^{-2is'}_{-\frac{1}{2}+2i\lambda}\!\left(\tanh\textstyle{\frac{L}{2}}\right)
%         = \bigg( \cosh(4 \pi s) + \cosh(4 \pi \lambda) \bigg) \frac{\delta(s-s')}{4 s \sinh(2 \pi s)} + \\
%         &+ \frac{2 \pi \cosh(2 \pi \lambda)}{\Gamma\!\left(\frac{1}{2} - 2i(\lambda+s)\right) \Gamma\!\left(\frac{1}{2} + 2i(\lambda-s)\right)} \frac{\delta(s+s')}{4 s \sinh(2 \pi s)}.
%     \end{split}
% \end{equation}
% Using \eqref{eq:orthogonality_Legendre} one obtains the following orthogonality condition for the eigenfunctions
\begin{equation}
    \bra{s}\ket{s'} = \int_0^\infty \dd L \, \psi_s(L) \, \psi_{s'}^*(L) = \frac{\sinh(2 \pi s)}{s} \delta(s-s'),
\end{equation}
meaning that we have the following resolution of the identity
\begin{equation}
    1 = \int_0^\infty \dd s \, \frac{s}{\sinh(2 \pi s)} \ket{s}\bra{s}.
\end{equation}
Furthermore, in order to apply our prescription \eqref{eq:prescription_L=0}, we mention that near $L=0$
\begin{equation}
    \psi_s(L) = L \frac{i 2^{2 i s}}{\pi^{3/2}} \sinh(2 \pi s) \, \Gamma\!\left( \textstyle{\frac{3}{4}} + i\lambda +is \right) \Gamma\!\left( \textstyle{\frac{3}{4}} - i\lambda +is \right) + O(L^2).
\end{equation}

As a sanity check of our computation, we can take the large cutoff limit of \eqref{eq:eigenfunction_Legendre}: sending $\lambda \to \infty$ while keeping the renormalized length \eqref{eq:geodesic_length_renormalized} fixed yields
\begin{equation}
    \psi_s(L)
    \overset{\Phi_b \to \infty}{\sim}
    \frac{2i}{\pi}
    (2 \lambda)^{2is}
    \sinh(2 \pi s) \,
    K_{2is}\!\left(2 e^{-L_{\text{ren}}/2}\right)
\end{equation}
where $L_{\text{ren}}\in(-\infty,+\infty)$. This precisely reproduces the large cutoff wavefunction of the universe obtained from the Hamiltonian quantization of ordinary JT gravity \cite{Harlow:2018tqv}.

Finally, employing the prescription introduced in Section \ref{sec:quantum_observables}, we obtain the following expression for the finite cutoff partition function:\footnote{Here we are keeping the vanishing prefactor $\lim_{L \to 0} L^2$, yet it is not physically relevant for the same arguments presented in Section \ref{sec:quantum_observables} and can thus be ignored.}
\begin{equation}\label{eq:partition_function2}
\begin{split}
        Z(\beta)
        &=
        \left( \lim_{L \to 0} \bra{L} \right) \hat{\Pi} e^{2 \pi \nu - \beta \hat{H}} \hat{\Pi} \left( \lim_{L' \to 0} \ket{L'} \right)\\
        % &=
        % \int_0^{\Phi_h} \dd s \,
        % \lim_{L \to 0}\abs{\psi_s(L)}^2 \, \frac{s}{\sinh(2 \pi s)}
        % e^{-\beta (\sqrt{\lambda^2 + \frac{1}{16}} - \sqrt{\lambda^2 - s^2 + \frac{1}{16}})}\\
        &=
        \left(\lim_{L\to 0} L^2 \right) \, e^{2 \pi \nu}
        \int_0^{\nu} \diff{s}\,
        \frac{s\sinh(2\pi s)}{\pi^3} \,
        \Gamma\!\left(\textstyle{\frac{3}{4}} \pm i \sqrt{\nu^2 - \frac{1}{16}} \pm is\right)
        e^{-\beta ( \nu - \sqrt{\nu^2 - s^2})},
        \end{split}
\end{equation}
where we wrote $\lambda$ in terms of $\nu$ using \eqref{eq:s_nu_def} and \eqref{eq:parameters_s_delta}.

Now we shall compare this result \eqref{eq:partition_function2} with the non-trivial ordering partition function \eqref{eq:partition_function_Gamma}.
The Boltzmann weights are identical, and the only relevant difference comes from the $\Gamma$ functions appearing in the respective spectral densities.
In particular, since the $\Gamma$ functions in \eqref{eq:partition_function2} depend on $\nu$ only through the square root term, a new scale $\Phi_b = \hbar/4$ with no clear physical interpretation 
seems to be introduced.

Moreover, the partition function \eqref{eq:partition_function2} is compatible with the known result \eqref{eq:partition_function_large_cutoff} in the large cutoff limit \eqref{eq:large_cutoff_limit}, but its semiclassical approximation differs from that of \eqref{eq:partition_function_Gamma} at the one-loop level. Indeed, let us consider the generalized expression
\begin{equation}\label{eq:partition_function_generalized}
    Z(\beta) = e^{2 \pi \nu} \int_0^\nu \text{d}s \, \frac{s \sinh(2 \pi s)}{\pi^2} \, \Gamma\!\left( \delta \pm i \sqrt{\nu^2 - c^2} \pm i s \right) \, e^{-\beta (\nu - \sqrt{\nu^2 - s^2})},
\end{equation}
which reduces to \eqref{eq:partition_function2} for $\delta=3/4$ and $c = 1/4$, and to \eqref{eq:partition_function} for $\delta = 1/2$ and $c=0$.
Reintroducing $\hbar$ by sending $s \to s/\hbar$ and $\nu \to \nu/\hbar$ and taking $\hbar \to 0$, we can perform a saddle point approximation of \eqref{eq:partition_function_generalized} obtaining
\begin{equation}
    Z(\beta) \overset{\hbar \to 0}{\sim}
    2 \nu^{4 \delta - \frac{1}{2}} \left(\frac{\beta}{2 \pi}\right)^{4 \delta - 1} \left[ 1 + \left( \frac{\beta}{2 \pi} \right)^2\right]^{-2 \delta - \frac{1}{4}} \, e^{\nu (\sqrt{4 \pi^2 + \beta^2} - \beta)}.
\end{equation}
Of course, this one-loop result is independent of $c$, whereas its dependence on $\beta$ changes for different values of $\delta$. Consequently, we conclude that the trivial ordering \eqref{eq:Poschl-Teller2} produces a partition function that is not semiclassically compatible with the expected behavior obtained independently in \cite{Griguolo:2025kpi}, which would instead require $\delta = 1/2$ consistently with the operator ordering discussed in Section \ref{sec:canonical_quantization}.

This consideration, together with the interpretative issues mentioned above, suggests that we should instead focus on the non-trivial operator ordering \eqref{eq:hamiltonian_Phi_h^2} presented in the main text.

\section{Solutions of the Schr\"odinger equation}\label{app:solutions_of_the_Schroedinger_equation}
In this Appendix, we analytically solve the Schr\"odinger equation \eqref{eq:Schroedinger_equation} from the main text, that we reproduce here for the reader's convenience
\begin{equation}\label{eq:Schroedinger_equation_appendix}
    -\dv[2]{\psi_s(L)}{L} + \bigg( \frac{\nu^2}{\cosh^2(L/2)} - \frac{1}{4 \sinh^2(L)} - s^2 \bigg) \, \psi_s(L) = 0.
\end{equation}
Here the parameter $\nu^2 > 0$ is positive because of its relation to the boundary value of the dilaton $\nu=\Phi_b/\hbar$, while we allow the eigenvalue $s^2 \in \mathbb{R}$ to take both positive and negative values, corresponding to scattering and bound states respectively.

First, we massage Equation \eqref{eq:Schroedinger_equation_appendix} into a more convenient form by introducing the variable
\begin{equation}
    u \equiv \tanh^2(L/2),
\end{equation}
yielding $\partial_L = \sqrt{u} (1-u) \partial_u$ and
\begin{equation}\label{eq:Schroedinger_equation_u}
    -\sqrt{u} (1-u) \partial_u\!\left( \sqrt{u} (1-u) \partial_u \psi_s \right) + \left( (1-u) \nu^2 - \frac{(1-u)^2}{16 u} - s^2 \right) \psi_s = 0.
\end{equation}
Furthermore, by employing the ansatz
\begin{equation}
    \psi_s(L(u)) \equiv u^{1/4} (1-u)^{is} \, y(u),
\end{equation}
Equation \eqref{eq:Schroedinger_equation_appendix} reduces to a hypergeometric equation for $y(u)$
\begin{equation}
    u (1-u) \partial_u^2 y(u) + \big( c - (a + b + 1) u \big) \partial_u y(u) - ab \, y(u) = 0
\end{equation}
with parameters
\begin{equation}\label{eq:hypergeometric_parameters}
    a = \frac{1}{2} + is + i\nu, \qquad b = \frac{1}{2} + is - i\nu, \qquad c = 1.
\end{equation}
Sticking with the notation found in \cite{Haraoka2022ConnectionRelations}, when the hypergeometric parameter $c$ is an integer as in our case, two linearly independent solutions of the hypergeometric equation are given by
\begin{equation}
    \begin{split}
        y_1(u) &= {}_2F_1( a, b, c; u), \\
        \hat{y}_2(u) &= \log(u) \, {}_2F_1(a, b, c; u) + \sum_{n=0}^\infty \frac{\Gamma(a+n)}{\Gamma(a)} \frac{\Gamma(b+n)}{\Gamma(b)} \frac{\Gamma(c)}{\Gamma(c+n)} \frac{u^n}{n!} \times \\
        &\quad \times \bigg( \psi(a+n) - \psi(a) + \psi(b+n) - \psi(b) - 2 \psi(c+n) + 2 \psi(c) \bigg).
    \end{split}
\end{equation}
Equivalently, one can choose an alternative basis for the eigenspace of $s^2$ given by the two functions
\begin{equation}
    \begin{split}
        y_3(u) &= {}_2F_1(a,b,a+b-c+1;1-u), \\
        y_4(u) &= (1-u)^{c-a-b} {}_2F_1(c-a,c-b,c-a-b+1;1-u),
    \end{split}
\end{equation}
which are related to $y_1(u)$ and $\hat{y}_2(u)$ by the connection identities
\begin{equation}\label{eq:connection_identity}
    \begin{split}
        y_1(u) &= \frac{\Gamma(c) \Gamma(c-a-b)}{\Gamma(c-a) \Gamma(c-b)} y_3(u) + \frac{\Gamma(c) \Gamma(a+b-c)}{\Gamma(a) \Gamma(b)} y_4(u), \\
        \hat{y}_2(u) &= \frac{\Gamma(c) \Gamma(c-a-b)}{\Gamma(c-a) \Gamma(c-b)} \left( \psi(1) + \psi(c) - \psi(1-a) - \psi(1-b) \right) y_3(u) + \\
        &\quad + \frac{\Gamma(c) \Gamma(a+b-c)}{\Gamma(a) \Gamma(b)} \left( \psi(1) + \psi(c) - \psi(a) - \psi(b) \right) y_4(u).
    \end{split}
\end{equation}
The basis formed by $y_3(u)$ and $y_4(u)$ comes in handy when discussing the behavior of $y_1(u)$ and $\hat{y}_2(u)$ close to $u=1$, i.e. as $L \to \infty$.

Putting these considerations together, we learn that two linearly independent solutions of the original equation \eqref{eq:Schroedinger_equation_appendix} are given by
\begin{equation}\label{eq:eigenfunctions_non_normalized}
    \mathcal{A}_s(L) = \frac{\sqrt{\tanh(L/2)}}{\cosh(L/2)^{2is}} \, y_1\!\left(\tanh^2(L/2)\right), \qquad \mathcal{B}_s(L) = \frac{\sqrt{\tanh(L/2)}}{\cosh(L/2)^{2is}} \, \hat{y}_2\!\left(\tanh^2(L/2)\right).
\end{equation}
Using the Euler transformation for hypergeometric functions
\begin{equation}
    {}_2F_1(a,b,c;u) = (1-u)^{c-a-b} {}_2F_1(c-a,c-b,c;u),
\end{equation}
one can show that the eigenfunctions above are real
\begin{equation}
    \mathcal{A}_s^*(L) = \mathcal{A}_s(L), \qquad \mathcal{B}_s^*(L) = \mathcal{B}_s(L).
\end{equation}

Next, we discuss specific features of the solutions \eqref{eq:eigenfunctions_non_normalized}. Since their behavior changes qualitatively depending on the sign of $s^2$, i.e. on whether they are scattering or bound states, we discuss these two cases separately in what follows.

\subsection{Scattering states}
We start off by analyzing scattering solutions of the Schr\"odinger equation \eqref{eq:Schroedinger_equation_appendix}, i.e. we specialize to $s^2 \geq 0$. We label these states with $s \geq 0$. From their definition in terms of hypergeometric functions, we learn that near $L=0$ the solutions behave as
\begin{equation}\label{eq:eigenfunctions_L=0}
        \mathcal{A}_s(L) \overset{L \to 0}{\sim} \sqrt{L/2}, \qquad
        \mathcal{B}_s(L) \overset{L \to 0}{\sim} \sqrt{L/2} \, \log(L/2),
\end{equation}
implying that they both vanish at the origin, consistently with the interpretation of $L$ as a geodesic length. To discuss their behavior as $L \to \infty$, we take advantage of the connection identities \eqref{eq:connection_identity} and find
\begin{equation}\label{eq:eigenfunctions_L=infty}
    \begin{split}
        \mathcal{A}_s(L) \overset{L \to \infty}{\sim} C_{s} \, e^{isL} + C_{s}^* \, e^{-isL}, \qquad
        \mathcal{B}_s(L) \overset{L \to \infty}{\sim} \widetilde{C}_{s} \, e^{isL} + \widetilde{C}_{s}^* \, e^{-isL},
    \end{split}
\end{equation}
where we introduced the quantities
\begin{equation}
    \begin{split}
        C_{s} &\equiv \frac{4^{-is} \, \Gamma(2is)}{\Gamma(\frac{1}{2} + is + i\nu) \Gamma(\frac{1}{2} + is - i\nu)}, \\
        \widetilde{C}_{s} &\equiv \bigg( 2 \psi(1) - \psi( \textstyle{\frac{1}{2} + is + i\nu}) - \psi( \textstyle{\frac{1}{2} + is - i\nu}) \bigg) C_{s}.
    \end{split}
\end{equation}
The asymptotic plane-wave behavior \eqref{eq:eigenfunctions_L=infty} is a reflection of the potential \eqref{eq:potential_quantum} vanishing exponentially fast at large $L$. This also means that the eigenfunctions are, at most, delta-normalizable and we now aim to compute the inner product
\begin{equation}\label{eq:eigenfunctions_orthogonality_condition}
    \int_0^\infty \dd L \, \psi_s(L) \psi_{s'}(L),
\end{equation}
where $\psi_s = \mathcal{A}_s$ or $\psi_s = \mathcal{B}_s$.
Since we know that both $\mathcal{A}_s$ and $\mathcal{B}_s$ are solutions of the second order Equation \eqref{eq:Schroedinger_equation_appendix}
\begin{equation}
    D_{s} \psi_s(L) = 0, \qquad D_{s} \equiv -\frac{\dd^2}{\dd L^2} + \frac{\nu^2}{\cosh^2(L/2)} - \frac{1}{4 \sinh^2(L)} - s^2,
\end{equation}
then the following identity holds
\begin{equation}\label{eq:orthogonality_sturm_liouville}
    0 = \int_0^\infty \dd L \, \bigg( \psi_{s'}(L) \, D_{s} \psi_s(L) - \psi_s(L) \, D_{s'} \psi_{s'}(L) \bigg).
\end{equation}
Integrating by parts in Equation \eqref{eq:orthogonality_sturm_liouville}, we obtain
\begin{equation}\label{eq:orthogonality_sturm_liouville_2}
    \begin{split}
        \int_0^\infty \dd L \, \psi_{s}(L) \psi_{s'}(L)
        &= - \lim_{L \to \infty} \frac{1}{s^2-s'^2} \bigg( \psi_{s'}(L) \dv{}{L} \psi_s(L) - \psi_s(L) \dv{}{L} \psi_{s'}(L)  \bigg),
    \end{split}
\end{equation}
where only the $L\to\infty$ contribution to the integral survives since, from Equation \eqref{eq:eigenfunctions_L=0}, we know that the behavior of $\psi_s$ near the origin does not depend on $s$.
Next, we specify to the case $\psi_s = \mathcal{A}_s$. Plugging the asymptotic behavior \eqref{eq:eigenfunctions_L=infty} into the right-hand-side of \eqref{eq:orthogonality_sturm_liouville_2}, we find
\begin{equation}\label{eq:orthogonality_limit}
    \begin{split}
        \int_0^\infty \dd L \, \mathcal{A}_{s}(L) \mathcal{A}_{s'}(L)
        &= -i \lim_{L \to \infty} \bigg( \frac{e^{i(s-s')L} C_{s} C^*_{s'} - e^{-i(s-s')L} C^*_{s} C_{s'}}{s-s'} + \\
        &\quad + \frac{e^{i(s+s')L} C_{s} C_{s'} - e^{-i(s+s')L} C^*_{s} C^*_{s'}}{s+s'} \bigg).
    \end{split}
\end{equation}
Since we consider $s,s' \geq 0$, the second term on the right-hand-side of Equation \eqref{eq:orthogonality_limit} vanishes in the distributional sense due to the presence of complex exponentials. Instead, the first term reduces to a delta function because of the identity
\begin{equation}
    \lim_{L \to \infty} \frac{\sin((s-s')L)}{s-s'} = \pi \, \delta(s-s').
\end{equation}
In summary, the regular eigenfunctions satisfy
\begin{equation}
    \int_0^\infty \dd L \, \mathcal{A}_{s}(L) \mathcal{A}_{s'}(L) = \frac{\delta(s-s')}{\rho_\mathcal{A}(s)}, \qquad  \rho_\mathcal{A}(s) = \frac{1}{2 \pi \abs{C_s}^2} = \frac{2 s \sinh(2 \pi s)}{\cosh(2 \pi s) + \cosh(2 \pi \nu)}.
\end{equation}
Replacing $C_s$ with $\widetilde{C}_s$ leads to an analogous result for the non-regular eigenfunctions:
\begin{equation}
    \int_0^\infty \dd L \, \mathcal{B}_{s}(L) \mathcal{B}_{s'}(L) = \frac{\delta(s-s')}{\rho_\mathcal{B}(s)},
\end{equation}
where
\begin{equation}
    \rho_\mathcal{B}(s) = \frac{1}{2 \pi \abs{\widetilde{C}_s}^2} = \frac{1}{\abs{2 \psi(1) - \psi( \textstyle{\frac{1}{2} + is + i\nu}) - \psi( \textstyle{\frac{1}{2} + is - i\nu)}}^2} \frac{2 s \sinh(2 \pi s)}{\cosh(2 \pi s) + \cosh(2 \pi \nu)}.
\end{equation}

\subsection{Bound states}\label{app:bound_states}
The Schr\"odinger equation \eqref{eq:Schroedinger_equation_appendix} also admits bound states solutions corresponding to negative values of $s^2 < 0$. We label these states as $s = -ik$, with $k \geq 0$.
From their definitions \eqref{eq:eigenfunctions_non_normalized} we see that the behavior at the origin does not depend on $s$ being real or imaginary, hence bound state eigenfunctions vanish at $L=0$ as well
\begin{equation}
    \begin{split}
        \mathcal{A}_{-ik}(L) &\overset{L \to 0}{\sim} \sqrt{L/2}, \\
        \mathcal{B}_{-ik}(L) &\overset{L \to 0}{\sim} \sqrt{L/2} \, \log(L/2).
    \end{split}
\end{equation}
Borrowing the result for the asymptotic behavior from the previous section \eqref{eq:eigenfunctions_L=infty} and setting $s=-ik$, we find that as $L \to \infty$ the eigenfunctions exhibit both a decaying and a growing mode
\begin{equation}\label{eq:A_bound_state_L_infty}
    \mathcal{A}_{-ik}(L) \overset{L \to \infty}{\sim} C_{-ik} \, e^{kL} + C_{ik} \, e^{-kL},
\end{equation}
\begin{equation}\label{eq:B_bound_state_L_infty}
    \mathcal{B}_{-ik}(L) \overset{L \to \infty}{\sim} \widetilde{C}_{-ik} \, e^{kL} + \widetilde{C}_{ik} \, e^{-kL}.
\end{equation}
Requiring normalizability thus forces us to suppress the growing modes, which can be achieved only when the label $k$ is such that the corresponding coefficients $C_{-ik}$ or $\widetilde{C}_{-ik}$ vanish.\footnote{One might think that requiring normalizability forces us to pick the linear combination of eigenfunctions fine-tuned to kill the exponentially growing mode at $L=\infty$. Actually, we do not have this freedom since the choice of self-adjoint extension of the operator already fixes the relative coefficient to $\mathcal{A}_{-ik}$ and $\mathcal{B}_{-ik}$.}

For the eigenfunctions $\mathcal{A}_{-ik}$ \eqref{eq:A_bound_state_L_infty}, the only way to meet this condition would be for one of the Gamma functions in the denominator of $C_{-ik}$ to have a pole: since $\Gamma$'s only have poles at non-positive integers and for us $k \geq 0$, there is no way to do so. We thus conclude that this class of eigenfunctions cannot describe bound states.

Instead, for the eigenfunctions $\mathcal{B}_{-ik}$ \eqref{eq:B_bound_state_L_infty}, requiring $\widetilde{C}_{-ik}$ to vanish translates into setting to zero the combination of digamma functions that appears in its definition, i.e. normalizable bound states only exist for values of $k$ such that
\begin{equation}\label{eq:bound_states_condition}
    \text{Re}\bigg[ \psi( \textstyle{\frac{1}{2} + k + i\nu}) \bigg] - \psi(1) = 0.
\end{equation}
The number of solutions of \eqref{eq:bound_states_condition} depends on $\nu$: via numerical analysis, we checked that a bound state is present only if $\nu < \nu_0$ for $\nu_0 \approx 0.638$, whereas no bound states are found if $\nu$ is larger than that threshold. We thus conclude that only the eigenfunctions $\mathcal{B}_{-ik}$, which behave as $\mathcal{B}_{-ik} \overset{L \to 0}{\sim} \sqrt{L/2} \log(\sqrt{L/2})$ near the origin, are compatible with the presence of a bound state in the spectrum of the theory, and only for small enough $\nu$.

\section{Asymptotics of the wavefunction}
In this Appendix, we compute the asymptotic forms of the eigenfunctions \eqref{eq:eigenfunction}, which we reproduce here for convenience
\begin{equation}\label{eq:eigenfunction_appendix}
\begin{split}
    \psi_{s}(L)
    = \frac{\sqrt{\tanh(L/2)}}{\cosh(L/2)^{2is}} \, _2F_1\!\left( \frac{1}{2} + is+ i\nu, \frac{1}{2} + is - i\nu, 1 ; \tanh^2(L/2)  \right).
\end{split}
\end{equation}

\subsection{Large cutoff asymptotics}\label{app:large_cutoff_asymptotics}
We now discuss the behavior of the eigenfunction \eqref{eq:eigenfunction_appendix} in the limit where $\Phi_b \to \infty$ while the renormalized length $L_\text{ren} \equiv L - 2 \log(2 \Phi_b)$ is kept finite.
We start by employing the Mellin--Barnes representation of the hypergeometric function (see e.g. \cite{whittakerwatson1927}, Section 14.35)
\begin{equation}\label{eq:2F1_Mellin_Barnes}
    _2F_1(a,b,c;u) = \frac{\Gamma(c)}{\Gamma(a) \Gamma(b) \Gamma(c-a) \Gamma(c-b)} \int_{-i\infty}^{+i\infty} \frac{\dd t}{2 \pi i} (1-u)^{-t} \Gamma(t) \Gamma(c-a-b+t) \Gamma(a-t) \Gamma(b-t).
\end{equation}
Applying Equation \eqref{eq:2F1_Mellin_Barnes} to our values of the parameters
\begin{equation}
    \begin{split}
        a &= \frac{1}{2} + is + i\nu, \quad b = \frac{1}{2} + is - i\nu, \quad c = 1,\\
        1-u &= 1-\tanh^2{L/2} = \frac{1}{\cosh^2(L/2)} \overset{\Phi_b \to \infty}{\sim} \left( \Phi_b \, e^{L_\text{ren}/2} \right)^{-2}
    \end{split}
\end{equation}
and shifting the integration variable $t \to t + is$, we obtain
\begin{equation}
    \begin{split}
        _2F_1(a,b,c;u)
        = \frac{\cosh(L/2)^{2is}}{\Gamma(\textstyle{\frac{1}{2}} \pm is \pm i \nu)} \int_{-i\infty}^{+i\infty} \frac{\dd t}{2 \pi i} \cosh(L/2)^{2t} \, \Gamma(t \pm is) \, \Gamma(\textstyle{\frac{1}{2}} - t \pm i\nu).
    \end{split}
\end{equation}
Using Stirling's approximation for the Gamma function $\Gamma(z) \overset{\abs{z}\to \infty}{\sim} \sqrt{2\pi} z^{z-1/2} e^{-z}$, we have
\begin{equation}\label{eq:eigenfunction_large_cutoff_pre}
    \psi^\text{JT}_s(L_\text{ren})
    \equiv
    \lim_{\Phi_b \to \infty} \psi_s(L)
    =
    \frac{e^{\pi\nu}}{2 \pi} \int_{-i\infty}^{+i\infty} \frac{\dd t}{2 \pi i} \left( e^{-L_\text{ren}/2} \right)^{-2t} \, \Gamma(t \pm is).
\end{equation}
Finally, comparing the asymptotic expression \eqref{eq:eigenfunction_large_cutoff_pre} to the Mellin--Barnes representation of the modified Bessel function of the second kind of purely imaginary order
\begin{equation}
    K_{2is}(z) = \frac{1}{2} \int_{-i\infty}^{+i\infty} \frac{\dd t}{2 \pi i} \left( \frac{z}{2} \right)^{-2t} \Gamma(t \pm is),
\end{equation}
we conclude that
\begin{equation}\label{eq:eigenfunction_large_cutoff}
    \psi^\text{JT}_s(L_\text{ren}) = \frac{e^{\pi\nu}}{\pi} K_{2is}\!\left( 2 e^{-L_\text{ren}/2} \right).
\end{equation}
We conclude by mentioning that the large cutoff eigenfunctions \eqref{eq:eigenfunction_large_cutoff} are real and obey the orthogonality relation \cite{Yakubovich2006}
\begin{equation}\label{eq:orthogonality_large_cutoff}
    \int_{-\infty}^{+\infty} \dd L_\text{ren} \, \psi^\text{JT}_{s'}(L_\text{ren}) \, \psi^\text{JT}_{s}(L_\text{ren}) = \frac{\delta(s-s')}{\rho_\text{JT}(s)},
    \qquad
    \rho_\text{JT}(s) = 4 s \, e^{-2 \pi \nu} \sinh(2 \pi s),
\end{equation}
meaning that we have the following resolution of the identity at large cutoff
\begin{equation}
    1 = \int_0^\infty \dd s \, \rho_\text{JT}(s) \ket{s}\bra{s}.
\end{equation}
As a consistency check, we notice that the large cutoff eigenfunction \eqref{eq:eigenfunction_large_cutoff} solves the eigenvalue equation corresponding to the Hamiltonian \eqref{eq:hamiltonian_Liouville}
\begin{equation}
    \hat{H}_\text{JT} \, \psi^\text{JT}_s(L_\text{ren}) = \frac{\Phi_h^2}{2 \Phi_b} \, \psi^\text{JT}_s(L_\text{ren}).
\end{equation}
Furthermore, the quantity $\rho_\text{JT}(s)$ appearing here matches the leading contribution in $\nu$ of its finite cutoff counterpart \eqref{eq:orthogonality} when expanded as in \eqref{eq:spectralexpansion}.

\subsection{Semiclassical asymptotics}\label{app:semiclassical_asymptotics}
We now discuss the asymptotic behavior of the eigenfunction \eqref{eq:eigenfunction_appendix} in the regime where both $s$ and $\nu$ are large while their ratio\footnote{In this Appendix we consider both cases $\alpha < 1$ and $\alpha > 1$. In Section \ref{subsec:boundary_behind_horizon} we discuss why the latter can be relevant, nevertheless we will learn that it does not contribute semiclassically.}
\begin{equation}
    \alpha \equiv \frac{s}{\nu} = \frac{\Phi_h}{\Phi_b}
\end{equation}
is kept finite. This regime corresponds to $\hbar \to 0$, or equivalently $G_\text{N} \to 0$.

\begin{figure}
    \centering
    \includegraphics{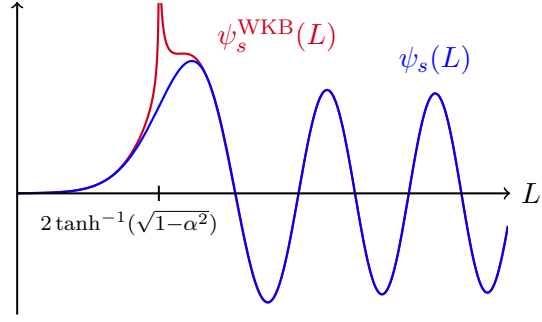}
    \caption{Comparison between the exact wavefunction \eqref{eq:eigenfunction} (blue curve) and its WKB approximation \eqref{eq:eigenfunction_WKB} (red curve). The divergence at the turning point $L = 2 \tanh^{-1}(\sqrt{1-\alpha^2})$ is an artifact due to the approximation method.}
    \label{fig:eigenfunction_WKB}
\end{figure}

To study this asymptotics, we perform a WKB approximation starting from the differential equation satisfied by the eigenfunctions \eqref{eq:Schroedinger_equation}
\begin{equation}\label{eq:Schroedinger_equation_WKB}
    \dv[2]{}{L}\psi_s(L) = \left[ \nu^2 \big( 1 - \alpha^2 - \tanh^2(L/2) \big) - \frac{1}{4 \sinh^2(L)} \right] \psi_s(L),
\end{equation}
where we conveniently expressed $s=\nu\alpha$ everywhere in order to have only one large parameter $\nu$.
As long as we are away from the turning point, i.e.
\begin{equation}\label{eq:turning_point}
    \tanh^2(L/2) \neq 1 - \alpha^2,
\end{equation}
Equation \eqref{eq:Schroedinger_equation_WKB} in the $\nu \to \infty$ limit simplifies to
\begin{equation}\label{eq:Schroedinger_equation_approximate}
    \dv[2]{}{L}\psi^\text{WKB}_s(L) = \nu^2 \big( 1 - \alpha^2 - \tanh^2(L/2) \big) \, \psi^\text{WKB}_s(L).
\end{equation}
Approximate solutions of \eqref{eq:Schroedinger_equation_approximate} at first order are of the form
\begin{equation}\label{eq:eigenfunction_approximate}
    \psi^\text{WKB}_s(L) = \frac{c_{+} \, e^{+\nu \xi_\alpha(L)} + c_{-} \, e^{-\nu \xi_\alpha(L)}}{(1 - \alpha^2 - \tanh^2(L/2))^{1/4}}, \qquad \xi_\alpha(L) \equiv \int \dd L \sqrt{1 - \alpha^2 - \tanh^2(L/2)},
\end{equation}
where the coefficients $c_\pm$ are to be fixed and the explicit form of $\xi_\alpha(L)$ reads
\begin{equation}
\begin{split}
    \xi_{\alpha<1}(L) &=
    \begin{cases}
        \displaystyle{
        2 \left( \alpha \cos^{-1}\!\left( \frac{\alpha \sinh(L/2)}{\sqrt{1-\alpha^2}} \right) - \cos^{-1}\!\left( \frac{\tanh(L/2)}{\sqrt{1-\alpha^2}} \right) \right)
        }, &\; \tanh^2(L/2)<1-\alpha^2,\\
        \displaystyle{
        2i \left( \alpha \cosh^{-1}\!\left( \frac{\alpha \sinh(L/2)}{\sqrt{1-\alpha^2}} \right) - \cosh^{-1}\!\left( \frac{\tanh(L/2)}{\sqrt{1-\alpha^2}} \right) \right)
        }, &\; \tanh^2(L/2)>1-\alpha^2,
    \end{cases}\\
    \xi_{\alpha>1}(L) &=
    2i \left( \alpha \sinh^{-1}\!\left( \frac{\alpha \sinh(L/2)}{\sqrt{\alpha^2-1}} \right) - \sinh^{-1}\!\left( \frac{\tanh(L/2)}{\sqrt{\alpha^2-1}} \right) \right).
\end{split}
\end{equation}
To conclude, we fix the coefficients $c_\pm$ in \eqref{eq:eigenfunction_approximate} by matching the behavior at $L \to \infty$ with the one of the exact eigenfunction \eqref{eq:eigenfunctions_L=infty} and by connecting the two regions across the turning point. The resulting WKB-approximate eigenfunction is plotted in Figure \ref{fig:eigenfunction_WKB} and reads
\begin{equation}\label{eq:eigenfunction_WKB}
    \psi^\text{WKB}_s(L)
    =
    \frac{\Theta(1-\alpha) e^{\pi \nu (1-\alpha)} + \Theta(\alpha-1)}{\sqrt{\pi \nu} \abs{\tanh^2(L/2) - 1 + \alpha^2}^{1/4}}
    \begin{cases}
    \displaystyle{
        \frac{e^{\nu \xi_\alpha(L)}}{2}
    }, &\; \tanh^2(L/2)<1-\alpha^2,\\
    \displaystyle{
        \cos\!\left( -i\nu\xi_\alpha(L) - \frac{\pi}{4} \right)
    }, &\; \tanh^2(L/2)>1-\alpha^2,
    \end{cases}
\end{equation}
where $\Theta$ is the Heaviside step function.
In order to compute the semiclassical approximation of observables, we also need to evaluate \eqref{eq:eigenfunction_WKB} at $L=0$, which reads
\begin{equation}\label{eq:eigenfunction_WKB_L=0}
    \psi^\text{WKB}_s(0) = \frac{\Theta(1-\alpha)}{2 \sqrt{\pi \nu} (1-\alpha^2)^{1/4}} + \frac{\Theta(\alpha-1)}{\sqrt{2 \pi \nu} (\alpha^2-1)^{1/4}}.
\end{equation}
Finally, the spectral density defined in Equation \eqref{eq:orthogonality} has the following semiclassical limit:
\begin{equation}\label{eq:spectral_density_WKB}
    \rho_\text{WKB}(s) = 2 \alpha \nu \bigg(e^{-2 \pi \nu (1-\alpha)} \Theta(1-\alpha) + \Theta(\alpha-1) \bigg).
\end{equation}

\section{Semiclassical observables}
The results from Appendix \ref{app:semiclassical_asymptotics} can be used to compute the semiclassical limit of any observable in the theory. In this Appendix, we perform a detailed computation of the semiclassical partition function and two-point function.

\subsection{Semiclassical partition function}\label{app:semiclassical_partition_function}
First, we compute the semiclassical approximation of the partition function \eqref{eq:partition_function}. To do so, we first recall that $\hbar$ can be reintroduced by sending $s \to s/\hbar$ and $\nu \to \nu/\hbar$. Then, we take $\hbar \to 0$ and find\footnote{Here the energy $E(s)$ is given by the piecewise proposal \eqref{eq:energy_piecewise}. This allows us to treat both regimes $\alpha < 1$ and $\alpha > 1$ at once.}
\begin{equation}\label{eq:partition_function_semiclassical_intermediate}
    \begin{split}
        Z(\beta)
        &= e^{2 \pi \nu} \bigg[ \int_0^\nu \dd s \, \frac{2 s \sinh(2 \pi s)}{\cosh(2 \pi s) + \cosh(2 \pi \nu)} \, e^{-\beta (\nu - \sqrt{\nu^2 - s^2})} +\\
        &\qquad\;\; + \int_\nu^\infty \dd s \, \frac{2 s \sinh(2 \pi s)}{\cosh(2 \pi s) +\cosh(2 \pi \nu)} \, e^{-\beta (\nu + \sqrt{s^2 - \nu^2})} \bigg]\\
        &\overset{\hbar \to 0}{\sim} 2 \nu^{2} \left[ \int_0^1 \dd\alpha \, \alpha \, e^{- \nu ( \beta (1 - \sqrt{1 - \alpha^2}) - 2 \pi \alpha)} + e^{2 \pi \nu} \int_1^\infty \dd\alpha \, \alpha \, e^{- \nu  \beta (1 + \sqrt{\alpha^2-1})} \right],
    \end{split}
\end{equation}
where we introduced the integration variable $\alpha=s/\nu$.
Next, we perform the saddle-point approximation of Equation \eqref{eq:partition_function_semiclassical_intermediate} as $\nu \to \infty$. Looking at the second branch, we see that the quantity in the exponent is monotonic in $\alpha$, hence its main contribution comes from $\alpha = 1$ and it is subleading with respect to the first branch.

Hence, the $\hbar \to 0$ limit of the partition function reads
\begin{equation}
    Z(\beta) \overset{\hbar \to 0}{\sim} 2 \nu^2 \int_0^1 \dd\alpha \, \alpha \, e^{-\nu f(\alpha)}, \qquad f(\alpha) = \beta (1 - \sqrt{1 - \alpha^2}) - 2 \pi \alpha.
\end{equation}
The corresponding saddle point equation $f'(\alpha_\star) = 0$ reads
\begin{equation}\label{eq:saddle_point_solution_partition_function}
    \alpha_\star = \frac{1}{\sqrt{1 + \left(  \frac{\beta}{2 \pi} \right)^2}},
\end{equation}
while the one-loop determinant is given by
\begin{equation}
    \sqrt{\frac{2 \pi}{\nu f''(\alpha_\star)}} = \nu^{-1/2} \beta \left[ 1 + \left( \frac{\beta}{2 \pi} \right)^2 \right]^{-3/4}.
\end{equation}
Putting everything together, we end up with the semiclassical partition function
\begin{equation}
    Z(\beta) \overset{\hbar \to 0}{\sim}
    2 \nu^{3/2} \frac{\beta}{2 \pi} \left[ 1 + \left( \frac{\beta}{2 \pi} \right)^2 \right]^{-5/4} \, e^{\nu \left( \sqrt{4 \pi^2 + \beta^2} - \beta \right)}
\end{equation}
at leading order.

\subsection{Semiclassical two-point function}\label{app:semiclassical_2pt_function}
In Section \ref{sec:2_point_function} we propose to compute the two-point function of JT gravity at finite cutoff as follows
\begin{equation}\label{eq:2pt_function_appendix}
\begin{split}
    \langle \mathcal{O}_\Delta(\tau_1,\tau_2) \rangle_\beta
    &=
    \frac{
    \bra{L=0}
    e^{-(\beta-\tau_2+\tau_1)\hat{H}}\,
    G_{\Delta}(\hat{L})\,
    e^{-(\tau_2 - \tau_1)\hat{H}}
    \ket{L=0}
    }
    {
    \bra{L=0} e^{-\beta\hat{H}} \ket{L=0}
    }\\
    &= \frac{e^{2 \pi \nu}}{Z(\beta)} \int_0^\infty \dd L \, G_\Delta(L) \, J_{\beta-\tau_2+\tau_1}(L) \, J^*_{\tau_2-\tau_1}(L),
\end{split}
\end{equation}
where\footnote{The integration range in \eqref{eq:2pt_function_integral} includes also the regime $s > \nu$. If we had not considered it, the conclusions would be left unchanged.}
\begin{equation}\label{eq:2pt_function_integral}
J_\tau(L) \equiv \int_0^\infty \dd s \, \rho(s) \, \psi_s(L \to 0) \, \psi^*_s(L) \, e^{-\tau E(s)}.
\end{equation}
In this appendix, we want to discuss which value $L$ provides the main contribution to the two-point function \eqref{eq:2pt_function_appendix} in the $G_\text{N} \to 0$ limit. Luckily, this task does not require us to specify the form of the operator $G_\Delta(\hat{L})$ as long as it does not depend exponentially on $1/G_\text{N}$.
The $s,\nu \to \infty$ behavior of every quantity appearing in the integral \eqref{eq:2pt_function_integral} was computed in Appendix \ref{app:semiclassical_asymptotics}: using \eqref{eq:eigenfunction_WKB}, \eqref{eq:eigenfunction_WKB_L=0} and \eqref{eq:spectral_density_WKB} leads us to\footnote{The WKB approximation used to compute the large-$\nu$ asymptotics of the eigenfunctions and the spectral density breaks down at the turning point $L=2 \cosh^{-1}(1/\alpha)$. For this reason, here we work under the assumption that the main contribution to the integral \eqref{eq:2pt_function_integral} does not come from a value of $\alpha$ close to the turning point. This assumption will only be justified a posteriori.}
\begin{equation}
    \begin{split}
        J_\tau(L) &\overset{\nu \to \infty}{\sim} \frac{\nu}{\pi} \int_0^{\frac{1}{\cosh(L/2)}} \frac{\dd\alpha \, \alpha \, e^{-\nu [\tau (1-\sqrt{1-\alpha^2}) + \pi (1-\alpha)]}}{(1-\alpha^2)^{1/4} (\frac{1}{\cosh^2(L/2)} - \alpha^2)^{1/4}} \frac{e^{\nu \xi_\alpha(L)}}{2} +\\
        &\qquad + \frac{\nu}{\pi} \int_{\frac{1}{\cosh(L/2)}}^1 \frac{\dd\alpha \, \alpha \, e^{-\nu [\tau (1-\sqrt{1-\alpha^2}) + \pi (1-\alpha)]}}{(1-\alpha^2)^{1/4} (\alpha^2 - \frac{1}{\cosh^2(L/2)})^{1/4}} \cos\!\left( -i \nu \xi_\alpha(L) - \frac{\pi}{4} \right) +\\
        &\qquad + \frac{\sqrt{2} \nu}{\pi} \int_1^\infty \frac{\dd\alpha \, \alpha \, e^{-\nu \tau (1+\sqrt{\alpha^2-1})}}{(\alpha^2-1)^{1/4} (\alpha^2 - \frac{1}{\cosh^2(L/2)})^{1/4}} \cos\!\left( -i \nu \xi_\alpha(L) - \frac{\pi}{4} \right).
    \end{split}
\end{equation}
Now we perform the saddle-point approximation of each contribution to the integral above: the saddle-point equation of the first two integrals is identical and reads
\begin{equation}\label{eq:2pt_function_saddle_point}
    \sinh(L/2) = \frac{\sqrt{1 - \alpha^2_\sharp}}{\alpha_\sharp} \abs{ \sin\!\left( \frac{\tau}{2} \frac{\alpha_\sharp}{\sqrt{1 - \alpha_\sharp^2}} \right) },
\end{equation}
whereas the one corresponding to the third integral is
\begin{equation}
    \sinh(L/2) = \pm i \frac{\sqrt{\alpha^2_\sharp-1}}{\alpha_\sharp} \abs{ \sin\!\left( \frac{\tau}{2} \frac{\alpha_\sharp}{\sqrt{\alpha_\sharp^2-1}} \right) }.
\end{equation}
Similarly to what happens in the case of the partition function \eqref{eq:partition_function_semiclassical_intermediate}, the branch $\alpha>1$ does not contribute semiclassically, as the corresponding saddle-point equation has no real solutions. Instead, the main contribution to \eqref{eq:2pt_function_integral} in the $G_\text{N} \to 0$ limit comes from the value of $\alpha = \alpha_\sharp$ implicitly defined via Equation \eqref{eq:2pt_function_saddle_point}. This means that $J^*_{\tau_2 - \tau_1}(L)$ and $J_{\beta - \tau_2 + \tau_1}(L)$ take their main semiclassical contribution from the values of $\alpha=\alpha_\sharp$ and $\alpha=\alpha_\flat$ defined, respectively, by
\begin{equation}\label{eq:saddle_solution_1}
    L = 2 \sinh^{-1}\!\left( \frac{\sqrt{1 - \alpha^2_\sharp}}{\alpha_\sharp} \abs{ \sin\!\left( \frac{\tau_2 - \tau_1}{2} \frac{\alpha_\sharp}{\sqrt{1 - \alpha_\sharp^2}} \right) } \right),
\end{equation}
\begin{equation}\label{eq:saddle_solution_2}
    L = 2 \sinh^{-1}\!\left( \frac{\sqrt{1 - \alpha^2_\flat}}{\alpha_\flat} \abs{ \sin\!\left( \frac{\beta - \tau_2 + \tau_1}{2} \frac{\alpha_\flat}{\sqrt{1 - \alpha_\flat^2}} \right) } \right).
\end{equation}
However, by looking at the full integral \eqref{eq:2pt_function_appendix}, one knows that the main contribution comes from the value of $L=L_\text{E}(\tau_2-\tau_1,\beta)$ for which both $J^*_{\tau_2-\tau_1}(L)$ and $J_{\beta-\tau_2+\tau_1}(L)$ can be made stationary simultaneously, i.e. for the same value of $L$. This requirement can be satisfied only if \eqref{eq:saddle_solution_1} and \eqref{eq:saddle_solution_2} hold at the same time, forcing upon us the condition
\begin{equation}\label{eq:saddle_condition}
	\frac{\sqrt{1-\alpha_\sharp^2}}{\alpha_\sharp} \abs{ \sin\!\left( \frac{\tau_2-\tau_1}{2} \frac{\alpha_\sharp}{\sqrt{1-\alpha_\sharp^2}} \right) } = \frac{\sqrt{1-\alpha_\flat^2}}{\alpha_\flat} \abs{ \sin\!\left( \frac{\beta-\tau_2+\tau_1}{2} \frac{\alpha_\flat}{\sqrt{1-\alpha_\flat^2}} \right) },
\end{equation}
which implies that $\alpha_\sharp$ and $\alpha_\flat$ must coincide and are given by\footnote{For $n=1$, this condition matches the saddle point equation found in the semiclassical computation of the partition function \eqref{eq:saddle_point_solution_partition_function}. For $n \geq 2$, these saddles correspond to the necklace configurations discussed in Section \ref{sec:Poschl_Teller_instantons}.}
\begin{equation}\label{eq:saddle_alpha}
	\frac{\alpha_\sharp}{\sqrt{1-\alpha_\sharp^2}} = \frac{\alpha_\flat}{\sqrt{1-\alpha_\flat^2}} = \frac{2 \pi n}{\beta}, \qquad n = 1,2,\dots.
\end{equation}
Putting all of the above considerations together and plugging \eqref{eq:saddle_alpha} into \eqref{eq:saddle_solution_1} and \eqref{eq:saddle_solution_2}, we conclude that the integral over $L$ in the $G_\text{N} \to 0$ limit picks up contributions from the saddles
\begin{equation}
	L_n(\tau_2-\tau_1,\beta) = 2 \sinh^{-1}\!\left( \frac{\beta}{2 \pi n} \abs{ \sin\!\left( \frac{\pi n}{\beta} (\tau_2-\tau_1) \right) } \right).
\end{equation}
Among all these saddles, the leading contribution comes from $n=1$, hence we conclude that the two-point function integral is dominated by
\begin{equation}
    L_\text{E}(\tau_2-\tau_1,\beta) = L_{n=1}(\tau_2-\tau_1,\beta) = 2 \sinh^{-1}\!\left( \frac{\beta}{2 \pi} \abs{ \sin\!\left( \pi \frac{\tau_2-\tau_1}{\beta} \right) } \right).
\end{equation}

\section{The Wilson function}\label{appendix:exact2point}
In this Appendix, we leverage some results from the literature \cite{Groenevelt:2003,Groenevelt:2005} in order to express the integral \eqref{eq:Delta_integral} in closed-form. As discussed in Section \ref{sec:correlation_functions}, this integral is a fundamental building block appearing in $n$-point correlation functions of JT gravity at finite cutoff.

We start by recalling the expression \eqref{eq:Delta_integral} for the reader's convenience
\begin{equation}\label{eq:2pt_function_I}
    \gamma_\Delta(s_1,s_2) = \Phi_b^{2\Delta} \int_0^\infty \dd L \, \cosh(L/2)^{-2\Delta} \, \psi_{s_1}(L) \psi_{s_2}^*(L).
\end{equation}
We now aim to massage the integral \eqref{eq:2pt_function_I} into a more convenient form. First, recall the hypergeometric function identity \begin{equation}
    {}_2F_1(a,b,c;u)=(1-u)^{-a} {}_2F_1\!\left(a,c-b,c;\frac{u}{u-1}\right).
\end{equation}
This relation allows us to express the eigenfunction \eqref{eq:eigenfunction} as
\begin{equation}
	\begin{split}
    \psi_{s}(L) = \sinh(L/2)^{1/2} \cosh(L/2)^{1/2+2i\nu} {}_2F_1(\textstyle{\frac{1}{2}} + is + i\nu, \textstyle{\frac{1}{2}} - is + i\nu, 1; -\sinh^2(L/2)).
	\end{split}
\end{equation}
Then, if we introduce the integration variable $z = \sinh^2(L/2)$, the integral \eqref{eq:2pt_function_I} becomes
\begin{equation}
    \begin{split}
	\gamma_\Delta(s_1,s_2)
    &= \Phi_b^{2\Delta} \int_0^\infty \dd z \, (1+z)^{-\Delta+2i\nu} \, {}_2F_1(\textstyle{\frac{1}{2}} + is_1 + i\nu, \textstyle{\frac{1}{2}} - is_1 + i\nu, 1; -z) \times\\
    &\quad \times {}_2F_1(\textstyle{\frac{1}{2}} + is_2 + i\nu, \textstyle{\frac{1}{2}} - is_2 + i\nu, 1; -z).
    \end{split}
\end{equation}
Integrals of this type are known in the literature (see Proposition 6.1 and Equation 6.9 in \cite{Groenevelt:2003}) and evaluate to
\begin{equation}\label{eq:Wilson_integral}
    \begin{split}
        \int_0^\infty &\dd z \, z^{2\alpha-1} \, (1+z)^{\beta+\mu-\delta} \, {}_2F_1\!\left(\begin{array}{c}
        \alpha+\mu+\gamma,\alpha+\mu-\gamma \\ 2\alpha
        \end{array} ; -z \right) \, {}_2F_1\!\left(\begin{array}{c}
        \alpha+\beta+\rho,\alpha+\beta-\rho \\ 2\alpha
        \end{array} ; -z \right) \\
        &= \Gamma(2\alpha)^2 \, \Gamma(\delta \pm \gamma \pm \rho) \, \mathbb{W}(-i\rho,-i\mu; \alpha+\gamma,\alpha-\gamma,\delta-\beta,\delta+\beta).
    \end{split}
\end{equation}
Here $\mathbb{W}(\lambda,x;a,b,c,d)$ is the Wilson function introduced in \cite{Groenevelt:2003,Groenevelt:2005}\footnote{The function $\mathbb{W}(\lambda,x;a,b,c,d)$ in \cite{Mertens:2017mtv} is related to the function $\phi_\lambda(x;a,b,c,d)$ in \cite{Groenevelt:2003,Groenevelt:2005} as $\mathbb{W}(\lambda,x;a,b,c,d) = \phi_\lambda(x;a,b,c,1-d)$.}
\begin{equation}
    \mathbb{W}(\lambda,x;a,b,c,d) = \frac{\Gamma(d-a) \, {}_4F_3\!\left( \begin{array}{c}
        a+ix,a-ix,\tilde{a}+i\lambda,\tilde{a}-i\lambda \\ a+b,a+c,1+a-d
    \end{array} ; 1 \right)}{\Gamma(a+b)\Gamma(a+c)\Gamma(d \pm ix)\Gamma(\tilde{d} \pm i\lambda)}
    + (a \leftrightarrow d),
\end{equation}
with $\tilde{a}=(a+b+c-d)/2$ and $\tilde{d}=(b+c+d-a)/2$.
Applying the result \eqref{eq:Wilson_integral} to our values of the parameters
\begin{equation}
    \alpha=1/2, \quad \beta=\mu=i\nu, \quad \gamma = is_1, \quad \rho = is_2, \quad \delta = \Delta,
\end{equation}
we obtain
\begin{equation}\label{eq:2pt_function_I_computed}
    \gamma_\Delta(s_1,s_2) = \Phi_b^{2\Delta} \, \Gamma(\Delta \pm is_1 \pm is_2) \, \mathbb{W}(s_2,\nu;\textstyle{\frac{1}{2}}+is_1, \textstyle{\frac{1}{2}}-is_1, \Delta - i\nu, \Delta + i\nu).
\end{equation}
As a consistency check, we notice that the symmetry under $s_1 \leftrightarrow s_2$ that was manifest in the initial expression \eqref{eq:2pt_function_I} can also be appreciated in the final form \eqref{eq:2pt_function_I_computed} using the invariance
\begin{equation}
    \mathbb{W}(s_1, \nu; \textstyle{\frac{1}{2}} + i s_2, \textstyle{\frac{1}{2}} - i s_2, \Delta - i \nu, \Delta + i \nu) = \mathbb{W}(s_2, \nu; \textstyle{\frac{1}{2}} + i s_1, \textstyle{\frac{1}{2}} - i s_1, \Delta - i \nu, \Delta + i \nu)
\end{equation}
following from the properties listed in \cite{Mertens:2017mtv}.
\newpage

\bibliographystyle{JHEP}
\bibliography{biblio}

@article{Moitra:2021uiv,
    author = "Moitra, Upamanyu and Sake, Sunil Kumar and Trivedi, Sandip P.",
    title = "{Jackiw-Teitelboim gravity in the second order formalism}",
    eprint = "2101.00596",
    archivePrefix = "arXiv",
    primaryClass = "hep-th",
    reportNumber = "TIFR/TH/20-53",
    doi = "10.1007/JHEP10(2021)204",
    journal = "JHEP",
    volume = "10",
    pages = "204",
    year = "2021"
}

@article{Nayak:2018qej,
    author = "Nayak, Pranjal and Shukla, Ashish and Soni, Ronak M. and Trivedi, Sandip P. and Vishal, V.",
    title = "{On the Dynamics of Near-Extremal Black Holes}",
    eprint = "1802.09547",
    archivePrefix = "arXiv",
    primaryClass = "hep-th",
    reportNumber = "TIFR/TH/17-35, TIFR-TH-17-35",
    doi = "10.1007/JHEP09(2018)048",
    journal = "JHEP",
    volume = "09",
    pages = "048",
    year = "2018"
}

@article{Stanford:2017thb,
    author = "Stanford, Douglas and Witten, Edward",
    title = "{Fermionic Localization of the Schwarzian Theory}",
    eprint = "1703.04612",
    archivePrefix = "arXiv",
    primaryClass = "hep-th",
    doi = "10.1007/JHEP10(2017)008",
    journal = "JHEP",
    volume = "10",
    pages = "008",
    year = "2017"
}

@article{Saad:2019lba,
    author = "Saad, Phil and Shenker, Stephen H. and Stanford, Douglas",
    title = "{JT gravity as a matrix integral}",
    eprint = "1903.11115",
    archivePrefix = "arXiv",
    primaryClass = "hep-th",
    month = "3",
    year = "2019"
}

@article{Maldacena:2015waa,
    author = "Maldacena, Juan and Shenker, Stephen H. and Stanford, Douglas",
    title = "{A bound on chaos}",
    eprint = "1503.01409",
    archivePrefix = "arXiv",
    primaryClass = "hep-th",
    doi = "10.1007/JHEP08(2016)106",
    journal = "JHEP",
    volume = "08",
    pages = "106",
    year = "2016"
}

@article{Iliesiu:2019xuh,
    author = "Iliesiu, Luca V. and Pufu, Silviu S. and Verlinde, Herman and Wang, Yifan",
    title = "{An exact quantization of Jackiw-Teitelboim gravity}",
    eprint = "1905.02726",
    archivePrefix = "arXiv",
    primaryClass = "hep-th",
    reportNumber = "PUPT-2584",
    doi = "10.1007/JHEP11(2019)091",
    journal = "JHEP",
    volume = "11",
    pages = "091",
    year = "2019"
}

@article{Blommaert:2018oro,
    author = "Blommaert, Andreas and Mertens, Thomas G. and Verschelde, Henri",
    title = "{The Schwarzian Theory - A Wilson Line Perspective}",
    eprint = "1806.07765",
    archivePrefix = "arXiv",
    primaryClass = "hep-th",
    doi = "10.1007/JHEP12(2018)022",
    journal = "JHEP",
    volume = "12",
    pages = "022",
    year = "2018"
}

@article{Blommaert:2018iqz,
    author = "Blommaert, Andreas and Mertens, Thomas G. and Verschelde, Henri",
    title = "{Fine Structure of Jackiw-Teitelboim Quantum Gravity}",
    eprint = "1812.00918",
    archivePrefix = "arXiv",
    primaryClass = "hep-th",
    doi = "10.1007/JHEP09(2019)066",
    journal = "JHEP",
    volume = "09",
    pages = "066",
    year = "2019"
}

@article{Alfinito:2026cky,
    author = "Alfinito, Eleonora and Beccaria, Matteo",
    title = "{Higher-loop wormhole length in sine-dilaton gravity from DSSYK Krylov complexity}",
    eprint = "2606.20220",
    archivePrefix = "arXiv",
    primaryClass = "hep-th",
    month = "6",
    year = "2026"
}

@article{Berkooz:2018jqr,
    author = "Berkooz, Micha and Isachenkov, Mikhail and Narovlansky, Vladimir and Torrents, Genis",
    title = "{Towards a full solution of the large N double-scaled SYK model}",
    eprint = "1811.02584",
    archivePrefix = "arXiv",
    primaryClass = "hep-th",
    doi = "10.1007/JHEP03(2019)079",
    journal = "JHEP",
    volume = "03",
    pages = "079",
    year = "2019"
}

@article{Berkooz:2018qkz,
    author = "Berkooz, Micha and Narayan, Prithvi and Simon, Joan",
    title = "{Chord diagrams, exact correlators in spin glasses and black hole bulk reconstruction}",
    eprint = "1806.04380",
    archivePrefix = "arXiv",
    primaryClass = "hep-th",
    doi = "10.1007/JHEP08(2018)192",
    journal = "JHEP",
    volume = "08",
    pages = "192",
    year = "2018"
}

@article{Kruthoff:2024gxc,
    author = "Kruthoff, Jorrit and Levine, Adam",
    title = "{Semi-classical dilaton gravity and the very blunt defect expansion}",
    eprint = "2402.10162",
    archivePrefix = "arXiv",
    primaryClass = "hep-th",
    doi = "10.1007/JHEP07(2025)211",
    journal = "JHEP",
    volume = "07",
    pages = "211",
    year = "2025"
}

@article{Iliesiu:2020qvm,
    author = "Iliesiu, Luca V. and Turiaci, Gustavo J.",
    title = "{The statistical mechanics of near-extremal black holes}",
    eprint = "2003.02860",
    archivePrefix = "arXiv",
    primaryClass = "hep-th",
    doi = "10.1007/JHEP05(2021)145",
    journal = "JHEP",
    volume = "05",
    pages = "145",
    year = "2021"
}

@article{Blommaert:2025rgw,
    author = "Blommaert, Andreas and Levine, Adam",
    title = "{Sphere amplitudes and observing the universe's size}",
    eprint = "2505.24633",
    archivePrefix = "arXiv",
    primaryClass = "hep-th",
    month = "5",
    year = "2025"
}

@article{Fumagalli:2024msi,
    author = "Fumagalli, Alessandro and Gorbenko, Victor and Kames-King, Joshua",
    title = "{De Sitter Bra-Ket wormholes}",
    eprint = "2408.08351",
    archivePrefix = "arXiv",
    primaryClass = "hep-th",
    doi = "10.1007/JHEP05(2025)074",
    journal = "JHEP",
    volume = "05",
    pages = "074",
    year = "2025"
}

@article{Iliesiu:2024cnh,
    author = "Iliesiu, Luca V. and Levine, Adam and Lin, Henry W. and Maxfield, Henry and Mezei, M{\'a}rk",
    title = "{On the non-perturbative bulk Hilbert space of JT gravity}",
    eprint = "2403.08696",
    archivePrefix = "arXiv",
    primaryClass = "hep-th",
    doi = "10.1007/JHEP10(2024)220",
    journal = "JHEP",
    volume = "10",
    pages = "220",
    year = "2024"
}

@article{Emparan:2025qqf,
    author = "Emparan, Roberto and Trezzi, Stefano",
    title = "{Quantum transparency of near-extremal black holes}",
    eprint = "2507.03398",
    archivePrefix = "arXiv",
    primaryClass = "hep-th",
    doi = "10.1007/JHEP10(2025)023",
    journal = "JHEP",
    volume = "10",
    pages = "023",
    year = "2025"
}

@article{Aguilar-Gutierrez:2026ogo,
    author = "Aguilar-Gutierrez, Sergio E.",
    title = "{Deforming the Double-Scaled SYK {\&} Reaching the Stretched Horizon From Finite Cutoff Holography}",
    eprint = "2602.06113",
    archivePrefix = "arXiv",
    primaryClass = "hep-th",
    doi = "10.1002/prop.70112",
    journal = "Fortsch. Phys.",
    volume = "74",
    number = "5",
    pages = "e70112",
    year = "2026"
}

@article{Castro:2022cuo,
    author = "Castro, Alejandra and Mariani, Francesca and Toldo, Chiara",
    title = "{Near-extremal limits of de Sitter black holes}",
    eprint = "2212.14356",
    archivePrefix = "arXiv",
    primaryClass = "hep-th",
    doi = "10.1007/JHEP07(2023)131",
    journal = "JHEP",
    volume = "07",
    pages = "131",
    year = "2023"
}

@article{Brown:2024ajk,
    author = "Brown, Adam R. and Iliesiu, Luca V. and Penington, Geoff and Usatyuk, Mykhaylo",
    title = "{The evaporation of charged black holes}",
    eprint = "2411.03447",
    archivePrefix = "arXiv",
    primaryClass = "hep-th",
    doi = "10.1007/JHEP01(2026)109",
    journal = "JHEP",
    volume = "01",
    pages = "109",
    year = "2026"
}

@article{Emparan:2025sao,
    author = "Emparan, Roberto",
    title = "{Quantum cross-section of near-extremal black holes}",
    eprint = "2501.17470",
    archivePrefix = "arXiv",
    primaryClass = "hep-th",
    doi = "10.1007/JHEP04(2025)122",
    journal = "JHEP",
    volume = "04",
    pages = "122",
    year = "2025"
}

@article{Tietto:2025oxn,
    author = "Tietto, Damiano and Verlinde, Herman",
    title = "{A microscopic model of de Sitter spacetime with an observer}",
    eprint = "2502.03869",
    archivePrefix = "arXiv",
    primaryClass = "hep-th",
    month = "2",
    year = "2025"
}

@article{Blommaert:2026ofx,
    author = "Blommaert, Andreas and Tietto, Damiano and Verlinde, Herman",
    title = "{An observer's quantization of 3d de Sitter}",
    eprint = "2606.26241",
    archivePrefix = "arXiv",
    primaryClass = "hep-th",
    month = "6",
    year = "2026"
}

@article{Turiaci:2020fjj,
    author = "Turiaci, Gustavo J. and Usatyuk, Mykhaylo and Weng, Wayne W.",
    title = "{2D dilaton-gravity, deformations of the minimal string, and matrix models}",
    eprint = "2011.06038",
    archivePrefix = "arXiv",
    primaryClass = "hep-th",
    doi = "10.1088/1361-6382/ac25df",
    journal = "Class. Quant. Grav.",
    volume = "38",
    number = "20",
    pages = "204001",
    year = "2021"
}

@article{Blommaert:2025avl,
    author = "Blommaert, Andreas and Levine, Adam and Mertens, Thomas G. and Papalini, Jacopo and Parmentier, Klaas",
    title = "{Wormholes, branes and finite matrices in sine dilaton gravity}",
    eprint = "2501.17091",
    archivePrefix = "arXiv",
    primaryClass = "hep-th",
    doi = "10.1007/JHEP09(2025)123",
    journal = "JHEP",
    volume = "09",
    pages = "123",
    year = "2025"
}

@article{Blommaert:2025eps,
    author = "Blommaert, Andreas and Tietto, Damiano and Verlinde, Herman",
    title = "{SYK collective field theory as complex Liouville gravity}",
    eprint = "2509.18462",
    archivePrefix = "arXiv",
    primaryClass = "hep-th",
    month = "9",
    year = "2025"
}

@article{Blommaert:2026lvp,
    author = "Blommaert, Andreas and Chen, Chang-Han",
    title = "{Time in gravitational subregions and in closed universes}",
    eprint = "2602.22153",
    archivePrefix = "arXiv",
    primaryClass = "hep-th",
    month = "2",
    year = "2026"
}

@article{Lam_2018,
   title={Shockwave S-matrix from Schwarzian quantum mechanics},
   volume={2018},
   ISSN={1029-8479},
   url={http://dx.doi.org/10.1007/JHEP11(2018)182},
   DOI={10.1007/jhep11(2018)182},
   number={11},
   journal={Journal of High Energy Physics},
   publisher={Springer Science and Business Media LLC},
   author={Lam, Ho Tat and Mertens, Thomas G. and Turiaci, Gustavo J. and Verlinde, Herman},
   year={2018},
   month=nov }

@article{Jackiw:1984je,
    author = "Jackiw, R.",
    editor = "Baier, R. and Satz, H.",
    title = "{Lower Dimensional Gravity}",
    reportNumber = "MIT-CTP-1203",
    doi = "10.1016/0550-3213(85)90448-1",
    journal = "Nucl. Phys. B",
    volume = "252",
    pages = "343--356",
    year = "1985"
}

@article{Griguolo:2023aem,
    author = "Griguolo, Luca and Guerrini, Luigi and Panerai, Rodolfo and Papalini, Jacopo and Seminara, Domenico",
    title = "{Supersymmetric localization of (higher-spin) JT gravity: a bulk perspective}",
    eprint = "2307.01274",
    archivePrefix = "arXiv",
    primaryClass = "hep-th",
    doi = "10.1007/JHEP12(2023)124",
    journal = "JHEP",
    volume = "12",
    pages = "124",
    year = "2023"
}

@article{Harlow:2018tqv,
    author = "Harlow, Daniel and Jafferis, Daniel",
    title = "{The Factorization Problem in Jackiw-Teitelboim Gravity}",
    eprint = "1804.01081",
    archivePrefix = "arXiv",
    primaryClass = "hep-th",
    doi = "10.1007/JHEP02(2020)177",
    journal = "JHEP",
    volume = "02",
    pages = "177",
    year = "2020"
}

@article{Witten:2020ert,
    author = "Witten, Edward",
    title = "{Deformations of JT Gravity and Phase Transitions}",
    eprint = "2006.03494",
    archivePrefix = "arXiv",
    primaryClass = "hep-th",
    month = "6",
    year = "2020"
}

@article{Maldacena:2019cbz,
    author = "Maldacena, Juan and Turiaci, Gustavo J. and Yang, Zhenbin",
    title = "{Two dimensional Nearly de Sitter gravity}",
    eprint = "1904.01911",
    archivePrefix = "arXiv",
    primaryClass = "hep-th",
    doi = "10.1007/JHEP01(2021)139",
    journal = "JHEP",
    volume = "01",
    pages = "139",
    year = "2021"
}

@article{Brown:1992br,
    author = "Brown, J. David and York, Jr., James W.",
    title = "{Quasilocal energy and conserved charges derived from the gravitational action}",
    eprint = "gr-qc/9209012",
    archivePrefix = "arXiv",
    reportNumber = "IFP-423-UNC, TAR-009-UNC",
    doi = "10.1103/PhysRevD.47.1407",
    journal = "Phys. Rev. D",
    volume = "47",
    pages = "1407--1419",
    year = "1993"
}

@article{doi:10.1137/1009129,
author = {Kazarinoff, Nicholas D.},
title = {Formulas and Theorems for the Special Functions of Mathematical Physics (Wilhelm Magnus, Fritz Oberhettinger and Raj Pal Soni)},
journal = {SIAM Review},
volume = {9},
number = {4},
pages = {755-755},
year = {1967},
doi = {10.1137/1009129},
URL = {https://doi.org/10.1137/1009129},
eprint = {https://doi.org/10.1137/1009129}
}

@article{sbielski2013,
	title = {Orthogonality relations for the associated Legendre functions of imaginary order},
	year = {2013},
	author = {Bielski, Sebastian},
	journal = {INTEGRAL TRANSFORMS AND SPECIAL FUNCTIONS},
	volume = {24},
	pages = {331--337},
}

@misc{feldbrugge2023orthogonalityrelationsconicalfunctions,
      title={Orthogonality relations for conical functions of imaginary order}, 
      author={Job Feldbrugge and Nynke M. D. Niezink},
      year={2023},
      eprint={2309.05616},
      archivePrefix={arXiv},
      primaryClass={math.CA},
      url={https://arxiv.org/abs/2309.05616}, 
}

@book{bateman_1953_cnd32-h9x80,
  author       = "Bateman and Erdélyi and Magnus and Oberhettinger and Tricomi",
  title        = {Higher Transcendental Functions [Volumes I-III]},
  publisher    = {McGraw-Hill Book Company},
  year         = 1953,
}

@article{Iliesiu:2020zld,
    author = "Iliesiu, Luca V. and Kruthoff, Jorrit and Turiaci, Gustavo J. and Verlinde, Herman",
    title = "{JT gravity at finite cutoff}",
    eprint = "2004.07242",
    archivePrefix = "arXiv",
    primaryClass = "hep-th",
    doi = "10.21468/SciPostPhys.9.2.023",
    journal = "SciPost Phys.",
    volume = "9",
    pages = "023",
    year = "2020"
}

@article{Stanford:2020qhm,
    author = "Stanford, Douglas and Yang, Zhenbin",
    title = "{Finite-cutoff JT gravity and self-avoiding loops}",
    eprint = "2004.08005",
    archivePrefix = "arXiv",
    primaryClass = "hep-th",
    month = "4",
    year = "2020"
}

@article{Griguolo:2025kpi,
    author = "Griguolo, Luca and Papalini, Jacopo and Russo, Lorenzo and Seminara, Domenico",
    title = "{A new perspective on dilaton gravity at finite cutoff}",
    eprint = "2512.21774",
    archivePrefix = "arXiv",
    primaryClass = "hep-th",
    month = "12",
    year = "2025"
}

@article{Bagrets:2016cdf,
    author = "Bagrets, Dmitry and Altland, Alexander and Kamenev, Alex",
    editor = "Unno, Yoshinobu and Ohsugi, Takashi and Hou, Suen and Sadrozinski, Hartmut F. -W. and Lou, Xinchou and Zhu, Hongbo and Ouyang, Qun",
    title = "{Sachdev{\textendash}Ye{\textendash}Kitaev model as Liouville quantum mechanics}",
    eprint = "1607.00694",
    archivePrefix = "arXiv",
    primaryClass = "cond-mat.str-el",
    doi = "10.1016/j.nuclphysb.2016.08.002",
    journal = "Nucl. Phys. B",
    volume = "911",
    pages = "191--205",
    year = "2016"
}

@article{Bagrets:2017pwq,
    author = "Bagrets, Dmitry and Altland, Alexander and Kamenev, Alex",
    title = "{Power-law out of time order correlation functions in the SYK model}",
    eprint = "1702.08902",
    archivePrefix = "arXiv",
    primaryClass = "cond-mat.str-el",
    doi = "10.1016/j.nuclphysb.2017.06.012",
    journal = "Nucl. Phys. B",
    volume = "921",
    pages = "727--752",
    year = "2017"
}

@article{Lin:2022zxd,
    author = "Lin, Henry W. and Maldacena, Juan and Rozenberg, Liza and Shan, Jieru",
    title = "{Looking at supersymmetric black holes for a very long time}",
    eprint = "2207.00408",
    archivePrefix = "arXiv",
    primaryClass = "hep-th",
    doi = "10.21468/SciPostPhys.14.5.128",
    journal = "SciPost Phys.",
    volume = "14",
    number = "5",
    pages = "128",
    year = "2023"
}

@article{Saad:2019pqd,
    author = "Saad, Phil",
    title = "{Late Time Correlation Functions, Baby Universes, and ETH in JT Gravity}",
    eprint = "1910.10311",
    archivePrefix = "arXiv",
    primaryClass = "hep-th",
    month = "10",
    year = "2019"
}

@article{Gorbenko:2018oov,
    author = "Gorbenko, Victor and Silverstein, Eva and Torroba, Gonzalo",
    title = "{dS/dS and $ T\overline{T} $}",
    eprint = "1811.07965",
    archivePrefix = "arXiv",
    primaryClass = "hep-th",
    doi = "10.1007/JHEP03(2019)085",
    journal = "JHEP",
    volume = "03",
    pages = "085",
    year = "2019"
}

@article{Torroba:2022jrk,
    author = "Torroba, Gonzalo",
    title = "{$ T\overline{T} $ + {\ensuremath{\Lambda}}$_{2}$ from a 2d gravity path integral}",
    eprint = "2212.04512",
    archivePrefix = "arXiv",
    primaryClass = "hep-th",
    doi = "10.1007/JHEP01(2023)163",
    journal = "JHEP",
    volume = "01",
    pages = "163",
    year = "2023"
}

@article{Chang:2025ays,
    author = "Chang, Jing-Cheng and He, Yang and Liu, Yu-Xiao and Sun, Yuan",
    title = "{Toward a unified de Sitter holography: A composite $T\bar{T}$ and $T\bar{T}+\Lambda_{2}$ flow}",
    eprint = "2511.16098",
    archivePrefix = "arXiv",
    primaryClass = "hep-th",
    doi = "10.1007/s11433-025-2918-x",
    journal = "Sci. China Phys. Mech. Astron.",
    volume = "69",
    number = "5",
    pages = "250414",
    year = "2026"
}

@article{AliAhmad:2025kki,
    author = "Ali Ahmad, Shadi and Almheiri, Ahmed and Lin, Simon",
    title = "{$T\overline{T}$ and the black hole interior}",
    eprint = "2503.19854",
    archivePrefix = "arXiv",
    primaryClass = "hep-th",
    month = "3",
    year = "2025"
}

@article{Kitaev:2017hnr,
    author = "Kitaev, Alexei",
    title = "{Notes on $\widetilde{\mathrm{SL}}(2,\mathbb{R})$ representations}",
    eprint = "1711.08169",
    archivePrefix = "arXiv",
    primaryClass = "hep-th",
    month = "11",
    year = "2017"
}

@article{Blommaert:2023opb,
    author = "Blommaert, Andreas and Mertens, Thomas G. and Yao, Shunyu",
    title = "{Dynamical actions and q-representation theory for double-scaled SYK}",
    eprint = "2306.00941",
    archivePrefix = "arXiv",
    primaryClass = "hep-th",
    doi = "10.1007/JHEP02(2024)067",
    journal = "JHEP",
    volume = "02",
    pages = "067",
    year = "2024"
}

@article{Belaey:2024dde,
    author = "Belaey, Andreas and Mariani, Francesca and Mertens, Thomas G.",
    title = "{Gravitational wavefunctions in JT supergravity}",
    eprint = "2405.09289",
    archivePrefix = "arXiv",
    primaryClass = "hep-th",
    doi = "10.1007/JHEP10(2024)037",
    journal = "JHEP",
    volume = "10",
    pages = "037",
    year = "2024"
}

@article{Mertens:2020hbs,
    author = "Mertens, Thomas G. and Turiaci, Gustavo J.",
    title = "{Liouville quantum gravity -- holography, JT and matrices}",
    eprint = "2006.07072",
    archivePrefix = "arXiv",
    primaryClass = "hep-th",
    doi = "10.1007/JHEP01(2021)073",
    journal = "JHEP",
    volume = "01",
    pages = "073",
    year = "2021"
}

@article{futurepaper,
    author = "Griguolo, Luca and Papalini, Jacopo and Russo, Lorenzo and Seminara, Domenico and Tarana, Alex",
    title = "{Dirichlet walls beyond JT gravity, to appear}",
}

@article{Belaey:2026xvh,
    author = "Belaey, Andreas and Mertens, Thomas G. and Russo, Lorenzo and Tarana, Alex",
    title = "{Holographic duals of JT gravity at finite cutoff: the classical case}",
    eprint = "2610.40391",
    archivePrefix = "arXiv",
    primaryClass = "hep-th",
    month = "10",
    year = "2026"
}

@article{Fanaras:2021awm,
    author = "Fanaras, Georgios and Vilenkin, Alexander",
    title = "{Jackiw-Teitelboim and Kantowski-Sachs quantum cosmology}",
    eprint = "2112.00919",
    archivePrefix = "arXiv",
    primaryClass = "gr-qc",
    doi = "10.1088/1475-7516/2022/03/056",
    journal = "JCAP",
    volume = "03",
    number = "03",
    pages = "056",
    year = "2022"
}

@article{Buchmuller:2024ksd,
    author = "Buchmuller, Wilfried and Hebecker, Arthur and Westphal, Alexander",
    title = "{DeWitt wave functions for de Sitter JT gravity}",
    eprint = "2412.09211",
    archivePrefix = "arXiv",
    primaryClass = "hep-th",
    reportNumber = "DESY-24-105",
    doi = "10.1007/JHEP06(2025)049",
    journal = "JHEP",
    volume = "06",
    pages = "049",
    year = "2025"
}

@article{Aguilar-Gutierrez:2024oea,
    author = "Aguilar-Gutierrez, Sergio E.",
    title = "{$T^2$ deformations in the double-scaled SYK model: Stretched horizon thermodynamics}",
    eprint = "2410.18303",
    archivePrefix = "arXiv",
    primaryClass = "hep-th",
    month = "10",
    year = "2024"
}

@article{Aguilar-Gutierrez:2026jjv,
    author = "Aguilar-Gutierrez, Sergio E. and Das, Rathindra Nath and Erdmenger, Johanna and Xian, Zhuo-Yu",
    title = "{Probing the chaos to integrability transition in double-scaled SYK}",
    eprint = "2601.09801",
    archivePrefix = "arXiv",
    primaryClass = "hep-th",
    doi = "10.1007/JHEP06(2026)109",
    journal = "JHEP",
    volume = "06",
    pages = "109",
    year = "2026"
}

@article{Gao:2021uro,
    author = "Gao, Ping and Jafferis, Daniel Louis and Kolchmeyer, David K.",
    title = "{An effective matrix model for dynamical end of the world branes in Jackiw-Teitelboim gravity}",
    eprint = "2104.01184",
    archivePrefix = "arXiv",
    primaryClass = "hep-th",
    doi = "10.1007/JHEP01(2022)038",
    journal = "JHEP",
    volume = "01",
    pages = "038",
    year = "2022"
}

@article{Heller:2025ddj,
    author = "Heller, Michal P. and Ori, Fabio and Papalini, Jacopo and Schuhmann, Tim and Wang, Meng-Ting",
    title = "{De Sitter holographic complexity from Krylov complexity in DSSYK}",
    eprint = "2510.13986",
    archivePrefix = "arXiv",
    primaryClass = "hep-th",
    month = "10",
    year = "2025"
}

@article{Cotler:2024xzz,
    author = "Cotler, Jordan and Jensen, Kristan",
    title = "{Non-perturbative de Sitter Jackiw-Teitelboim gravity}",
    eprint = "2401.01925",
    archivePrefix = "arXiv",
    primaryClass = "hep-th",
    doi = "10.1007/JHEP12(2024)016",
    journal = "JHEP",
    volume = "12",
    pages = "016",
    year = "2024"
}

@article{McGough:2016lol,
    author = "McGough, Lauren and Mezei, M{\'a}rk and Verlinde, Herman",
    title = "{Moving the CFT into the bulk with $ T\overline{T} $}",
    eprint = "1611.03470",
    archivePrefix = "arXiv",
    primaryClass = "hep-th",
    doi = "10.1007/JHEP04(2018)010",
    journal = "JHEP",
    volume = "04",
    pages = "010",
    year = "2018"
}

@article{Gross:2019ach,
    author = "Gross, David J. and Kruthoff, Jorrit and Rolph, Andrew and Shaghoulian, Edgar",
    title = "{$T\overline{T}$ in AdS$_2$ and Quantum Mechanics}",
    eprint = "1907.04873",
    archivePrefix = "arXiv",
    primaryClass = "hep-th",
    doi = "10.1103/PhysRevD.101.026011",
    journal = "Phys. Rev. D",
    volume = "101",
    number = "2",
    pages = "026011",
    year = "2020"
}

@article{Gross:2019uxi,
    author = "Gross, David J. and Kruthoff, Jorrit and Rolph, Andrew and Shaghoulian, Edgar",
    title = "{Hamiltonian deformations in quantum mechanics, $T\bar T$, and the SYK model}",
    eprint = "1912.06132",
    archivePrefix = "arXiv",
    primaryClass = "hep-th",
    doi = "10.1103/PhysRevD.102.046019",
    journal = "Phys. Rev. D",
    volume = "102",
    number = "4",
    pages = "046019",
    year = "2020"
}

@article{Chakraborty_2020,
   title={$T\overline{T}$ and $J\overline{T}$ deformations in quantum mechanics},
   volume={2020},
   ISSN={1029-8479},
   url={http://dx.doi.org/10.1007/JHEP11(2020)099},
   DOI={10.1007/jhep11(2020)099},
   number={11},
   journal={Journal of High Energy Physics},
   publisher={Springer Science and Business Media LLC},
   author={Chakraborty, Soumangsu and Mishra, Amiya},
   year={2020},
   month=Nov }

@article{Kitaev:2018wpr,
    author = "Kitaev, Alexei and Suh, S. Josephine",
    title = "{Statistical mechanics of a two-dimensional black hole}",
    eprint = "1808.07032",
    archivePrefix = "arXiv",
    primaryClass = "hep-th",
    doi = "10.1007/JHEP05(2019)198",
    journal = "JHEP",
    volume = "05",
    pages = "198",
    year = "2019"
}

@article{Yang:2018gdb,
    author = "Yang, Zhenbin",
    title = "{The Quantum Gravity Dynamics of Near Extremal Black Holes}",
    eprint = "1809.08647",
    archivePrefix = "arXiv",
    primaryClass = "hep-th",
    doi = "10.1007/JHEP05(2019)205",
    journal = "JHEP",
    volume = "05",
    pages = "205",
    year = "2019"
}

@article{Teitelboim:1983ux,
  author        = {Teitelboim, Claudio},
  title         = {{Gravitation and Hamiltonian Structure in Two Space-Time Dimensions}},
  journal       = {Phys. Lett. B},
  volume        = {126},
  pages         = {41--45},
  year          = {1983},
  doi           = {10.1016/0370-2693(83)90012-6}
}

@article{Maldacena:2016upp,
  author        = {Maldacena, Juan and Stanford, Douglas and Yang, Zhenbin},
  title         = {{Conformal symmetry and its breaking in two dimensional Nearly Anti-de-Sitter space}},
  journal       = {PTEP},
  volume        = {2016},
  number        = {12},
  pages         = {12C104},
  year          = {2016},
  eprint        = {1606.01857},
  archivePrefix = {arXiv},
  primaryClass  = {hep-th},
  doi           = {10.1093/ptep/ptw124}
}

@article{Jensen:2016pah,
  author        = {Jensen, Kristan},
  title         = {{Chaos in AdS$_2$ Holography}},
  journal       = {Phys. Rev. Lett.},
  volume        = {117},
  number        = {11},
  pages         = {111601},
  year          = {2016},
  eprint        = {1605.06098},
  archivePrefix = {arXiv},
  primaryClass  = {hep-th},
  doi           = {10.1103/PhysRevLett.117.111601}
}

@article{Engelsoy:2016xyb,
  author        = {Engelsoy, Julius and Mertens, Thomas G. and Verlinde, Herman},
  title         = {{An investigation of AdS$_2$ backreaction and holography}},
  journal       = {JHEP},
  volume        = {07},
  pages         = {139},
  year          = {2016},
  eprint        = {1606.03438},
  archivePrefix = {arXiv},
  primaryClass  = {hep-th},
  doi           = {10.1007/JHEP07(2016)139}
}

@article{Sachdev:1992fk,
  author        = {Sachdev, Subir and Ye, Jinwu},
  title         = {{Gapless spin-fluid ground state in a random quantum Heisenberg magnet}},
  journal       = {Phys. Rev. Lett.},
  volume        = {70},
  pages         = {3339--3342},
  year          = {1993},
  eprint        = {cond-mat/9212030},
  archivePrefix = {arXiv},
  doi           = {10.1103/PhysRevLett.70.3339}
}

@article{Maldacena:2016hyu,
  author        = {Maldacena, Juan and Stanford, Douglas},
  title         = {{Remarks on the Sachdev-Ye-Kitaev model}},
  journal       = {Phys. Rev. D},
  volume        = {94},
  number        = {10},
  pages         = {106002},
  year          = {2016},
  eprint        = {1604.07818},
  archivePrefix = {arXiv},
  primaryClass  = {hep-th},
  doi           = {10.1103/PhysRevD.94.106002}
}

@article{Kitaev:2017awl,
  author        = {Kitaev, Alexei and Suh, S. Josephine},
  title         = {{The soft mode in the Sachdev-Ye-Kitaev model and its gravity dual}},
  journal       = {JHEP},
  volume        = {05},
  pages         = {183},
  year          = {2018},
  eprint        = {1711.08467},
  archivePrefix = {arXiv},
  primaryClass  = {hep-th},
  doi           = {10.1007/JHEP05(2018)183}
}

@article{Stanford:2019vob,
  author        = {Stanford, Douglas and Witten, Edward},
  title         = {{JT Gravity and the Ensembles of Random Matrix Theory}},
  journal       = {Adv. Theor. Math. Phys.},
  volume        = {24},
  number        = {6},
  pages         = {1475--1680},
  year          = {2020},
  eprint        = {1907.03363},
  archivePrefix = {arXiv},
  primaryClass  = {hep-th},
  doi           = {10.4310/ATMP.2020.v24.n6.a4}
}

@article{inoue2025scattering,
  title={Scattering theory and an index theorem on the radial part of SL (2, R)},
  author={Inoue, H and Richard, Serge},
  journal={Journal of Topology and Analysis},
  volume={17},
  number={05},
  pages={1445--1479},
  year={2025},
  publisher={World Scientific}
}

@article{Goel:2020yxl,
  author        = {Goel, Akash and Iliesiu, Luca V. and Kruthoff, Jorrit and Yang, Zhenbin},
  title         = {{Classifying boundary conditions in JT gravity: from energy-branes to $\alpha$-branes}},
  journal       = {JHEP},
  volume        = {04},
  pages         = {069},
  year          = {2021},
  eprint        = {2010.12592},
  archivePrefix = {arXiv},
  primaryClass  = {hep-th},
  doi           = {10.1007/JHEP04(2021)069}
}

@article{Zamolodchikov:2004ce,
  author        = {Zamolodchikov, A. B.},
  title         = {{Expectation value of composite field $T\bar T$ in two-dimensional quantum field theory}},
  year          = {2004},
  eprint        = {hep-th/0401146},
  archivePrefix = {arXiv},
  primaryClass  = {hep-th}
}

@article{Smirnov:2016lqw,
  author        = {Smirnov, F. A. and Zamolodchikov, A. B.},
  title         = {{On space of integrable quantum field theories}},
  journal       = {Nucl. Phys. B},
  volume        = {915},
  pages         = {363--383},
  year          = {2017},
  eprint        = {1608.05499},
  archivePrefix = {arXiv},
  primaryClass  = {hep-th},
  doi           = {10.1016/j.nuclphysb.2016.12.014}
}

@article{Cavaglia:2016oda,
  author        = {Cavagli\`a, Andrea and Negro, Stefano and Sz\'ecs\'enyi, Istv\'an M. and Tateo, Roberto},
  title         = {{$T\bar T$-deformed 2D Quantum Field Theories}},
  journal       = {JHEP},
  volume        = {10},
  pages         = {112},
  year          = {2016},
  eprint        = {1608.05534},
  archivePrefix = {arXiv},
  primaryClass  = {hep-th},
  doi           = {10.1007/JHEP10(2016)112}
}

@article{Mertens:2017mtv,
  author        = {Mertens, Thomas G. and Turiaci, Gustavo J. and Verlinde, Herman L.},
  title         = {{Solving the Schwarzian via the Conformal Bootstrap}},
  journal       = {JHEP},
  volume        = {08},
  pages         = {136},
  year          = {2017},
  eprint        = {1705.08408},
  archivePrefix = {arXiv},
  primaryClass  = {hep-th},
  doi           = {10.1007/JHEP08(2017)136}
}

@article{Mertens:2018fds,
  author        = {Mertens, Thomas G.},
  title         = {{The Schwarzian Theory -- Origins}},
  journal       = {JHEP},
  volume        = {05},
  pages         = {036},
  year          = {2018},
  eprint        = {1801.09605},
  archivePrefix = {arXiv},
  primaryClass  = {hep-th},
  doi           = {10.1007/JHEP05(2018)036}
}

@book{ReedSimonII,
  author        = {Reed, Michael and Simon, Barry},
  title         = {{Methods of Modern Mathematical Physics. II: Fourier Analysis, Self-Adjointness}},
  publisher     = {Academic Press},
  address       = {New York},
  year          = {1975}
}

@article{Case:1950an,
  author        = {Case, K. M.},
  title         = {{Singular Potentials}},
  journal       = {Phys. Rev.},
  volume        = {80},
  pages         = {797--806},
  year          = {1950},
  doi           = {10.1103/PhysRev.80.797}
}

@article{Griguolo:2021wgy,
  author        = {Griguolo, Luca and Panerai, Rodolfo and Papalini, Jacopo and Seminara, Domenico},
  title         = {{Nonperturbative effects and resurgence in JT gravity at finite cutoff}},
  journal       = {Phys. Rev. D},
  volume        = {105},
  number        = {4},
  pages         = {046015},
  year          = {2022},
  eprint        = {2106.01375},
  archivePrefix = {arXiv},
  primaryClass  = {hep-th},
  doi           = {10.1103/PhysRevD.105.046015}
}

@article{Groenevelt:2003,
  author        = {Groenevelt, Wolter},
  title         = {{The Wilson function transform}},
  journal       = {Int. Math. Res. Not.},
  volume        = {2003},
  number        = {52},
  pages         = {2779--2817},
  year          = {2003},
  eprint        = {math/0306424},
  archivePrefix = {arXiv},
  doi           = {10.1155/S1073792803212128},
  url           = {https://arxiv.org/abs/math/0306424}
}

@article{Groenevelt:2005,
  author        = {Groenevelt, Wolter},
  title         = {{The Wilson function transform Related to Racah Coefficients}},
  journal       = {Acta Applicandae Mathematica},
  year          = {2005},
  volume        = {91},
  pages         = {133--191},
  eprint        = {math/0501511},
  archivePrefix = {arXiv},
  doi           = {10.1007/s10440-006-9024-7},
  url           = {https://arxiv.org/abs/math/0501511}
}

@article{Blommaert:2023wad,
  author        = {Blommaert, Andreas and Mertens, Thomas G. and Yao, Shunyu},
  title         = {{The $q$-Schwarzian and Liouville gravity}},
  journal       = {JHEP},
  volume        = {11},
  pages         = {054},
  year          = {2024},
  eprint        = {2312.00871},
  archivePrefix = {arXiv},
  primaryClass  = {hep-th},
  doi           = {10.1007/JHEP11(2024)054}
}

@article{Blommaert:2024whf,
  author        = {Blommaert, Andreas and Mertens, Thomas G. and Papalini, Jacopo},
  title         = {{The dilaton gravity hologram of double-scaled SYK}},
  journal       = {JHEP},
  volume        = {06},
  pages         = {050},
  year          = {2025},
  eprint        = {2404.03535},
  archivePrefix = {arXiv},
  primaryClass  = {hep-th},
  doi           = {10.1007/JHEP06(2025)050}
}

@article{Bossi:2024tvh,
  author        = {Bossi, Leonardo and Griguolo, Luca and Papalini, Jacopo and Russo, Lorenzo and Seminara, Domenico},
  title         = {{Sine-dilaton gravity vs double-scaled SYK: exploring one-loop quantum corrections}},
  journal       = {JHEP},
  volume        = {06},
  pages         = {152},
  year          = {2025},
  eprint        = {2411.15957},
  archivePrefix = {arXiv},
  primaryClass  = {hep-th},
  doi           = {10.1007/JHEP06(2025)152}
}

@article{Haraoka2022ConnectionRelations,
  author  = {Yoshishige Haraoka},
  title   = {Complete List of Connection Relations for Gauss Hypergeometric Differential Equation},
  journal = {Kumamoto Journal of Mathematics},
  volume   = {35},
  year     = {2022},
  pages    = {1--60}
}

@article{Louis-Martinez:1993bge,
    author = "Louis-Martinez, D. and Gegenberg, J. and Kunstatter, G.",
    title = "{Exact Dirac quantization of all 2-D dilaton gravity theories}",
    eprint = "gr-qc/9309018",
    archivePrefix = "arXiv",
    reportNumber = "WIN-93-06, UNB-93-03",
    doi = "10.1016/0370-2693(94)90463-4",
    journal = "Phys. Lett. B",
    volume = "321",
    pages = "193--198",
    year = "1994"
}

@article{Yakubovich2006,
author = "Yakubovich, Semyon B.",
title = "A distribution associated with the Kontorovich-Lebedev transform",
journal = "Opuscula Math.",
year = "2006",
volume = "26",
number = "1",
pages = "161--172",
url = "https://www.opuscula.agh.edu.pl/vol26/1/art/opuscula_math_2612.pdf",
issn = "1232-9274"
}

@book{whittakerwatson1927,
  author    = {Whittaker, E. T. and Watson, G. N.},
  title     = {A Course of Modern Analysis},
  publisher = {Cambridge University Press},
  year      = {1927},
  edition   = {4th},
  note      = {Section 14.53}
}

@article{Oliver_Coussaert_1995,
doi = {10.1088/0264-9381/12/12/012},
url = {https://doi.org/10.1088/0264-9381/12/12/012},
year = {1995},
month = {dec},
publisher = {},
volume = {12},
number = {12},
pages = {2961},
author = {Oliver Coussaert and Marc Henneaux and Peter van Driel},
title = {The asymptotic dynamics of three-dimensional Einstein gravity with a negative cosmological constant},
journal = {Classical and Quantum Gravity}
}

@article{Berkooz:2022mfk,
    author = "Berkooz, Micha and Isachenkov, Misha and Isachenkov, Mikhail and Narayan, Prithvi and Narovlansky, Vladimir",
    title = "{Quantum groups, non-commutative AdS$_{2}$, and chords in the double-scaled SYK model}",
    eprint = "2212.13668",
    archivePrefix = "arXiv",
    primaryClass = "hep-th",
    doi = "10.1007/JHEP08(2023)076",
    journal = "JHEP",
    volume = "08",
    pages = "076",
    year = "2023"
}

@article{Lin:2022rbf,
    author = "Lin, Henry W.",
    title = "{The bulk Hilbert space of double scaled SYK}",
    eprint = "2208.07032",
    archivePrefix = "arXiv",
    primaryClass = "hep-th",
    doi = "10.1007/JHEP11(2022)060",
    journal = "JHEP",
    volume = "11",
    pages = "060",
    year = "2022"
}

@article{Goel:2023svz,
    author = "Goel, Akash and Narovlansky, Vladimir and Verlinde, Herman",
    title = "{Semiclassical geometry in double-scaled SYK}",
    eprint = "2301.05732",
    archivePrefix = "arXiv",
    primaryClass = "hep-th",
    doi = "10.1007/JHEP11(2023)093",
    journal = "JHEP",
    volume = "11",
    pages = "093",
    year = "2023"
}

@article{Shyam:2021ciy,
    author = "Shyam, Vasudev",
    title = "{$ \mathrm{T}\overline{\mathrm{T}} $ + {\ensuremath{\Lambda}}$_{2}$ deformed CFT on the stretched dS$_{3}$ horizon}",
    eprint = "2106.10227",
    archivePrefix = "arXiv",
    primaryClass = "hep-th",
    doi = "10.1007/JHEP04(2022)052",
    journal = "JHEP",
    volume = "04",
    pages = "052",
    year = "2022"
}

@article{Lin:2023trc,
    author = "Lin, Henry W. and Stanford, Douglas",
    title = "{A symmetry algebra in double-scaled SYK}",
    eprint = "2307.15725",
    archivePrefix = "arXiv",
    primaryClass = "hep-th",
    doi = "10.21468/SciPostPhys.15.6.234",
    journal = "SciPost Phys.",
    volume = "15",
    number = "6",
    pages = "234",
    year = "2023"
}

@article{Almheiri:2024ayc,
    author = "Almheiri, Ahmed and Popov, Fedor K.",
    title = "{Holography on the quantum disk}",
    eprint = "2401.05575",
    archivePrefix = "arXiv",
    primaryClass = "hep-th",
    doi = "10.1007/JHEP06(2024)070",
    journal = "JHEP",
    volume = "06",
    pages = "070",
    year = "2024"
}

@article{Berkooz:2024evs,
    author = "Berkooz, Micha and Brukner, Nadav and Jia, Yiyang and Mamroud, Ohad",
    title = "{From Chaos to Integrability in Double Scaled Sachdev-Ye-Kitaev Model via a Chord Path Integral}",
    eprint = "2403.01950",
    archivePrefix = "arXiv",
    primaryClass = "hep-th",
    doi = "10.1103/PhysRevLett.133.221602",
    journal = "Phys. Rev. Lett.",
    volume = "133",
    number = "22",
    pages = "221602",
    year = "2024"
}

@article{Belaey:2025ijg,
    author = "Belaey, Andreas and Mertens, Thomas G. and Tappeiner, Thomas",
    title = "{Quantum group origins of edge states in double-scaled SYK}",
    eprint = "2503.20691",
    archivePrefix = "arXiv",
    primaryClass = "hep-th",
    month = "3",
    year = "2025"
}

@article{vanderHeijden:2025zkr,
    author = "van der Heijden, Jeremy and Verlinde, Erik and Xu, Jiuci",
    title = "{Quantum Symmetry and Geometry in Double-Scaled SYK}",
    eprint = "2511.08743",
    archivePrefix = "arXiv",
    primaryClass = "hep-th",
    month = "11",
    year = "2025"
}

@article{Schouten:2025tvn,
    author = "Schouten, Koen and Isachenkov, Mikhail",
    title = "{The von Neumann algebraic quantum group $\mathrm{SU}_q(1,1) \times \mathbb{Z}_2$ and the DSSYK model}",
    eprint = "2512.10101",
    archivePrefix = "arXiv",
    primaryClass = "math-ph",
    month = "12",
    year = "2025"
}

@article{Abdalla:2025gzn,
    author = "Abdalla, Ahmed I. and Antonini, Stefano and Iliesiu, Luca V. and Levine, Adam",
    title = "{The gravitational path integral from an observer{\textquoteright}s point of view}",
    eprint = "2501.02632",
    archivePrefix = "arXiv",
    primaryClass = "hep-th",
    doi = "10.1007/JHEP05(2025)059",
    journal = "JHEP",
    volume = "05",
    pages = "059",
    year = "2025"
}

@article{Ebert:2022ehb,
    author = "Ebert, Stephen and Ferko, Christian and Sun, Hao-Yu and Sun, Zhengdi",
    title = "{$T\bar{T}$ in JT Gravity and BF Gauge Theory}",
    eprint = "2205.07817",
    archivePrefix = "arXiv",
    primaryClass = "hep-th",
    doi = "10.21468/SciPostPhys.13.4.096",
    journal = "SciPost Phys.",
    volume = "13",
    number = "4",
    pages = "096",
    year = "2022"
}

@article{Aguilar-Gutierrez:2024nst,
    author = "Aguilar-Gutierrez, Sergio E. and Svesko, Andrew and Visser, Manus R.",
    title = "{$ \textrm{T}\overline{\textrm{T}} $ deformations from AdS$_{2}$ to dS$_{2}$}",
    eprint = "2410.18257",
    archivePrefix = "arXiv",
    primaryClass = "hep-th",
    doi = "10.1007/JHEP01(2025)120",
    journal = "JHEP",
    volume = "01",
    pages = "120",
    year = "2025"
}

@article{Callebaut:2025thw,
    author = "Callebaut, Nele and Selle, Matteo",
    title = "{Setting $T^2$ free for braneworld holography}",
    eprint = "2510.01099",
    archivePrefix = "arXiv",
    primaryClass = "hep-th",
    month = "10",
    year = "2025"
}

@article{Blacker:2024rje,
    author = "Blacker, Matthew J. and Callebaut, Nele and Hergueta, Blanca and Ning, Sirui",
    title = "{Radial canonical AdS$_{3}$ gravity and $ T\overline{T} $}",
    eprint = "2406.02508",
    archivePrefix = "arXiv",
    primaryClass = "hep-th",
    doi = "10.1007/JHEP01(2025)092",
    journal = "JHEP",
    volume = "01",
    pages = "092",
    year = "2025"
}

@article{Morone:2024ffm,
    author = "Morone, Tommaso and Negro, Stefano and Tateo, Roberto",
    title = "{Gravity and $\bar{T}$T flows in higher dimensions}",
    eprint = "2401.16400",
    archivePrefix = "arXiv",
    primaryClass = "hep-th",
    doi = "10.1016/j.nuclphysb.2024.116605",
    journal = "Nucl. Phys. B",
    volume = "1005",
    pages = "116605",
    year = "2024"
}

@article{Ferrari:2024kpz,
    author = "Ferrari, Frank",
    title = "{Jackiw-Teitelboim gravity, random disks of constant curvature, self-overlapping curves, and Liouville CFT1}",
    eprint = "2402.08052",
    archivePrefix = "arXiv",
    primaryClass = "hep-th",
    doi = "10.1103/PhysRevD.111.L061901",
    journal = "Phys. Rev. D",
    volume = "111",
    number = "6",
    pages = "L061901",
    year = "2025"
}

@article{Chaudhuri:2024yau,
    author = "Chaudhuri, Soumyadeep and Ferrari, Frank",
    title = "{Finite cut-off JT and Liouville quantum gravities on the disk at one loop}",
    eprint = "2404.03748",
    archivePrefix = "arXiv",
    primaryClass = "hep-th",
    doi = "10.1007/JHEP01(2025)160",
    journal = "JHEP",
    volume = "01",
    pages = "160",
    year = "2025"
}

@article{Betzios:2025sct,
    author = "Betzios, Panos and Papadoulaki, Olga and Zhou, Yanjun",
    title = "{Near-extremal quantum cross-section for charged fields and superradiance}",
    eprint = "2507.13896",
    archivePrefix = "arXiv",
    primaryClass = "hep-th",
    doi = "10.1007/JHEP11(2025)114",
    journal = "JHEP",
    volume = "11",
    pages = "114",
    year = "2025"
}

@misc{dhoker2002supersymmetricgaugetheoriesadscft,
      title={Supersymmetric Gauge Theories and the AdS/CFT Correspondence}, 
      author={Eric D'Hoker and Daniel Z. Freedman},
      year={2002},
      eprint={hep-th/0201253},
      archivePrefix={arXiv},
      primaryClass={hep-th},
      url={https://arxiv.org/abs/hep-th/0201253}, 
}

@article{Bhattacharyya_2026,
   title={Finite Cutoff Jackiw–Teitelboim (JT) Gravity: Baby Universes, Matrix Dual, and (Krylov) Complexity},
   volume={2026},
   ISSN={2050-3911},
   url={http://dx.doi.org/10.1093/ptep/ptag083},
   DOI={10.1093/ptep/ptag083},
   number={7},
   journal={Progress of Theoretical and Experimental Physics},
   publisher={Oxford University Press (OUP)},
   author={Bhattacharyya, Arpan and Ghosh, Saptaswa and Pal, Sounak and Vinod, Anandu},
   year={2026},
   month=May }

@article{Correa_2009,
   title={Aharonov–Bohm effect on AdS2 and nonlinear supersymmetry of reflectionless Pöschl–Teller system},
   volume={324},
   ISSN={0003-4916},
   url={http://dx.doi.org/10.1016/j.aop.2009.01.009},
   DOI={10.1016/j.aop.2009.01.009},
   number={5},
   journal={Annals of Physics},
   publisher={Elsevier BV},
   author={Correa, Francisco and Jakubský, Vít and Plyushchay, Mikhail S.},
   year={2009},
   month=May, pages={1078–1094} }
\end{document}